\documentclass[a4paper,11pt]{article}
\pdfoutput=1
\usepackage{jheppub} % for details on the use of the package, please
\usepackage{etoolbox}% http://ctan.org/pkg/etoolbox
    \makeatletter
    \patchcmd{\maketitle}{\@fpheader}{}{}{}
    \makeatother
\usepackage{amsmath,amssymb,amsthm,amscd,graphicx,amsfonts}
\usepackage{enumerate}
\usepackage{ytableau}
\usepackage{mathrsfs}
\usepackage{amsmath,amsfonts,amssymb,amsthm,graphics,graphicx,epsfig,bbm,verbatim}
\input epsf.sty
\ytableausetup{mathmode, boxsize=0.4em}
\newtheorem{theorem}{Theorem}

\theoremstyle{definition}

\newcommand{\CO}{{\cal O}}
\newcommand{\CP}{{\cal P}}

\def\cO{\mathcal{O}}
\def\IZ{{\mathbb Z}}
\def\IR{{\mathbb R}}
\def\IC{{\mathbb C}}

\def\IP{{\mathbb P}}

\def\IF{{\mathbb F}}
\def\IQ{{\mathbb Q}}

\def\eq{{\epsilon_1}}
\def\et{{\epsilon_2}}
\def\frakq{{\mathfrak{q}}}

\newcommand{\linf}{\ell_\infty}

\newcommand{\bp}{{\widehat\IP}^2}

\newcommand{\jj}{{j_L,j_R}}
\newcommand{\bd}{{\bf d}}
\newcommand{\bt}{{\bf t}}
\newcommand{\N}{{N^{\bf d}_{\jj}}}
\newcommand{\Nd}{{N^{d}_{\jj}}}
\newcommand{\aijk}{{a_{ijk}}}
\newcommand{\bins}{{b_i^{\rm NS}}}
\newcommand{\bi}{{b_i}}
\newcommand{\bR}{{\md R}}
\newcommand{\br}{{\md r}}
\newcommand{\ep}{\epsilon}

\newcommand{\ve}{\epsilon}

\newcommand{\eee}{{\ep_1+\ep_2}}
\newcommand{\re}{{\rm e}}
\newcommand{\ri}{\mathsf{i}}
\newcommand{\rd}{{\rm d}}

\renewcommand{\d}{\partial}

\def\half {{1\over 2}}

\newcommand{\mA}{\mathsf{A}}

\newcommand{\mO}{{\mathsf{O}}}
\newcommand{\mP}{\mathsf{P}}

\newcommand{\mx}{\mathsf{x}}

\newcommand{\im}{\mathsf{i}}

\newcommand{\mb}{{\mathsf{b}}}
\newcommand{\be}{\begin{equation}}
\newcommand{\ee}{\end{equation}}
\newcommand{\ba}{\begin{aligned}}
\newcommand{\ea}{\end{aligned}}
\newcommand{\ben}{\begin{eqnarray}\displaystyle}
\newcommand{\een}{\end{eqnarray}}

\newdimen\tableauside\tableauside=1.0ex
\newdimen\tableaurule\tableaurule=0.4pt
\newdimen\tableaustep
\def\phantomhrule#1{\hbox{\vbox to0pt{\hrule height\tableaurule width#1\vss}}}
\def\phantomvrule#1{\vbox{\hbox to0pt{\vrule width\tableaurule height#1\hss}}}
\def\sqr{\vbox{%
  \phantomhrule\tableaustep
  \hbox{\phantomvrule\tableaustep\kern\tableaustep\phantomvrule\tableaustep}%
  \hbox{\vbox{\phantomhrule\tableauside}\kern-\tableaurule}}}
\def\squares#1{\hbox{\count0=#1\noindent\loop\sqr
  \advance\count0 by-1 \ifnum\count0>0\repeat}}
\def\tableau#1{\vcenter{\offinterlineskip
  \tableaustep=\tableauside\advance\tableaustep by-\tableaurule
  \kern\normallineskip\hbox
    {\kern\normallineskip\vbox
      {\gettableau#1 0 }%
     \kern\normallineskip\kern\tableaurule}%
  \kern\normallineskip\kern\tableaurule}}
\def\gettableau#1{\ifnum#1=0\let\next=\null\else
\squares{#1}\let\next=\gettableau\fi\next}

\def\({\left(}
\def\){\right)}

\usepackage{amsmath,amssymb,graphicx,epstopdf,subfigure}
\newcommand{\bm}[1]{\mbox{\boldmath{$#1$}}}
\numberwithin{equation}{section}

\usepackage{mathtools}

\usepackage{caption}
\usepackage{tabulary}
\usepackage{longtable}
\usepackage{breqn}
\usepackage{xcolor}
\usepackage{indentfirst}
\usepackage{setspace}

\newcommand{\bs}{\begin{split}}
\newcommand{\es}{\end{split}}

\newcommand{\cz}{\mathbb{C}^3/\mathbb{Z}_5}

\newcommand{\bdm}{\begin{dmath*}}
\newcommand{\edm}{\end{dmath*}}
\def\mb{\mathbb}

\def\md{\mathbf}

\def\({\left(}
\def\){\right)}
\def\l{\left\{}
\def\r{\right\}}
\newcommand{\np}{{\mathrm{np}}}

\newcommand{\dB}{\mathbf{B}}

\newcommand{\dt}{\mathbf{t}}

\graphicspath{{figs/}}
\def\fq{\mathfrak{q}}
\def\q{\mathfrak{q}}

\def\({\left(}
\def\){\right)}

\def\mbb{\mathbb}
\def\mbf{\mathbf}

\newcommand{\nn}{\nonumber \\}

\allowdisplaybreaks

\newcommand{\xdownarrow}[1]{%
  {\left\downarrow\vbox to #1{}\right.\kern-\nulldelimiterspace}
}

\usepackage{tikz}
\usetikzlibrary{decorations.pathmorphing}
\usetikzlibrary{arrows,decorations.markings}
\usetikzlibrary{calc}

\preprint{USTC-ICTS-17-13}

\title{Blowup Equations for Refined Topological Strings}
\author[\S]{Min-xin Huang,}
\author[\dagger]{Kaiwen Sun,}
\author[\S]{Xin Wang}
\affiliation[\S]{Interdisciplinary Center for Theoretical Study, Department of Modern Physics, University of Science and Technology of China, 96 Jinzhai Road, Hefei, Anhui 230026, China}
\affiliation[\dagger]{International School of Advanced Studies (SISSA), via Bonomea 265, 34136 Trieste, Italy}

\emailAdd{minxin@ustc.edu.cn}
\emailAdd{ksun@sissa.it}
\emailAdd{wxin@mail.ustc.edu.cn}

\abstract{G\"{o}ttsche-Nakajima-Yoshioka K-theoretic blowup equations characterize the Nekrasov partition function of five dimensional $\mathcal{N}=1$ supersymmetric gauge theories compactified on a circle, which via geometric engineering correspond to the refined topological string theory on $SU(N)$ geometries. In this paper, we study the K-theoretic blowup equations for general local Calabi-Yau threefolds. We find that both vanishing and unity blowup equations exist for the partition function of refined topological string, and the crucial ingredients are the $\bf r$ fields introduced in our previous paper. These blowup equations are in fact the functional equations for the partition function and each of them results in infinite identities among the refined free energies. Evidences show that they can be used to determine the full refined BPS invariants of local Calabi-Yau threefolds. This serves an independent and sometimes more powerful way to compute the partition function other t
 han the refined topological vertex in the A-model and the refined holomorphic anomaly equations in the B-model. We study the modular properties of the blowup equations and provide a procedure to determine all the vanishing and unity $\bf r$ fields from the polynomial part of refined topological string at large radius point. We also find that certain form of blowup equations exist at generic loci of the moduli space.

\rule{0pt}{0pt}
\\
\rule{0pt}{0pt}

\rightline{\emph{To Sheldon Katz on his 60th anniversary}}}
\begin{document}
\maketitle
\bibliographystyle{unsrt}
%\tableofcontents
%\newpage
\section{Introduction}\label{sec:intro}
Blowup formulae originated from the attempt to understand the relation between the Donaldson invariants of a four-manifold $X$ and those of its blowup $\widehat{X}=X\#\,\overline{\mathbb{P}}^2$. Based on the pioneering works of Kronheimer-Mrowka \cite{KM} and Taubes \cite{Taubes} (see also \cite{Bryan}\cite{Ozsvath}), Fintushel and Stern proposed a concise form of the blowup formulae for the $SU(2)$ and $SO(3)$ Donaldson invariants in \cite{FS}. It is well known in Donaldson-Witten theory that the Donaldson polynomial invariants are realized as the correlation functions of certain observables in the topological twisted $\mathcal{N}=2$ supersymmetric Yang-Mills theory \cite{Witten:1988ze}. After the breakthrough of Seiberg-Witten on $\mathcal{N}=2$ gauge theories \cite{Seiberg:1994rs}\cite{Seiberg:1994aj}, the generating function for these correlators can be computed by using the low-energy exact solutions \cite{Witten:1994cg}. Therefore, the blowup formulae can be regarded as certain
  universal property of the $\mathcal{N}=2$ theories. This was extensively studied by Moore-Witten using the technique of $u$-plane integral \cite{Moore:1997pc} and soon was generalized to $SU(N)$ cases \cite{Marino:1998bm}\cite{Edelstein:2000aj}. In fact, the relation can already be seen in \cite{FS} that the Seiberg-Witten curve naturally appears in the setting of blowup formulae. Besides, the blowup formulae are also closely related to the wall-crossing of Donaldson invariants \cite{Gottsche:1996aa}\cite{Gottsche:1996aoa}, integrable (Whitham) hierarchies \cite{Takasaki:1999nv}\cite{Takasaki:1999zq}\cite{Marino:1999qk} and contact term equations \cite{Losev:1997tp}\cite{Losev:1997wp}.

In \cite{Nekrasov:2002qd}, Nekrasov formulated the four-dimensional $\mathcal{N}=2$ gauge theory on the so called Omega background $\Omega(\epsilon_1,\epsilon_2)$, which is a two-parameter deformation of $\IR^4\simeq\IC^2$. In physics, this means to turn on the graviphoton background field and $\epsilon_{R/L}=\frac{1}{2}(\eq\pm\et)$ denote the self-dual and anti-self-dual parts of the graviphoton field strength respectively. Such background breaks the Poincare symmetry but maximally preserves the supersymmetry. The partition function computable from the localization on instanton moduli space can reproduce the Seiberg-Witten prepotential at the limit $\epsilon_1,\epsilon_2\to 0$, which was conjectured by Nekrasov and independently proved by Nakajima-Yoshioka \cite{Nakajima:2003pg}, Nekrasov-Okounkov \cite{Nekrasov:2003rj} and Braverman-Etingof \cite{Braverman:2004cr} from different viewpoints. In the first approach, a generalization of the blowup formulae containing the two deformatio
 n parameters were proposed and proved, which played a crucial role to confirm Nekrasov's conjecture (see also \cite{Nakajima:2003uh}). Mathematically, the Nekrasov instanton partition function for gauge group $SU(N)$ is defined as the generating function of the integral of the equivariant cohomology class $1$ of the framed moduli space $M(N,n)$ of torsion free sheaves $E$ of $\IP^2$ with rank $N$, $c_2=n$:
\be
Z_{\rm Nek}^{\rm inst}(\eq,\et,\vec{a};\mathfrak{q})=\sum_{n=0}^{\infty}\frakq^n\int_{M(N,n)}1,
\ee
where the framing is a trivialization of the restriction of $E$ at the line at infinity $\ell_{\infty}$. On the blowup $\widehat{\IP}^2$ with exceptional divisor $C$, one can define similar partition function via the frame moduli space $\widehat{M}(N,k,n)$, where $\langle c_1(E),[C]\rangle = -k$ and $\langle c_2(E) - \frac{N-1}{2N} c_1(E)^2, [\bp]\rangle = n$. Based on the localization computation on the fixed point set of $\IC^*\times\IC^*$ in $\widehat{\IC}^2=\widehat{\IP}^2\backslash\linf$, such partition function can be represented in terms of the original Nekrasov partition function. Combining with the well-known vanishing theorem in Donaldson theory, the blowup formulae emerge as a system of equations satisfied by the Nekrasov partition function.

Lifted by a circle,  $\mathcal{N}=1$ supersymmetry gauge theories on the five-dimensional Omega background exhibit similar phenomena. The partition function here becomes K-theoretic and relates to the equivariant Donaldson invariants. The K-theoretic Nekrasov partition function is defined mathematically by replacing the integration in the equivariant cohomology by one in equivariant K-theory:
\be
Z_{\rm Nek}^{\rm inst}(\eq,\et,\vec{a};\mathfrak{q},\beta)=\sum_{n}\(\frakq\beta^{2N}e^{-N\beta(\eq+\et)/2}\)^n\sum_i(-1)^i\mathrm{ch}H^i\({M(N,n)},\mathcal{O}\),
\ee
where $\beta$ is the radius of the circle. When $\beta\to 0$, the K-theoretic partition function becomes the homological one.  It was proved in \cite{Nakajima:2005fg} that such partition function also satisfies certain blowup formulae. Besides, one can also consider the partition function with five-dimensional Chern-Simons term of which the coefficient $m=0,1,\dots,r$ \cite{Intriligator:1997pq}\cite{Tachikawa:2004ur}. The corresponding blowup formulae were conjectured in \cite{Gottsche:2006bm} and proved in \cite{Nakajima:2009qjc}, which we call the G\"{o}ttsche-Nakajima-Yoshioka K-theoretic blowup equations. Such equations are one of our starting points in this paper.

Geometric engineering connects certain supersymmetric gauge theories with the topological string theory on local Calabi-Yau manifolds,  see e.g. \cite{Katz:1996fh}\cite{Katz:1997eq}.  Such correspondence can be established on classical level $(\epsilon_1,\epsilon_2\to 0)$, unrefined level ($\eq+\et\to 0,\ \eq=g_{\rm s}$), quantum level ($\epsilon_1\to 0,\ \et=\hbar$) and refined level (generic $\epsilon_1,\epsilon_2$) \cite{Iqbal:2007ii}. Each level contains rich structures in mathematical physics. The typical example on refined level is the correspondence between the five-dimensional  $\mathcal{N}=1$ $SU(N)$ gauge theory with Chern-Simons coefficient $m$ on Omega background and the refined topological string theory on local toric Calabi-Yau threefold $X_{N,m}$, which is the resolution of the cone over the $Y^{N,m}$ singularity. The description of such geometries can be found in \cite{Brini:2008rh}. Physically, one can consider M-theory compactified on local Calabi-Yau threefold $X$ 
 with K\"{a}hler moduli $\mathbf{t}$, then the BPS particles in the five dimensional supersymmetric gauge theory arising from M2-branes wrapping the holomorphic curves within $X$. Besides the homology class $\beta\in H_2(X,\IZ)$ which can be represented by a degree vector $\bf d$, these particles are in addition classified by their spins $(j_L,j_R)$ under the five-dimensional little group $SU(2)_L\times SU(2)_R$. The multiplicities $N^{\bf d}_{\jj}$ of the BPS particles are called the refined BPS invariants. The instanton partition function of refined topological string can be obtained from the refined Schwinger-loop calculation \cite{Iqbal:2007ii}
\be\label{eq:Zref1}
 Z_{\rm ref}^{\rm inst}(\eq,\et,{\bf t})=\prod_{\jj,{\bf d}}\prod_{m_L=-J_L}^{J_L}\prod_{m_R=-J_R}^{J_R}\prod_{m_1,m_2=1}^{\infty}\(1-q_L^{m_L}q_R^{m_R}q_1^{m_1-\frac{1}{2}}q_2^{m_2-\frac{1}{2}}e^{-\bd\cdot\bt}\)^{(-1)^{2(j_L+j_R)}\N},
\ee
where $q_{1,2}=e^{\epsilon_{1,2}}$ and $q_{R/L}=e^{\epsilon_{R/L}}$. With appropriate identification of parameters, this is equivalent to the refined Pandharipande-Thomas partition function, which is rigorously defined in mathematics as the generating function of the counting of refined stable pairs on $X$ \cite{Choi:2012jz}. Recently, Maulik and Toda proposed the refined BPS invariants can also be defined using perverse sheaves of vanishing cycles \cite{Maulik:2016rip}. The basic result of geometric engineering is the equivalence between the K-theoretic Nekrasov partition function and the partition function of refined topological string, with appropriate identification among the Coulomb parameters $\vec{a}$ and the K\"{a}hler moduli $\mathbf{t}$. Therefore, the blowup formulae satisfied by the K-theoretic Nekrasov partition function can also be regarded as the functional equations of the partition function of refined topological string, at least for those local Calabi-Yau which can 
 engineer suitable supersymmetry gauge theories. The main purpose of this paper is to generalize such functional equations to arbitrary local Calabi-Yau threefolds. The geometric engineering relation between four-dimensional supersymmetric gauge theory and local Calabi-Yau is also interesting, if we consider the superstring theory compactification instead of M-theory compactification. In such cases,  one mainly deal with the Dijkgraaf-Vafa geometries \cite{Dijkgraaf:2002fc}. We expect certain blowup formulae exist as well for those geometries, but that won't be addressed in the current paper.

Another clue of the blowup formulae for general local Calabi-Yau came from the recent study on the exact quantization of mirror curves, which is within the framework of B model of topological strings. It is well known that the information of the mirror of a local Calabi-Yau threefold is encoded in a Riemann surface, called mirror curve \cite{Chiang:1999tz}. On the classical level, the B-model topological string is governed by the special geometry on the mirror curve. All physical quantities in the geometric engineered supersymmetric gauge theory such as Seiberg-Witten differential, prepotential, periods and dual periods have direct correspondences in the special geometry. On the quantum level, the high genus free energy of topological string can be computed by the holomorphic anomaly equations \cite{Bershadsky:1993cx}.  For compact Calabi-Yau threefolds, the holomorphic anomaly equations are normally not enough to determine the full partition function due to the holomorphic ambiguiti
 es, while for local Calabi-Yau, new symmetry emerges whose Ward identities are sufficient to completely determine the partition function at all genera. This is based on the observation on the relations among quantum mirror curves, topological strings and integrable hierarchies \cite{Aganagic:2003qj}. The appearance of integrable hierarchies here is not surprising since the correspondence between the four-dimensional $\mathcal{N}=2$ gauge theories and integrable systems have been proposed in \cite{Gorsky:1995zq}\cite{Martinec:1995by} and well studied in 1990s, see for example \cite{DHoker:1999yni}. One can regard the relation web in the context of local Calabi-Yau as certain generalization. In mathematics, the using of mirror curve to construct the B-model partition function on local Calabi-Yau is usually called Eynard-Orantin topological recursion \cite{Eynard:2007kz} or BKMP remodeling conjecture \cite{Bouchard:2007ys}, which was rigorously proved in \cite{Eynard:2012nj}.

In \cite{Nekrasov:2009rc}, Nekrasov-Shatashvili studied the chiral limit ($\epsilon_1\to 0,\ \et=\hbar$) and found that the quantization of the underlying integrable systems is governed by the supersymmetric gauge theories under such limit. Here the Nekrasov-Shatashvili free energy (effective twisted superpotential) which is the chiral limit of Nekrasov partition function serves as the Yang-Yang function of the quantum integrable systems while the supersymmetric vacua become the eigenstates and the supersymmetric vacua equations become the thermodynamic Bethe ansatz. Mathematically, this equates quantum K-theory of a Nakajima quiver variety with Bethe equations for a certain quantum affine Lie algebra. Via geometric engineering, such correspondence can be rephrased as a direct relation between the quantum phase volumes of the mirror curve of a local Calabi-Yau and the Nekrasov-Shatashvili free energy of topological string. Now the Bethe ansatz is just the traditional Bohr-Sommerfeld 
 or EBK quantization conditions for the mirror curves \cite{Aganagic:2011mi}. For certain local toric Calabi-Yau, the topological string theory is directly related to five-dimensional gauge theory. In five dimension, certain non-perturbative contributions need to be added to the original formalism of NS quantization conditions. This was first noticed in \cite{Kallen:2013qla}. The exact NS quantization conditions were proposed in \cite{Wang:2015wdy} for toric Calabi-Yau with genus-one mirror curve, and soon were generalized to arbitrary toric cases in \cite{Franco:2015rnr}. The exact NS quantization conditions were later derived in \cite{Sun:2016obh}\cite{Hatsuda:2015fxa} by replacing the original partition function to the Lockhart-Vafa partition function of non-perturbative topological string \cite{Lockhart:2012vp} with the knowledge of quantum mirror map \cite{Aganagic:2011mi} and the property of $\bf B$ field \cite{Hatsuda:2013oxa}. On the other hand, Grassi-Hatsuda-Mari\~no propos
 ed an entirely different approach to exactly quantize the mirror curve \cite{Grassi:2014zfa}. This approach takes root in the study on the non
 -perturbative effects in ABJM theories on three sphere, which is dual to topological string on local Hirzebruch surface $\mathbb{F}_0=\mathbb{P}^1\times \mathbb{P}^1$ \cite{Marino:2009jd}. The equivalence between the two quantization approaches was established in \cite{Sun:2016obh} by introducing the $\bf r$ fields and certain compatibility formulae which are constraint equations for the refined free energy of topological string. It was later realized in \cite{Grassi:2016nnt} that for $SU(N)$ geometries $X_{N,m}$ such compatibility formulae were exactly the Nekrasov-Shatashvili limit of some G\"{o}ttsche-Nakajima-Yoshioka K-theoretic blowup equations. This inspired that the constraint equations in \cite{Sun:2016obh} should be able to generalize to refined level, as was proposed in \cite{Gu:2017ccq} and called generalized blowup equations.  It was also shown in \cite{Gu:2017ccq} that the partition function of E-string theory which is equivalent to the refined topological string on lo
 cal half K3 satisfies the generalized blowup equations. This suggests blowup formulae should exist for non-toric Calabi-Yau as well.

The G\"{o}ttsche-Nakajima-Yoshioka K-theoretic blowup equations can be divided to two sets of equations. Roughly speaking, the equations in one set indicate that certain infinite bilinear summations of Nekrasov partition function vanish, those in the other set indicate that certain other infinite bilinear summations result in the Nekrasov partition function itself. The former set of equations was generalized to the refined topological string on generic local Calabi-Yau in \cite{Gu:2017ccq}, which we call the vanishing blowup equations in this paper. The latter set of equations will be generalized in this paper, which we call the unity blowup equations. The main goal of this paper is to detailedly study these two types of blowup equations.

Now we are on the edge to present the blowup equations for general local Calabi-Yau threefolds. The full partition function of refined topological string $Z_{\rm ref}(\eq,\et;{\md t})$ is the product of the instanton partition function (\ref{eq:Zref1}) and the perturbative contributions which will be given in (\ref{eq:F-pert}). To make contact with the quantization of mirror curve, we also need to make certain twist to the original partition function, denoted as $\widehat{Z}_{\rm ref}(\eq,\et;{\md t})$. Such twist which will be defined in (\ref{eq:twist}) does not lose any information of the partition function, in particular the refined BPS invariants. The blowup equations are the functional equations of the twisted partition function of refined topological string.

The main result of this paper is as follows: \emph{For an arbitrary local Calabi-Yau threefold $X$ with mirror curve of genus $g$, suppose there are $b={\rm dim}H_2(X,\IZ)$ irreducible curve classes corresponding to K\"{a}hler moduli $\mathbf{t}$ in which $b-g$ classes correspond to mass parameters $\md m$, and denote $\bf C$ as the intersection matrix between the $b$ curve classes and the $g$ irreducible compact divisor classes, then there exist infinite constant integral vectors ${\bf r}\in\IZ^{b}$ such that the following functional equations for the twisted partition function of refined topological string on $X$ hold:}
\be\label{eq:blowupZ}
\ba
	\sum_{\md n \in \mb Z^{g}}  (-1)^{|\md n|}\ &\widehat{Z}_{\rm ref}\(\eq,\et-\eq; \md t+ \eq\bR\)\cdot\widehat{Z}_{\rm ref}\(\eq-\et,\et; \md t+ \et\bR\) \\
&=\begin{cases}
	0, & \text{for}\ \md r\in\mathcal{S}_{\mathrm{vanish}}, \\
	\Lambda(\eq,\et;{\bf m},{\bf r})\widehat{Z}_{\rm ref}\(\eq,\et;\md t\), \quad\quad & \text{for}\ \md r\in\mathcal{S}_{\mathrm{unity}}, \\
\end{cases}
\ea
\ee
%\emph{Here the $\lambda$ is simple normalization factor of form}
%\be
%\sum_i\exp\(c_i(\eq+\et)+c_{ij}t_j\)
%\ee
\emph{where $|\md n|=\sum_{i=1}^{g} n_i$, $\bR=\md C\cdot \md n + \mathbf{r}/2$ and $\Lambda$ is a simple factor that is independent from the true moduli and purely determined by the polynomial part of the refined free energy. In addition, all the vector $\mathbf{r}$ are the representatives of the $\mathbf{B}$ field of $X$, which means for all triples of degree ${\bf d}$, spin $j_L$ and $j_R$ such that the refined BPS invariants $N^{{\bf d}}_{j_L, j_R}(X) $ is non-vanishing, they must satisfy}
\be
\label{eq:rcondition}
(-1)^{2j_L + 2 j_R-1}= (-1)^{{\bf r} \cdot {\bf d}}.
\ee
\emph{Besides, both sets $\mathcal{S}_{\mathrm{vanish}}$ and $\mathcal{S}_{\mathrm{unity}}$ are finite under the quotient of shift $2\md C\cdot \md n$ symmetry.} We further conjecture that \emph{with the classical information of an arbitrary local Calabi-Yau threefold, the blowup equations combined together can uniquely determine its refined partition function, in particular all the refined BPS invariants}. 

Let us make a few remarks here. The true moduli and masses are also called the normalizable and non-normalizable K\"{a}hler parameters, details for which can be found in for example \cite{Huang:2013yta}\cite{Klemm:2015iya}. For local toric Calabi-Yau, the matrix $\md C$ is just part of the well known charge matrix of the toric action. The factor $\Lambda(\eq,\et;{\bf m},{\bf r})$ in the unity blowup equations normally has very simple expression and can be easily determined, as will be shown in section \ref{sec:blowup}. We also propose a procedure to determine all vanishing and unity $\md r$ fields from the polynomial part of refined topological string and the modular invariance of $\Lambda$ in section \ref{sec:blowup}. This is understandable since the classical information already fixes a Calabi-Yau. It is important that factor $\Lambda$ only depends on the mass parameters, but not on true moduli. However, there is no unique choice for the mass parameters. One simple way to distingui
 sh mass parameters and true moduli is that the $\md C$ components for mass parameters are zero while those for true moduli are nonzero. In addition, the blowup equations (\ref{eq:blowupZ}) is invariant under the shift $\md t\to\md t +2\md C\cdot \md n$, thus we only need to consider the equivalent classes of the $\md r$ fields. Let us denote the corresponding symmetry group as $\Gamma_{\md C}$. The condition \ref{eq:rcondition} can be derived from the blowup equations, as well be shown \ref{sec:constrain}. Such condition was known as the $\md B$ field condition which was established in \cite{Hatsuda:2013oxa} for local del Pezzoes and in \cite{Sun:2016obh} for arbitrary local toric Calabi-Yau. Denote the set of all $\md r$ fields satisfying the condition \ref{eq:rcondition} as $\IZ^{b}_{\md B}$, then all possible non-equivalent $\md r$ fields of blowup equations are in the quotient $\IZ^{b}_{\md B}/\Gamma_{\md C}$. Normally, this quotient set is not finite. Therefore, the vanishing a
 nd unity $\md r$ fields are in generally rare in $\IZ^{b}_{\md B}/\Gamma_{\md C}$.

Previously, the partition function of refined topological string theory on local Calabi-Yau can be obtained by using the refined topological vertex in the A-model side \cite{Iqbal:2007ii}\cite{Taki:2007dh}\cite{Iqbal:2012mt}, or refined holomorphic anomaly equations in the B-model side \cite{Huang:2010kf}\cite{Huang:2013yta}\cite{Klemm:2015iya}. We will use those results to check the validity of blowup equations. Moreover, we can also do reversely. Assume the correctness of blowup equations and use them to determine the refined partition function. We show evidences that these blowup equations all together are strong enough to determine the full partition function of refined topological string. While the holomorphic anomaly equations are directly related to the worldsheet physics and Gromov-Witten formulation, the blowup equations on the other hand are directly related to the target physics and Gopakumar-Vafa (BPS) formulation. Therefore, if the refined BPS invariants are the main con
 cern, the blowup equations will be a more effective technique to determine them. Besides, unlike the holomorphic anomaly equations which are differential equations and suffer from the holomorphic ambiguities, the blowup equations are functional equations which are expected to be able to fully determine the partition function. As for the rigorous proof of the blowup equations for general local Calabi-Yau, it seems still beyond the current reach. However, it should be possible to give a physical proof based on the brane picture in \cite{Aganagic:2011mi}.

This paper is organized as follows. In section \ref{sec:quant}, we introduce the background of current paper, including local Calabi-Yau, refined topological string and quantum mirror curves. In section \ref{sec:blowup}, we detailedly analyze the vanishing and unity blowup equations, including their expansion, relation with GNY blowup equations, modular property, constraints on the refined BPS invariants and non-perturbative formulation. In section \ref{sec:ex}, we study a various of examples including resolved conifold, local $\mathbb{P}^2$, $\mathbb{F}_n$, $\mathfrak{B}_3$ and resolved $\mathbb{C}^3/\mathbb{Z}_5$ orbifold. In section \ref{sec:nontoric}, we study the blowup equations for E-string theory, which is equivalent to refined topological string theory on local half K3, a typical non-toric example. We check the blowup equations in terms of $E_8$ Weyl-invariant Jacobi forms, which is different from the method in \cite{Gu:2017ccq}. In section \ref{sec:solve}, we reversely use 
 the blowup equations to solve the refined free energy. We show several supports to our conjecture that blowup equations can uniquely determine the full refined BPS invariants, including a count of the independent component equations at $\eq,\et$ expansion, a strict proof for resolved conifold and a test for local $\IP^2$. In section \ref{sec:generic}, we study the blowup equations on the other points in the moduli space including the conifold points and orbifold points. In section \ref{sec:outlook}, we conclude with some interesting questions for future study.
%%%%%%%%%%%%%%%%%%%%%%%%%%%%%%%%%%%%%%%%%%%%%%%%%%%%%%%%%%%%

\section{Quantum mirror curve and refined topological string}\label{sec:quant}

\subsection{Refined A model}\label{sec:A}
To give the basic components of blowup equation, here we first briefly review some well-known definitions in (refined) topological string theory. We follow the notion in \cite{Codesido:2015dia}. The Gromov-Witten invariants of a Calabi-Yau $X$ are encoded in the partition function $Z({\bf t})$ of topological string on $X$. It has a genus expansion $Z({\bf t})=\re^{\sum_{g=0}^{\infty}g_s^{2g-2}F_g({\bf t})}$ in terms of genus $g$ free energies $F_g({\bf t})$. At genus zero,
\be
\label{gzp}
F_0({\bf t})={1\over 6} a_{ijk} t_i t_j t_k  +P_2(t)+ \sum_{{\bf d}} N_0^{ {\bf d}} \re^{-{\bf d} \cdot {\bf t}},
\ee
where $a_{ijk}$ denotes the classical intersection and $P_2(t)$ is ambiguous which is irrelevant in our current discussion. At genus one, one has
\be
\label{gop}
F_1({\bf t})=b_i t_i + \sum_{{\bf d}} N_1^{ {\bf d}} \re^{-{\bf d} \cdot {\bf t}},
\ee
At higher genus, one has
\be
\label{genus-g}
F_g({\bf t})= C_g+\sum_{{\bf d}} N_g^{ {\bf d}} \re^{-{\bf d} \cdot {\bf t}}, \qquad g\ge 2,
\ee
where $C_g$ is the constant map contribution to the free energy.
The total free energy of the topological string is formally defined as the sum,
\be
\label{tfe}
F^{\rm WS}\left({\bf t}, g_s\right)= \sum_{g\ge 0} g_s^{2g-2} F_g({\bf t})=F^{({\rm p})}({\bf t}, g_s)+ \sum_{g\ge 0} \sum_{\bf d} N_g^{ {\bf d}} \re^{-{\bf d} \cdot {\bf t}} g_s^{2g-2},
\ee
where
\be
F^{({\rm p})}({\bf t}, g_s)= {1\over 6 g_s^2} a_{ijk} t_i t_j t_k +b_i t_i + \sum_{g \ge 2}  C_g g_s^{2g-2}.
\ee
The BPS part of partition function (\ref{tfe}) can be resumed with a new set of enumerative invariants, called Gopakumar-Vafa (GV)
invariants $n^{\bf d}_g$, as \cite{Gopakumar:1998jq}
\be
\label{GVgf}
F^{\rm GV}\left({\bf t}, g_s\right)=\sum_{g\ge 0} \sum_{\bf d} \sum_{w=1}^\infty {1\over w} n_g^{ {\bf d}} \left(2 \sin { w g_s \over 2} \right)^{2g-2} \re^{-w {\bf d} \cdot {\bf t}}.
\ee
Then,
\be
\label{gv-form}
F^{\rm WS}\left({\bf t}, g_s\right)=F^{({\rm p})}({\bf t}, g_s)+F^{\rm GV}\left({\bf t}, g_s\right).
\ee
For local Calabi-Yau threefold, topological string have a refinement correspond to the supersymmetric gauge theory in the omega background. In refined topological string, the Gopakumar-Vafa invariants can be generalized to the refined BPS invariants  $N^{\bf d}_{j_L, j_R}$ which depend on the degrees ${\bf d}$ and spins, $j_L$, $j_R$ \cite{Iqbal:2007ii}\cite{Choi:2012jz}\cite{Nekrasov:2014nea}. Refined BPS invariants are positive integers and are closely related with the Gopakumar-Vafa invariants,
\be
\label{ref-gv}
\sum_{j_L, j_R} \chi_{j_L}(q) (2j_R+1) N^{\bf d} _{j_L, j_R} = \sum_{g\ge 0} n_g^{\bf d} \left(q^{1/2}- q^{-1/2} \right)^{2g},
\ee
where $q$ is a formal variable and
\be
\chi_{j}(q)= {q^{2j+1}- q^{-2j-1} \over q-q^{-1}}
\ee
is the $SU(2)$ character for the spin $j$. Using these refined BPS invariants, one can define the NS free energy as%
\be
\label{NS-j}
F^{\rm NS}({\bf t}, \hbar) ={1\over 6 \hbar} a_{ijk} t_i t_j t_k +b^{\rm NS}_i t_i \hbar +\sum_{j_L, j_R} \sum_{w, {\bf d} }
N^{{\bf d}}_{j_L, j_R}  \frac{\sin\frac{\hbar w}{2}(2j_L+1)\sin\frac{\hbar w}{2}(2j_R+1)}{2 w^2 \sin^3\frac{\hbar w}{2}} \re^{-w {\bf d}\cdot{\bf  t}}.
\ee
in which $b_i^{\rm NS}$ can be obtained by using mirror symmetry as in \cite{Huang:2010kf}. By expanding (\ref{NS-j}) in powers of $\hbar$, we find the NS free energies at order $n$,
\be
\label{ns-expansion}
F^{\rm NS}({\bf t}, \hbar)=\sum_{n=0}^\infty  F^{\rm NS}_n ({\bf t}) \hbar^{2n-1}.
\ee
The BPS part of free energy of refined topological string is defined by refined BPS invariants as
\be
\ba
F_{\text{ref}}^{\text{BPS}}(\mathbf{t},\epsilon_1,\epsilon_2)
=\sum_{j_L,j_R}\sum_{w, d_j \geq 1}
\frac{1}{w} N_{j_L,j_R}^{\bf{d}}
\frac{\chi_{j_L}(q_L^w)\chi_{j_R}(q_R^w)}{(q_1^{w/2}-q_1^{-w/2})(q_2^{w/2}-q_2^{-w/2})}
\re^{-w \bf{d} \cdot \bf{t}},
\ea
\label{eq:F-ref}
\ee
where
\be
\ba
\epsilon_j=2\pi \tau_j,\quad q_j=\re^{2\pi \ri \tau_j},\quad (j=1,2),\qquad q_L=\re^{\pi \ri (\tau_1-\tau_2)},\qquad
q_R=\re^{\pi \ri (\tau_1+\tau_2)}.
\ea
\ee
The refined topological string free energy can also be defined by refined Gopakumar-Vafa invariants as
\be
\label{ref-free-2}
F_{\text{ref}}^{\rm BPS}(\mathbf{t}, \epsilon_1, \epsilon_2)=
\sum_{g_L, g_R \ge 0 }\sum_{w\ge 1}\sum_{{\bf d}}
{1\over w} n_{g_L, g_R}^{\bf d}
{ \left(q_L^{w /2} - q_L^{-w /2} \right)^{2g_L} \over q^{w/2} - q^{-w/2} }
{ \left(q_R^{w /2} - q_R^{-w /2} \right)^{2g_R} \over t^{w/2} -t^{-w/2} }
e^{-w\mathbf{d}\cdot\mathbf{t}},
\ee
where
\be
\label{qt}
q=\re^{ \im \epsilon_1},   \qquad t=\re^{-\im \epsilon_2}.
\ee
The refined Gopakumar-Vafa invariants are related with refined BPS invariants,
\be
\label{change-basis}
\ba
\sum_{j_L, j_R\ge 0}  N_{j_L, j_R}^{\bf d}
\chi_{j_L}(q_L) \chi_{j_R}(q_R)=\sum_{g_L, g_R \ge 0 }  n_{g_L, g_R}^{\bf d}
\left(q_L^{1 /2} - q_L^{-1 /2} \right)^{2g_L}
\left(q_R^{1 /2} - q_R^{-1 /2} \right)^{2g_R}.
\ea
\ee
The refined topological string free energy can be expand as
\be
F(\mathbf{t},\epsilon_1,\epsilon_2)=\sum_{n,g=0}^{\infty}(\epsilon_1+\epsilon_2)^{2n}(\epsilon_1\epsilon_2)^{g-1}\mathcal{F}^{(n,g)}(\mathbf{t})
\ee
where $\mathcal{F}^{(n,g)}(\mathbf{t})$ can be determined recursively using the refined holomorphic anomaly equations.

With the refined free energy, the traditional topological string free energy can be obtained by taking the unrefined limit,
\be
\epsilon_1=-\epsilon_2=g_s.
\ee
Therefore,
\be
F_{\rm GV}\left({\bf t}, g_s\right)=F(\mathbf{t},g_s,-g_s).
\ee
The NS free energy can be obtained by taking the NS limit in refined topological string,
\be
F^{\rm NS}({\bf t}, \hbar)=\lim_{\epsilon_1\rightarrow 0}\epsilon_1F(\mathbf{t},\epsilon_1,\hbar).
\ee
	
	We will also need to specify an $s$ dimensional integral vector $\md B$ such that non-vanishing BPS invariants $N^{\md d}_{j_L,j_R}$ occur only at
	\begin{equation} \label{condB}
	2j_L + 2j_R + 1 \equiv \md B \cdot \md d \quad \mod 2 \, .
	\end{equation}		
	This condition specifies $\md B$ only mod 2. The existence of such a vector $\md B$ is guaranteed by the fact that the non-vanishing BPS invariants follow a so-called checkerboard pattern, as first observed in \cite{Choi:2012jz}, and is also important in the pole cancellation in the non-perturbative completion\cite{Hatsuda:2013oxa} .
	
	We define the twisted refined free energies $\widehat{F}_{\rm ref}(\md t, \epsilon_1,\epsilon_2)$ via
	\begin{equation}\label{eq:twist}
	\widehat{F}_{\rm ref}(\md t;\epsilon_1,\epsilon_2) = F^{\rm pert}_{\rm ref}(\md t;\epsilon_1,\epsilon_2) + F^{\rm inst}_{\rm ref}(\md t+\pi\ri \md B; \epsilon_1,\epsilon_2) \ .
	\end{equation}	
	Here, the perturbative contributions are given by
	\begin{equation}\label{eq:F-pert}
	F_{\rm ref}^{\rm pert}(\md t;\epsilon_1, \epsilon_2) = \frac{1}{\epsilon_1\epsilon_2}\( \frac{1}{6} \sum_{i,j,k=1}^{s} a_{ijk} t_i t_j t_k + 4\pi^2 \sum_{i=1}^{s} b_i^{\rm NS}t_i\)  + \sum_{i=1}^{s} b_i t_i - \frac{(\epsilon_1+\epsilon_2)^2}{\epsilon_1\epsilon_2} \sum_{i=1}^{s} b_i^{\rm NS} t_i \ ,
	\end{equation}
	where $a_{ijk}$ and $b_i$ are related to the topological intersection numbers in $X$, and $b_i^{\rm NS}$ can be obtained from the refined genus one holomorphic anomaly equation. The ambiguity factor is fixed here from the S-dual fact of quantization condition of the mirror curve \cite{Wang:2015wdy} \cite{Hatsuda:2015fxa}. The instanton contributions are given by the refined Gopakumar-Vafa formula,
\begin{equation}\label{eq:F-inst}
F_{\rm ref}^{\rm inst}(\md t,\epsilon_1,\epsilon_2) = \sum_{j_L, j_R \ge 0} \sum_{\md d} \sum_{w=1}^{\infty}  (-1)^{2j_L+2j_R} N^{\md d}_{j_L,j_R} \frac{\chi_{j_L}(q_L^w) \chi_{j_R}(q_R^w)}{w(q_1^{w/2}-q_1^{-w/2})(q_2^{w/2} - q_2^{-w/2})} e^{-w \md d\cdot \md t } \ ,
\end{equation}
where
\begin{equation}
q_{1,2} = e^{\epsilon_{1,2}}\ ,\quad q_{L,R} = e^{(\epsilon_1 \mp \epsilon_2)/2} \ ,
\end{equation}
and
\begin{equation}
\chi_j(q) = \frac{q^{2j+1} - q^{-2j-1}}{q - q^{-1}} \ .
\end{equation}
%%%%%%%%%%%%%%%%%%%%%%%%%%%%%%%%%%%%
Apparently, both $ F^{\rm pert}_{\rm ref}(\md t;\epsilon_1,\epsilon_2) $ and $F^{\rm inst}_{\rm ref}(\md t; \epsilon_1,\epsilon_2)$ are invariant under the $\epsilon_{1,2}\to {-\epsilon_{1,2}}$, thus
\be\label{eq:con1}
\widehat{F}_{\rm ref}(\md t;\epsilon_1,\epsilon_2) =\widehat{F}_{\rm ref}(\md t;-\epsilon_1,-\epsilon_2).
\ee
%%%%%%%%%%%%%%%%%%%%%%%%%%%%%%%%%%%%%%%%%%%%%%%%%%%%%%%%%%%%%%%%%%
\subsection{Refined B model}\label{sec:B}
A toric Calabi-Yau threefold is a toric variety given by the quotient,
\be
M=(\mathbb{C}^{k+3}\text{\textbackslash} \mathcal{SR})/G,
\ee
where $G=(\mathbb{C}^*)^k$ and $\mathcal{SR}$ is the Stanley-Reisner ideal of $G$. The quotient is specified by a matrix of charges $Q_i^\alpha$, $i=0, \cdots, k+2$, $\alpha=1, \cdots, k$. The group $G$ acts on the homogeneous coordinates $x_i$ as
\be
x_i \rightarrow \lambda_\alpha^{Q^{\alpha}_i} x_i,\quad\quad i=0, \cdots, k+2,
\ee
where $\alpha=1,\ldots,k$,  $\lambda_\alpha \in \mathbb{C}^*$ and $Q^{\alpha}_i \in \mathbb{Z}$.

The toric variety $M$ has a physical understanding, it is the vacuum configuration of a two-dimensional abelian $(2,2)$ gauged linear sigma model. Then $G$ is gauge group $U(1)^k$. The vacuum configuration is constraint by
\be
\sum_{i=1}^{k+3}Q_{i}^\alpha |x_i|^2=r^\alpha,\ \ \ \alpha=1,\ldots,k,
\ee
where $r^{\alpha}$ is the K\"{a}hler class. In general, the K\"{a}hler class is complexified by adding a theta angel $t^{\alpha}=r^{\alpha}+i\theta^\alpha$. Mirror symmetry of A,B model need existence of R symmetry. In order to avoid R symmetry anomaly, one has to put condition
\be
\sum_{i=1}^{k+3}{Q^{\alpha}_i}=0,\ \ \ \alpha=1,\ldots,k.
\ee
This is the Calabi-Yau condition in the geometry side.

The mirror to toric Calabi-Yau was constructed in \cite{Chiang:1999tz}. We define the Batyrev coordinates
\begin{equation}\label{w_1}
z_\alpha=\prod_{i=1}^{k+3} x_i^{Q_i^{\alpha}},\ \ \ \alpha=1,\ldots,k,
\end{equation}
and
\begin{equation}\label{w_2}
H=\sum_{i=1}^{k+3}x_i.
\end{equation}
The homogeneity allows us to set one of $x_i$ to be one. Eliminate all the $x_i$ in (\ref{w_2}) by using (\ref{w_1}), and choose other two as $e^x$ and $e^{p}$, then the mirror geometry is described by
\begin{equation}\label{w_3}
uv=H(e^x,e^p;z_{\alpha}),\ \ \ \alpha=1,\ldots,k,
\end{equation}
where $x,p,u,v\in\mathbb{C}$. Now we see that all the information of mirror geometry is encoded in the function $H$. The equation
\be
H(e^x,e^p;z_{\alpha})=0
\ee
defines a Riemann surface $\Sigma$, which is called the mirror curve to a toric Calabi-Yau. We denote $g_\Sigma$ as the genus of the mirror curve. 

The form of mirror curve can be written down specifically with the vectors in the toric diagram. Given the matrix of charges $Q^\alpha_i$, we introduce the vectors,
\be
 \nu^{(i)}=\left(1, \nu^{(i)}_1, \nu^{(i)}_2\right),\qquad i=0, \cdots, k+2,
\ee
satisfying the relations
\be
\sum_{i=0}^{k+2} Q^\alpha_i \nu^{(i)}=0.
\ee
In terms of these vectors, the mirror curve can be written as
\be
\label{coxp}
H(\re^x, \re^p)=\sum_{i=0}^{k+2}x_i  \exp\left( \nu^{(i)}_1 x+  \nu^{(i)}_2 p\right).
 \ee
It is easy to construct the holomorphic 3-form for mirror Calabi-Yau as
\be
\Omega=\frac{du}{u}\wedge dx \wedge dp
\ee
At classical level, the periods of the holomorphic 3-form are
\be
t_i=\oint_{A_i}\Omega,\ \ \ \ \ \mathcal{F}_i=\oint_{B_i}\Omega.
\ee
If we integrate out the non-compact directions, the holomorphic 3-forms become meromorphic 1-form on the mirror curve \cite{Katz:1996fh}\cite{Chiang:1999tz}:
\be
\label{diff-la}
\lambda=p\, \rd x.
\ee
The mirror maps and the genus zero free energy $F_0 ({\bf t})$ are determined by making an
appropriate choice of cycles on the curve, $\alpha_i$, $\beta_i$, $i=1, \cdots,s$, then we have
\be
t_i = \oint_{\alpha_i} \lambda, \qquad \qquad  {\partial F_0 \over \partial t_i} = \oint_{\beta_i} \lambda, \qquad i=1, \cdots, s.
\ee
In general, $s\ge g_\Sigma$, where $g_\Sigma$ is the genus of the mirror curve. The $s$ complex moduli can be divided into two classes, which are $g_\Sigma$ true moduli, $\kappa_i$, $i=1, \cdots, g_\Sigma$, which is Coulomb branch parameter in gauge theory, and
$r_\Sigma$ mass parameters, $\xi_j$, $j=1, \cdots, r_\Sigma$, where $r_\Sigma=s- g_\Sigma$. Complex parameter $z_i$ of B model is the product of true moduli and mass parameters. The true moduli can be also be expressed with the chemical potentials $\mu_i$,
\be
\kappa_i =\re^{\mu_i}, \qquad i=1, \cdots, g_{\Sigma}.
\ee
Among the K\"{a}hler parameters, there are $g_\Sigma$ of them which correspond to the true moduli, and their mirror map at large $\mu_i$ is of the form
\be
\label{tmu}
t_i \approx \sum_{j=1}^{g_\Sigma} C_{ji} \mu_j + \sum_{j=1}^{r_\Sigma} \alpha_{ij} t_{\xi_j}, \qquad i=1, \cdots, g_\Sigma,
\ee
where $t_{\xi_j}$ is the flat coordinate associated to the mass parameter $\xi_j$ by an algebraic mirror map. For toric case, the $g_\Sigma\times s$ matrix $C_{ij}$ can be read off from the toric data of $X$ directly. For generic local case \cite{Gu:2017ccq}, $C_{ij}$ should be understood as the intersection number of K\"ahler class $C_i$ to the $g_X$ irreducible compact divisor classes $D_j$ in the geometry,
\be
C_{ij} = D_i  \cdot  C_j \, 
\ee
It was shown in \cite{Aganagic:2011mi} that the classical mirror maps can be promoted to quantum mirror maps. Note only the mirror maps for the true moduli have quantum deformation, while the mirror maps for mass parameters remain the same. Since the blowup equation depends on K\"ahler parameters in A-model directly, the form of mirror map does not have effects on blowup equation, so we will not care about quantum mirror map in our paper.

In previous section, we also define twisted refined free energy. It is noted in \cite{Sun:2016obh} the $\mathbf{B}$ field has a representation denote as $\mathbf{r}$ field, and the existence $\mathbf{r}$ field can be cancelled by a sign changing of complex parameter $z_i$, this in principle is a change of variables and do not effect B model but only with some extra constant terms appear in genus 0,1 free energy. In the following discussion, we will not mention this twisted form in B model.

\subsection{Nekrasov-Shatashvili quantization}\label{sec:ns}
In this and following sections, we review two quantization conditions of mirror curves, and their equivalence condition, which promote the study of blowup equation of refined topological string. The study of this paper will also leads to new results to the equivalence condition.
In 2009, Nekrasov and Shatashvili promoted this correspondence between $N=2$ supersymmetric gauge theories and integrable systems to the quantum level \cite{Nekrasov:2009rc}, see also \cite{Nekrasov:2009uh}\cite{Nekrasov:2009ui}. The correspondence is usually called Bethe/Gauge correspondence. As it is not proved in full generality yet, we also refer the correspondence on quantum level as Nekrasov-Shatashvili conjecture. The $SU(2)$ and $SU(N)$ cases are soon checked for the first few orders in \cite{Mironov:2009uv}\cite{Mironov:2009dv}. For a proof of the conjecture for $SU(N)$ cases, see \cite{Kozlowski:2010tv}\cite{Meneghelli:2013tia}. This conjecture as a 4D/1D correspondence is also closely related to the AGT conjecture \cite{Alday:2009aq}, which is a 4D/2D correspondence. In fact, the duality web among $N=2$ gauge theories, matrix model, topological string and integrable systems (CFT) can be formulated in generic Nekrasov deformation \cite{Dijkgraaf:2009pc}, not just in NS limi
 t. See \cite{Rim:2015aha} for a proof of Nekrasov-Shatashvili conjecture for $SU(N)$ quiver theories based on AGT correspondence.\\
\indent It is well-known the prepotential of Seiberg-Witten theory can be obtained from the Nekrasov partition function \cite{Nekrasov:2002qd},
\be
{\mathcal{F}}(\vec {a})=\lim_{\epsilon_1,\epsilon_2\rightarrow0} \epsilon_1\epsilon_2\log Z_{\text{Nek}}(\vec{a}; \epsilon_1,\epsilon_2),
\ee
where $\vec{a}$ denotes the collection of all Coulomb parameters. In \cite{Nekrasov:2009rc}, Nekrasov and Shatashvili made an interesting observation that in the limit where one of the deformation parameters is sent to zero while the other is kept fixed ($\epsilon_1\rightarrow 0,\,\epsilon_2 = \hbar$), the partition function is closely related to certain quantum integrable systems. The limit is usually called Nekrasov-Shatashvili limit in the context of refined topological string, or classical limit in the context of AGT correspondence. To be specific, the Nekrasov-Shatashvili conjecture says that the supersymmetric vacua equation
\be\label{qc1}
\exp {(\partial_{a_I}{\mathcal{W}}(\vec{a};\hbar))}=1\,,
\ee
of the Nekrasov-Shatashvili free energy
\be\label{NSfree1}
\mathcal{W}(\vec{a};\hbar)=\lim_{\epsilon_1\rightarrow0} \epsilon_1\log Z_{\text{Nek}}(\vec{a}; \epsilon_1,\epsilon_2=\hbar)
\ee
gives the Bethe ansatz equations for the corresponding integrable system. The Nekrasov-Shatashvili free energy (\ref{NSfree1}) is also called effective twisted superpotential in the context of super gauge theory. According to the NS conjecture, it serves as the Yang-Yang function of the integrable system. The quantized/deformed Seiberg-Witten curve becomes the quantized spectral curve and the twisted chiral operators become the quantum Hamiltonians. Since it is usually difficult to written down the Bethe ansatz for general integrable systems, this observation provide a brand new perspective to study quantum integrable systems.\\
\indent The physical explanations of Bethe/Gauge correspondence was given in \cite{Nekrasov:2010ka}\cite{Aganagic:2011mi}. Let us briefly review the approach in \cite{Aganagic:2011mi}, which is closely related to topological string. In the context of geometric engineering, the NS free energy of supersymmetric gauge theory is just the NS limit of the partition function of topological string \cite{Katz:1996fh},
\be\label{FNS5}
\mathcal{W}(\vec{a};\hbar)=F^{\rm NS}({\bf t}, \hbar).
\ee
See \cite{Huang:2012kn} for the detailed study on the relation between gauge theory and topological string in Nekrasov-Shatashvili limit.

Consider the branes in unrefined topological string theory, it is well known \cite{Aganagic:2003qj} that for B-model on a local Calabi-Yau given by
\be
uv+H(x,p)=0
\ee
the wave-function $\Psi(x)$ of a brane whose position is labeled by a point $x$ on the Riemann
surface $H(x,p)=0$ classically, satisfies an operator equation
\be
H(\mx, \mathsf{p})\Psi=0,
\ee
with the Heisenberg relation\footnote{In general, this relation only holds up to order $g_s$ correction.}
\be
[\mx, \mathsf{p}]= \im g_s.
\ee
In the refined topological string theory, the brane wave equation is generalized to a multi-time dependent Schr\"odinger equation,
\be
H(x,p)\Psi =\epsilon_1\epsilon_2\sum f_i(t){\partial \Psi \over \partial t_i},
\ee
where $f_i(t)$ are some functions of the K\"{a}hler moduli $t_i$ and the momentum operator is given by either $p =i \epsilon_1 \partial_x$ or $p= i\epsilon_2 \partial_x$, depending on the type of brane under consideration.\\
\indent In the NS limit $\epsilon_1\rightarrow0,\,\epsilon_2=\hbar$, the time dependence vanishes, and we simply obtain the time-independent Schr\"odinger equation
\be
H(\mx, \mathsf{p})\Psi=0,
\ee
with
\be
[\mx, \mathsf{p}]= \im \hbar.
\ee
To have a well-defined wave function we need the wave function to be single-valued under monodromy. In unrefined topological string, the monodromy is characterized by taking branes around the cycles of a Calabi-Yau shifts the dual periods in units of $g_s$. While in the NS limit, the shifts becomes derivatives. Therefore, the single-valued conditions now are just the supersymmetric vacua equation (\ref{qc1}). In the context of topological string, we expect to have

\be\label{qc2}
C_{ij}\frac{\partial{F^{\rm NS}({\bf t}, \hbar)}}{\partial{t_j}}=2\pi \left(n_i+\frac{1}{2}\right),\ i=1,\cdots,g.
\ee
In fact these conditions are just the concrete form of Einstein-Brillouin-Keller (EBK) quantization, which is the generalization of Bohr-Sommerfeld quantization for high-dimensional integrable systems. Therefore, we can also regard the left side of (\ref{qc2}) as phase volumes corresponding to each periods of the mirror curve,
\be\label{vol}
\mathrm{Vol}_i({\bf t}, \hbar)=\hbar C_{ij}\frac{{F^{\rm NS}({\bf t}, \hbar)}}{\partial{t_j}},\ i=1,\cdots,g.
\ee
Now the NS quantization conditions for the mirror curve are just the EBK quantization conditions,
\be\label{qc3}
\mathrm{Vol}_i({\bf t}, \hbar)=2\pi\hbar \left(n_i+\frac{1}{2}\right),\ i=1,\cdots,g.
\ee
As we mentioned, these NS quantization conditions need non-perturbative completions \cite{Kallen:2013qla}\cite{Wang:2015wdy}\cite{Hatsuda:2015qzx}\cite{Franco:2015rnr} and such completion can be obtained by simply substituting Lockhart-Vafa free energy $F_{\text{LV}}$ into (\ref{vol}).

Based on the localization calculation on the partition function of superconformal theories on squashed $S^5$, a non-perturbative definition of refined topological string was proposed by Lockhart and Vafa in \cite{Lockhart:2012vp}. Roughly speaking, the proposal for the non-perturbative topological string partition function take the form
\be
Z_{\mathrm{LV}}(t_i,m_j,\tau_1,\tau_2)={Z_{\text{ref}}(t_i,m_j;\tau_1,\tau_2)\over Z_{\text{ref}}(t_i/\tau_1,m_j/\tau_1; -1/\tau_1,\tau_2/\tau_1)\cdot Z_{\text{ref}}(t_i/\tau_2,m_j/\tau_2; \tau_1/\tau_2,-1/\tau_2)}
\ee
where $t_i, m_j$ are normalizable and non-normalizable K\"{a}hler classes, and $\tau_1,\tau_2$
are the two couplings of the refined topological strings with relation $\epsilon_{1,2}=2\pi\tau_{1,2}$. By carefully considering the spin structure in gauge precise and make the K\"{a}hler parameters suitable for the current context, the precise form of Lockhart-Vafa partition function should be
\be \label{eq:Z-LV}
Z_{\mathrm{LV}}(\mathbf{t},\tau_1,\tau_2)={Z_{\text{ref}}(\mathbf{t},\tau_1+1,\tau_2)\over Z_{\text{ref}}(\mathbf{t}/\tau_1,-1/\tau_1,\tau_2/\tau_1+1)\cdot Z_{\text{ref}}(\mathbf{t}/\tau_2. \tau_1/\tau_2+1,-1/\tau_2)}
\ee
Here we do not bother to distinguish the mass parameters from the true K\"{a}hler moduli. Then the non-perturbative free energy of refined topological string is given by
\be
F_{\text{LV}}({\bf{t}},\tau_1,\tau_2)=F_{\text{ref}}({\bf{t}},\tau_1+1,\tau_2)-F_{\text{ref}} \( \frac{{\bf{t}}}{\tau_1},-\frac{1}{\tau_1},\frac{\tau_2}{\tau_1}+1 \)
-F_{\text{ref}} \( \frac{{\bf{t}}}{\tau_2},\frac{\tau_1}{\tau_2}+1,-\frac{1}{\tau_2} \).
\label{eq:F-LV}
\ee
%%%%%%%%%%%%%%%%%%%%%%%%%%%%%%%%%%%%%%%%%%%%%%%%%%%%%%%%%%%%%%%%%%%
\subsection{Grassi-Hatsuda-Mari\~no conjecture}\label{sec:ghm}
For a mirror curve $\Sigma$ with genus $g_{\Sigma}$, there are $g_{\Sigma}$ different canonical forms for the curve,
\be
\CO_i (x,y) + \kappa_i=0, \qquad i=1, \cdots, g_{\Sigma}.
\ee
Here, $\kappa_i$ is normally a true modulus $x_i$ of $\Sigma$. The different canonical forms of the curves are related by reparametrizations and overall factors,
\be
\label{o-rel}
\CO_i +\kappa_i  = \CP_{ij} \left( \CO_j +\kappa_j\right), \qquad i,j=1, \cdots, g_{\Sigma},
\ee
where $\CP_{ij}$ is of form $\re^{ \lambda x + \mu y}$. Equivalently, we can write
\be
\label{oi-exp}
\CO_i = \CO_i^{(0)}+ \sum_{j \not=i} \kappa_j \CP_{ij}.
\ee
Perform the Weyl quantization of the operators $\CO_i(x,y)$, we obtain
$g_{\Sigma}$ different Hermitian operators $\mO_i$, $i=1, \cdots, g_{\Sigma}$,
\be
\mO_i=\mO_i^{(0)}+ \sum_{j \not=i} \kappa_j \mP_{ij}.
\ee
The operator $\mO_i^{(0)}$ is the unperturbed operator, while the moduli $\kappa_j$ encode different
perturbations. It turns out that the most interesting operator was not $\mO$, but its inverse
$\rho$. This is because $\rho$ is expected to be of trace class and
positive-definite, therefore it has a discrete, positive spectrum, and its Fredholm (or spectral) determinant is well-defined. We have

\be
\label{Xops}
\rho_i =\mO_i^{-1}, \qquad i=1, \cdots, g_\Sigma,
\ee
and
\be
\label{unp-inv}
\rho_i^{(0)}=\left( \mO_i^{(0)}\right)^{-1}, \qquad i=1, \cdots, g_{\Sigma}.
\ee
For the discussion on the eigenfunctions of $\rho$, see a recent paper \cite{Marino:2016rsq}. In order to construct the generalized spectral determinant, we need to introduce the following operators,
\be
\label{ajl}
\mA_{jl}= \rho_j^{(0)} \mP_{jl}, \quad j, l=1, \cdots, g_{\Sigma}.
\ee
Now the generalized spectral determinant is defined as
\be
\label{gsd}
\Xi_X ( {\boldsymbol \kappa}; \hbar)= {\rm det} \left( 1+\kappa_1 \mA_{j1} +\cdots+ \kappa_{g_{\Sigma}} \mA_{j g_{\Sigma}}  \right).
\ee
It is easy to prove this definition does not depend on the index $j$.

This completes the definitions on quantum mirror curve from the quantum-mechanics side. Let us now turn to the topological string side. The total modified grand potential for CY with arbitrary-genus mirror curve is defined as
\be
\label{jtotal}
\mathsf{J}_{X}(\boldsymbol{\mu}, \boldsymbol{\xi},\hbar) = \mathsf{J}^{\rm WKB}_X (\boldsymbol{\mu}, \boldsymbol{\xi},\hbar)+ \mathsf{J}^{\rm WS}_X
(\boldsymbol{\mu},  \boldsymbol{\xi} , \hbar),
\ee
where
\be
\label{jm2}
\mathsf{J}^{\rm WKB}_X(\boldsymbol{\mu}, \boldsymbol{\xi}, \hbar)= {t_i(\hbar) \over 2 \pi}   {\partial F^{\rm NS}({\bf t}(\hbar), \hbar) \over \partial t_i}
+{\hbar^2 \over 2 \pi} {\partial \over \partial \hbar} \left(  {F^{\rm NS}({\bf t}(\hbar), \hbar) \over \hbar} \right) + {2 \pi \over \hbar} b_i t_i(\hbar) + A({\boldsymbol \xi}, \hbar).
\ee
and
\be
\label{jws}
\mathsf{J}^{\rm WS}_X(\boldsymbol{\mu}, \boldsymbol{\xi}, \hbar)=F^{\rm GV}\left( {2 \pi \over \hbar}{\bf t}(\hbar)+ \pi \ri {\bf B} , {4 \pi^2 \over \hbar} \right).
\ee
The modified grand potential has the following structure,
\be
\label{j-larget}
\mathsf{J}_{X}(\boldsymbol{\mu}, \boldsymbol{\xi},\hbar)= {1\over 12 \pi \hbar} a_{ijk} t_i(\hbar) t_j(\hbar) t_k(\hbar) + \left( {2 \pi b_i \over \hbar} + {\hbar b_i^{\rm NS} \over 2 \pi} \right) t_i(\hbar) +
\CO\left( \re^{-t_i(\hbar)}, \re^{-2 \pi t_i(\hbar)/\hbar} \right).
\ee
$A({\boldsymbol \xi}, \hbar)$ is some unknown function, which is relevant to the spectral determinant but does not affect the quantum Riemann theta function, therefore does not appear in the quantization conditions.

GHM conjecture says that the generalized spectral determinant (\ref{gsd}) is given by

\be
\label{our-conj}
\Xi_X({\boldsymbol \kappa}; \hbar)= \sum_{ {\bf n} \in \IZ^{g_{\Sigma}}} \exp \left( \mathsf{J}_{X}(\boldsymbol{\mu}+2 \pi \ri  {\bf n}, \boldsymbol{\xi}, \hbar) \right).
\ee
As a corollary, we have, the quantization condition for the mirror curve is given by
\be
\Xi_X({\boldsymbol \kappa}; \hbar)=0.
\ee

The $\mathbf{r}$ field characterizes the phase-changing of complex moduli in the way that when one makes a transformation for the Batyrev coordinates
\be
(z_1,\dots,z_n)\rightarrow(z_1e^{r_1\pi\ri},\dots,z_ne^{r_n\pi\ri}),
\ee
equivalently, we have the following translation on the K\"{a}hler parameters:
\be
\mathbf{t}\rightarrow\mathbf{t}+\pi\ri\mathbf{r}.
\ee
This makes the effect of $\mathbf{r} $ field just like $\mathbf{B} $ field. For some specific choices of $\mathbf{r} $ fields, we have the generalized GHM conjecture as
\be
\Xi(\mathbf{t}+\pi\ri\mathbf{r},\hbar)=0.
\ee
%%%%%%%%%%%%%%%%%%%%%%%%%%%%%%%%%%%%%%%%%%%%%%%%%%%%%%%%%%%%%%
\subsection{Compatibility formulae}\label{sec:compat}

We can see now the main difference between quantization conditions is that NS quantization condition quantize $g_{\Sigma}$ particles of a integrable systems \cite{Franco:2015rnr}, there are $g_{\Sigma}$ constraint equations. But GHM quantization quantize the operators, and the number of operators is usually greater than $g_{\Sigma}$. In \cite{Sun:2016obh}, we introduce $\mathbf{r}$ fields. The value of $\mathbf{r}$ fields stand for the phase of complex parameters $z_i$. And different $\mathbf{r}$ fields quantize the operators in different phase. Because the definition of the generalized spectral determinant involves infinite sum, it is easy to see that different choices of $\mathbf{r} $ fields may result in the same functions. We define non-equivalent $\mathbf{r} $ fields as those which produce non-equivalent generalized spectral determinant. We denote the number of non-equivalent $\mathbf{r}$ fields as $w_{\Sigma}$. This lead to $w_{\Sigma}$ different quantization conditions. It is 
 conjectured in \cite{Sun:2016obh}:

{ \noindent\emph{The spectra of quantum mirror curve are solved by the simultaneous equations:
\be
\bigg\{\Theta(\mathbf{t}+\ri\pi\mathbf{r}^a,\hbar)=0,\ a=1,\cdots,w_{\Sigma}.\bigg\}
\ee}}
\emph{This spectra is the same as the spectra of NS quantization conditions:}
\be\label{equiv2}
\bigg\{\Theta(\mathbf{t}+\ri\pi\mathbf{r}^a,\hbar)=0,\ a=1,\cdots,w_{\Sigma}.\bigg\} \Leftrightarrow\left\{\mathrm{Vol}_i(\mathbf{t},\hbar)=2\pi\hbar \left(n_i+\frac{1}{2}\right),\ i=1,\cdots,g_{\Sigma}.\right\}
\ee
\emph{In addition, all the vector $\mathbf{r}^a$ are the representatives of the $\mathbf{B}$ field of $X$, which means for all triples of degree ${\bf d}$, spin $j_L$ and $j_R$ such that the refined BPS invariants $N^{{\bf d}}_{j_L, j_R}(X) $ is non-vanishing, they must satisfy}
\be
\label{ra-prop}
(-1)^{2j_L + 2 j_R-1}= (-1)^{{\bf r}^a \cdot {\bf d}},\quad a=1,\cdots,w_{\Sigma}.
\ee
The equivalence at perturbative level is studied in section \ref{sec:vanishingr}, we will also give the method to derive all $\mathbf{r}^a$ fields there. At nonperturbative level, we find a set of novel identities called compatibility formulae which guarantee the above equivalence in our previous paper \cite{Sun:2016obh}:
\be\label{conjecture}
\sum_{\mathbf{n}\in\mathbb{Z}^g}\exp\left(i\sum_{i=1}^{g}n_i \pi+F_{\text{unref}}\left(\mathbf{t}+ \ri\hbar \mathbf{n}\cdot\mathbf{C}+\frac{1}{2}\ri\hbar\mathbf{r},\hbar\right)
 -\ri n_jC_{ji}\frac{\partial }{\partial t_i}F_{\text{NS}}\left(\mathbf{t},\hbar\right)\right)=0.
\ee
in which $F_{\text{unref}}$ is the traditional topological string partition function, $F_{\text{NS}}$ is the Nekrasov-Shatashvili free energy, $C$ is the charge matrix of toric Calabi-Yau and $a=1,\cdots,w_{\Sigma}$. This identities is now known as NS limit of vanishing blowup equation \cite{Grassi:2016nnt}. The study of modular properties of blowup equation will also leads to an equivalence for quantization condition beyond large radius point, e.g. half orbifold points studied in \cite{Codesido:2015dia}.

%%%%%%%%%%%%%%%%%%%%%%%%%%%%%%%%%%%%%%%%%%%%%%%%%%%%%%%%%%%%%%%%
\section{K-theoretic blowup equations}\label{sec:blowup}
The compatibility formulae (\ref{conjecture})  between the quantization conditions in \cite{Wang:2015wdy} and the Grassi-Hatsuda-Marino quantization condition in \cite{Hatsuda:2015qzx} gives one inspiration for the new structures of refined topological string theory.  To distinguish the two formulations,  here we usually also refer to the formulation in \cite{Wang:2015wdy} as ``Nekrasov-Shatashvili quantization condition" because this formulation uses purely the NS limit of the topological string free energy to construct the non-perturbative contributions. The other inspiration comes from the blowup equations in supersymmetric gauge theories. In \cite{Nakajima:2003pg}, Nakajima and Yoshioka proposed the blowup equations for four-dimensional $\mathcal{N}=2$ $SU(N)$ gauge theories to prove Nekrasov's conjecture. It was soon realized by them that similar blowup equations can be established for five-dimensional $\mathcal{N}=1$ gauge theories \cite{Nakajima:2005fg}. Because the five-dimen
 sional spacetime here is the $S^1$ lift of the four-dimensional spacetime, the 5D equations can be seen as the equivariant version of the 4D equations, and are called the \emph{K-theoretic} blowup equations. In \cite{Gottsche:2006bm}, G\"{o}ttsche-Nakajima-Yoshioka further generalized the K-theoretic blowup equations to the gauge theories with 5D Chern-Simons term whose coefficient gives another integer $m$. As these are the most general cases at that time, we call such equations as \emph{G\"{o}ttsche-Nakajima-Yoshioka K-theoretic blowup equations}.

Rather surprisingly, the blowup equations did not draw much attention in the recent decade. Although it is long known that the five-dimensional $\mathcal{N}=1$ $SU(N)$ gauge theories with level $m$ Chern-Simons term can be engineered from M-theory compactified on $X_{N,m}$ Calabi-Yau geometries, thus are corresponding to the refined topological string theory on such geometries, yet till recently, to our knowledge, no effort has been made to study the blowup equations for the refined topological string on general local Calabi-Yau manifolds. The reason may be the G\"{o}ttsche-Nakajima-Yoshioka K-theoretic blowup equations are complicated enough and the proof seems to rely much on the gauge symmetry which does not have counterpart for general local Calabi-Yau, like local $\IP^2$. Nevertheless, we would like to show that the blowup equations indeed exist for the  partition function of refined topological string on general local Calabi-Yau threefolds.
%%%%%%%%%%%%%%%%%%%%%%%%%%%%%%%%%%%%%%%%%%%%%%%%%%%%%% 
\subsection{G\"{o}ttsche-Nakajima-Yoshioka K-theoretic blowup equations}\label{sec:gny}
In this section, we give a brief review for the K-theoretic blowup equations of 5D Nekrasov partition function. Our conventions are the same with \cite{Nakajima:2005fg} but slightly different from \cite{Gottsche:2006bm}. We begin with the pure $SU(N)$ gauge theory with no Chern-Simons term, i.e. $m=0$. As we mentioned in the introduction, the instanton part of K-theoretic Nekrasov partition function on $\IP^2$ is defined as
\begin{equation}
\begin{gathered}
	Z^{\rm inst}(\epsilon_1, \epsilon_2,\vec{a}; \fq, \beta) = \sum_{n=0}^\infty \left(\fq \beta^{2N} e^{-N\beta(\epsilon_1+\epsilon_2)/2} \right)^n Z_n(\epsilon_1,\epsilon_2,\vec{a};\beta),\\
	  Z_n(\epsilon_1,\epsilon_2,\vec{a};\beta) =\sum_i (-1)^i \mathrm{ch} H^i(M(N,n),\mathcal{O}).
	 \end{gathered}
\end{equation}
Here $\fq$ is the instanton counting parameter, and $\beta$ is the radius of the fifth dimension $S^1$. On the blowup $\widehat{\IP}^2$ with exceptional divisor $C$, one can define similar instanton partition function by
\begin{equation}
 Z^{\rm inst}_{k,d}(\epsilon_1, \epsilon_2,\vec{a}; \fq, \beta) = \sum_{n=0}^\infty \left(\fq \beta^{2N} e^{-N\beta(\epsilon_1+\epsilon_2)/2} \right)^n
  \left(\iota_{0*}\right)^{-1}\left( \widehat{\pi}_* ({\mathcal{O}}(d
  \mu(C))) \right),
\end{equation}
where $k$ is an integer characterizing the blowup with $\langle c_1(E),[C]\rangle = -k$. Using Atiyah-Bott-Lefschetz fixed
points formula, one can compute it as
\begin{multline}\label{Zinstkd}
 Z^{\rm inst}_{k,d}(\epsilon_1, \epsilon_2,\vec{a}; \fq, \beta) =
  \sum_{\{\vec{k}\}=-{k}/{r}}
   \frac{(e^{\beta(\ve_1+ \ve_2) (d - r/2)} \fq
   \beta^{2r})^{(\vec{k},\vec{k})/2}e^{\beta(\vec{k},\vec{a})d}}
   {\prod_{\vec{\alpha}\in\Delta} l^{\vec{k}}_{\vec{\alpha}}(\ve_1,\ve_2,\vec{a})} \times\\
% &
   Z^{\rm inst}(\ve_1,\ve_2-\ve_1,\vec{a}+\ve_1\vec{k};
    e^{\beta\ve_1 (d-r/2)} \fq,\beta)
   Z^{\rm inst}(\ve_1-\ve_2,\ve_2,\vec{a}+\ve_2\vec{k};
    e^{\beta\ve_2 (d-r/2)}\fq,\beta).
\end{multline}
Here $\Delta$ is the roots of $\mathfrak{su}(N)$. The vector $\vec{k}$ runs
over the coweight lattice
	\begin{equation} \label{kconst}
		\vec{k} \in\{ (k_\alpha) = (k_1,k_2,\ldots,k_N) \in \IQ^N \Big| \sum_{\alpha} k_\alpha = 0, \forall \alpha \, k_\alpha \equiv -k/N \mod \IZ \},
	\end{equation}
and
\begin{equation}\label{eq:l}
   l^{\vec{k}}_{\vec{\alpha}}(\ve_1,\ve_2,\vec{a}) = 
   \begin{cases}
     {\displaystyle
     \prod_{\substack{i,j\ge 0\\i+j \le -\langle\vec{k},
     \vec{\alpha}\rangle-1}}}
          (1-e^{\beta(i\ve_1 +j\ve_2 - \langle \vec{a}, \vec{\alpha}\rangle)})
       & \text{if $\langle \vec{k}, \vec{\alpha}\rangle < 0$}, \\
     {\displaystyle
     \prod_{\substack{i,j\ge 0\\i+j\le \langle \vec{k},
     \vec{\alpha}\rangle-2}}}
          \left(1-e^{\beta(-(i+1)\ve_1 - (j+1)\ve_2 - 
         \langle \vec{a},\vec{\alpha}\rangle)}\right)
       & \text{if $\langle\vec{k}, \vec{\alpha}\rangle > 1$}, \\
     1 & \text{otherwise}.
   \end{cases}
\end{equation}
From the geometric argument, Nakajima and Yoshioka established the following theorem in \cite{Nakajima:2005fg}
\begin{theorem}[Nakajima-Yoshioka]
\textup{(1)(}$d=0$ case\textup) 
\begin{equation}\label{eq:blow-up2}
Z^{\rm inst}_{k,0}(\ve_1,\ve_2,\vec{a};\q,\beta)
=
 (\q \beta^{2r} e^{-N\beta(\ve_1+\ve_2)/2})^{\frac{k(N-k)}{2N}}
Z^{\rm inst}(\ve_1,\ve_2,\vec{a};\q,\beta).
\end{equation}

\textup{(2)(}$0<d<N$ case\textup)
\begin{equation}\label{eq:blow-up3}
   Z^{\rm inst}_{k,d}(\ve_1,\ve_2,\vec{a};\q,\beta) = 
 \begin{cases}
   Z^{\rm inst}(\ve_1,\ve_2,\vec{a};\q,\beta) & \text{for $k = 0$,}\\
   0 & \text{for $0 < k < N$.}
 \end{cases}  
\end{equation}

\textup{(3)(}$d=N$ case\textup)
\begin{equation}
Z^{\rm inst}_{k,N}(\ve_1,\ve_2,\vec{a};\q,\beta)=
(-1)^{k(N-k)}(t_1 t_2)^{{k(N-k)}/{2}}
 (\q \beta^{2r} e^{-N\beta(\ve_1+\ve_2)/2})^{\frac{k(N-k)}{2N}}
Z^{\rm inst}(\ve_1,\ve_2,\vec{a};\q,\beta).
\end{equation}
\end{theorem}
Here $t_1=e^{\beta\ve_1}$, $t_2=e^{\beta\ve_2}$. Basically, the above theorem shows that for $0 < k < N,\ 0 < d < N$, $Z^{\rm inst}_{k,d}$ vanishes. We call this part of the theorem as vanishing blowup equations. While for $d=0,\ 0 < k < N$ and $k=0,\ 0 < d < N$ and $d=N,\ 0 < k < N$, $Z^{\rm inst}_{k,d}$ is proportional to $Z^{\rm inst}$, with the proportionality factor independent from $\vec{a}$! We call this part of theorem as unity blowup equations. In fact, if properly defined, $Z^{\rm inst}_{k,d}$ for $k=N,\ 0 < d < N$ is also proportional to $Z^{\rm inst}$ in this sense! In the context of refined topological strings, there is no difficulty to deal with such cases at all. Therefore, we can observe that the $(k,d)$ pair for vanishing cases are within a square and those for unity cases are exactly surrounding the square.

One can also combine the perturbative part and instanton part together to obtain the full 5D Nekrasov partition function. To defined the perturbative part, one need to introduce the following function
\begin{equation}
\begin{split}
  & \gamma_{\ve_1,\ve_2}(x|\beta;\Lambda) =
\frac{1}{2\ve_1\ve_2}\left(
-\frac{\beta}{6}\left(x+\frac{1}{2}(\ve_1+\ve_2)\right)^3 
+x^2\log(\beta\Lambda)\right)+\sum_{n \geq 1}\frac{1}{n}
\frac{e^{-\beta nx}}{(e^{\beta n\ve_1}-1)(e^{\beta n\ve_2}-1)},
\\
  & \widetilde{\gamma}_{\ve_1,\ve_2}(x|\beta;\Lambda)
\\
= \; & 
  \gamma_{\ve_1,\ve_2}(x|\beta;\Lambda)
  + \frac{1}{\ve_1 \ve_2} \left(\frac{\pi^2 x}{6
  \beta}-\frac{\zeta(3)}{\beta^2} \right)
  +\frac{\ve_1+\ve_2}{2\ve_1 \ve_2} 
  \left( x \log (\beta \Lambda)+\frac{\pi^2}{6\beta} \right)+
  \frac{\ve_1^2+\ve_2^2+3\ve_1 \ve_2}{12 \ve_1 \ve_2} \log(\beta\Lambda).
\end{split}
\end{equation}
Here $\Lambda = \q^{1/2N}$. Then the full Nekrasov partition function is defined by
	\begin{equation}
	\begin{aligned}
		Z(\epsilon_1,\epsilon_2,\vec{a};\fq,\beta)  &= \exp(-\sum_{\vec{\alpha}\in\Delta} \tilde{\gamma}_{\epsilon_1,\epsilon_2}(\langle \vec{a}, \vec{\alpha}\rangle | \beta;\Lambda)) Z^{\rm inst}(\epsilon_1,\epsilon_2,\vec{a};\fq,\beta) \ ,\\
		\widehat{Z}_{k,d}(\epsilon_1,\epsilon_2,\vec{a};\fq,\beta) &= \exp(-\sum_{\vec{\alpha}\in\Delta} \tilde{\gamma}_{\epsilon_1,\epsilon_2}(\langle \vec{a}, \vec{\alpha}\rangle | \beta;\Lambda)) \widehat{Z}^{\rm inst}_{k,d}(\epsilon_1,\epsilon_2,\vec{a};\fq,\beta) \ .
	\end{aligned}
	\end{equation}
Using formula \ref{Zinstkd}, one can obtain the blowup formula for the full partition function: 
	\begin{equation}\label{eq:Zhat-formula}
	\begin{aligned}
		\widehat{Z}_{k,d}(\epsilon_1,\epsilon_2,\vec{a};\fq,\beta) = \exp&\left[-\frac{(4d-N)(N-1)}{48} \beta(\epsilon_1+\epsilon_2)  \right] \\
		\times \sum_{\{\vec{k}\}=-k/N} & Z\left(\epsilon_1,\epsilon_2-\epsilon_1,\vec{a}+ \epsilon_1 \vec{k} ; \exp\left( \epsilon_1 (d - \tfrac{N}{2})\right) \fq ,\beta\right) \\
		\times & Z\left(\epsilon_1-\epsilon_2,\epsilon_2,\vec{a}+ \epsilon_2 \vec{k} ; \exp\left( \epsilon_2 (d- \tfrac{N}{2})\right) \fq ,\beta\right) \ .
	\end{aligned}	
	\end{equation}
Similar as the instanton partition function, the full partition function $\widehat{Z}_{k,d}$ also has simple relation with $Z$.  For $0 < k < N,\ 0 < d < N$, $\widehat{Z}_{k,d}$ just vanishes. For the unity $(k,d)$ pair in the instanton case, $\widehat{Z}_{k,d}$ is proportional to $Z$, with the proportionality factor independent from $\vec{a}$. These blowup equations for the full 5D Nekrasov partition function are in fact the special cases of the blowup equations for refined topological string theory, where the $(k,d)$ pair is generalized to non-equivalent $\br$ fields.

%Full Nekrasov partition functions defined by
%	\begin{equation}
%	\begin{aligned}
%		Z(\epsilon_1,\epsilon_2,\vec{a};\fq)  &= \exp(-\sum_{\vec{\alpha}\in\Delta} \tilde{\gamma}_{\epsilon_1,\epsilon_2}(\langle \vec{a}, \vec{\alpha}\rangle | \fq)) Z^{\rm inst}(\epsilon_1,\epsilon_2,\vec{a};\fq) \ ,\\
%		\widehat{Z}_{k,d}(\epsilon_1,\epsilon_2,\vec{a};\fq) &= \exp(-\sum_{\vec{\alpha}\in\Delta} \tilde{\gamma}_{\epsilon_1,\epsilon_2}(\langle \vec{a}, \vec{\alpha}\rangle | \fq)) \widehat{Z}^{\rm inst}_{k,d}(\epsilon_1,\epsilon_2,\vec{a};\fq) \ .
%	\end{aligned}
%	\end{equation}

%	\begin{equation}
%		\tilde{\gamma}_{\epsilon_1,\epsilon_2}(x;\fq) = \gamma^{\rm (cls)}_{\epsilon_1,\epsilon_2}(x;\fq) + \gamma^{\rm (1-loop)}_{\epsilon_1,\epsilon_2}(x) \ ,
%	\end{equation}
%	where
%	\begin{equation}
%	\begin{aligned}
%		\gamma^{\rm (cls)}_{\epsilon_1,\epsilon_2}(x;\fq) =& \frac{1}{24N\epsilon_1\epsilon_2}\log(\fq)(6x^2 + 6x(\epsilon_1+\epsilon_2) + \epsilon_1^2+\epsilon_2^2 + 3\epsilon_1\epsilon_2) \\
%		&+\frac{1}{12\epsilon_1\epsilon_2} \left( \pi^2(\epsilon_1+\epsilon_2) - \frac{1}{8}(2x+\epsilon_1+\epsilon_2)^3 + 2(\pi^2 x - 6\zeta(3))\right) \ ,
%	\end{aligned}	
%	\end{equation}
%	and
%	\begin{equation}
%		\gamma^{\rm (1-loop)}_{\epsilon_1,\epsilon_2}(x) = \sum_{n\geqslant 1} \frac{1}{n}\frac{e^{-n x}}{(e^{ n \epsilon_1}-1)( e^{ n \epsilon_2} - 1)} \ .
%	\end{equation}

Now we turn to the 5D $\mathcal{N}=1$ $SU(N)$ gauge theories theory with Chern-Simons term. Mathematically, one need to consider the line bundle ${\mathcal{L}}:=\lambda_{\mathcal{E}}(\mathcal{O}_{\IP^2}(-\ell_{\infty}))^{-1}$ where $\mathcal{E}$ be the universal sheaf on $\IP^2\times M(N,n)$. The instanton part of $K$-theoretic Nekrasov partition functions on $\IP^2$ with generic Chern-Simons level $m$ is defined by
\begin{equation}
Z^{\rm inst}_m(\ve_1,\ve_2,\vec{a};\Lambda,\beta)=\sum_{n=0}^\infty
((\beta\Lambda)^{2N}e^{-\beta(N+m)(\ve_1+\ve_2)/2})^n \sum_{i} (-1)^i \mathrm{ch} H^i(M(N,n),{\mathcal{L}}^{\otimes m}). 
\end{equation}
This instanton partition function is again computable using localization formula. On blowup $\widehat{\IP}^2$, again using Atiyah-Bott-Lefschetz fixed points formula, it was obtained in \cite{Gottsche:2006bm} that
{\allowdisplaybreaks
\begin{equation}\label{eq:blow-up2}
\begin{split}
  Z^{\rm inst}_{m,k,d}(&\ve_1,\ve_2,\vec{a};\Lambda,\beta) 
    =
  \exp(\frac{k^3m\beta}{6N^2}(\ve_1+\ve_2))
\\
  & \times
  \sum_{\{\vec{l}\} = -k/N}
   \frac{(\exp\left[{\beta(\ve_1+ \ve_2) 
       (d+m\left(-\frac12+\frac{k}N\right)-\frac{N}2)}\right]
   (\beta\Lambda)^{2N})^{(\vec{l},\vec{l})/2}}
   {\prod_{\vec\alpha\in\Delta}
     l^{\vec{l}}_{\vec{\alpha}}(\ve_1,\ve_2,\vec{a})}
\\
  & \times
  \exp\left[\beta(\vec{l},\vec{a})(d+m(-\frac12+\frac{k}N))\right]
\\
  & \times
  \exp\left[m\beta \left(\frac16(\ve_1+\ve_2)\sum_\alpha l_\alpha^3
      + \frac12\sum_\alpha l_\alpha^2 a_\alpha\right)\right]
\\
 &
   \times
  Z^{\rm inst}_m(\ve_1,\ve_2-\ve_1,\vec{a}+\ve_1\vec{l};
    \exp\left[{\frac{\beta\ve_1}{2r}
        \left\{d+m\left(-\frac12+\frac{k}N\right)-\frac{N}2\right\}}\right]
    \Lambda,\beta)
\\ 
  &
  \times
  Z^{\rm inst}_m(\ve_1-\ve_2,\ve_2,\vec{a}+\ve_2\vec{l};
    \exp\left[{\frac{\beta\ve_2}{2N}
        \left\{d+m\left(-\frac12+\frac{k}N\right)-\frac{N}2\right\}}\right]
    \Lambda,\beta).
\end{split}
\end{equation}
By} numerical computation, a conjectural blowup equation was proposed in \cite{Gottsche:2006bm}, which is
\begin{equation}
 Z^{\rm inst}_{m,0,d}(\ve_1,\ve_2,\vec{a};\Lambda,\beta) =
 Z^{\rm inst}_m(\ve_1,\ve_2,\vec{a};\Lambda,\beta)
\qquad\text{for $0\le d\le r$, $|m|\le r$.}
\end{equation}
These equations are in fact just some special cases of the unity blowup equations. By numerical computation, we conjecture the following blowup equations
\be\label{eq:instblm}
Z^{\rm inst}_{m,k,d}(\ve_1,\ve_2,\vec{a};\Lambda,\beta) =
\begin{cases}
   0 & \text{for $0 < k < N$, $0<d<N$,}\\
   f(m,k,d,N,\ve_1,\ve_2,\Lambda,\beta)Z^{\rm inst}_m(\ve_1,\ve_2,\vec{a};\Lambda,\beta) & \text{for $(k,d)\in S_{\mathrm{unity}}$,}
 \end{cases} 
\ee
where $S_{\mathrm{unity}}=\{(k,d)|d=0,0\le k<N\ \mathrm{or}\ 0<d<N, k=0\ \mathrm{or}\ d=N, 0\le k<r\ \mathrm{or}\ 0\le d\le N,k=N\}$. It is important that $f(m,k,d,\ve_1,\ve_2,\Lambda,\beta)$ does not depend on $\vec{a}$. In the context of refined topological string theory, this means that this the proportional factor does not depend on the true moduli of the Calabi-Yau.

It is also useful to introduce the full $K$-theoretic Nekrasov partition functions which are defined by
	\begin{equation}\label{eq:Zm}
	\begin{aligned}
		Z_m(\epsilon_1,\epsilon_2,\vec{a}; \fq,\beta) &= \exp\left(-\sum_{\vec{\alpha}\in \Delta} \tilde{\gamma}_{\epsilon_1,\epsilon_2}( \langle \vec{a} ,\vec{\alpha} \rangle\, | \beta,\Lambda ) - m\beta \sum_{\alpha=1}^{N} \frac{a_\alpha^3}{6\epsilon_1\epsilon_2}\right)  Z_m^{\rm inst}(\epsilon_1,\epsilon_2,\vec{a};\Lambda,\beta) \ , \\
		{Z}_{m,k,d}(\epsilon_1,\epsilon_2,\vec{a};\fq,\beta) &= \exp\left(-\sum_{\vec{\alpha}\in \Delta} \tilde{\gamma}_{\epsilon_1,\epsilon_2}( \langle \vec{a} ,\vec{\alpha} \rangle\, | \beta,\Lambda ) - m\beta \sum_{\alpha=1}^{N} \frac{a_\alpha^3}{6\epsilon_1\epsilon_2}\right)  {Z}_{m,k,d}^{\rm inst}(\epsilon_1,\epsilon_2,\vec{a};\Lambda,\beta).
	\end{aligned}
	\end{equation}
Then the blowup formula \eqref{eq:Zhat-formula} is generalized to
	\begin{equation}\label{eq:blowup-Zm}
	\begin{aligned}
		{Z}_{m,k,d}(\epsilon_1,\epsilon_2,\vec{a};\fq,\beta) =& \exp\left[\left( -\frac{(4(d+m(-\tfrac{1}{2}+\tfrac{k}{N}))-N)(N-1)}{48}  +\frac{k^3 m}{6N^2} \right) \beta(\epsilon_1+\epsilon_2)  \right] \\
		\times \sum_{\{\vec{k}\}=-k/N} &Z_m\left(\epsilon_1,\epsilon_2-\epsilon_1,\vec{a}+ \epsilon_1 \vec{k} ; \exp\left( \beta\epsilon_1 (d+m(-\tfrac{1}{2}+\tfrac{k}{N}) - \tfrac{N}{2})\right) \fq,\beta \right) \\
		\times &Z_m\left(\epsilon_1-\epsilon_2,\epsilon_2,\vec{a}+ \epsilon_2 \vec{k} ; \exp\left( \beta\epsilon_2 (d+m(-\tfrac{1}{2}+\tfrac{k}{N}) - \tfrac{N}{2})\right) \fq,\beta \right) \ .
	\end{aligned}	
	\end{equation}
The full partition function $\widehat{Z}_{m,k,d}$ satisfies similar equations like those for its instanton part (\ref{eq:instblm}), which we write as
\be\label{eq:fullblm}
Z_{m,k,d}(\ve_1,\ve_2,\vec{a};\fq,\beta) =
\begin{cases}
   0 & \text{for $0 < k < N$, $0<d<N$,}\\
   g(m,k,d,N,\ve_1,\ve_2,\fq,\beta)Z_m(\ve_1,\ve_2,\vec{a};\fq,\beta) & \text{for $(k,d)\in S_{\mathrm{unity}}$,}
 \end{cases} 
\ee
Note that it is important that $g(m,k,d,N,\ve_1,\ve_2,\Lambda,\beta)$ does not depend on $\vec{a}$. In Section \ref{sec:f1}, we will show the blowup equations for refined topological string on local $\mathbb{F}_1$, which is equivalent to the case $N=2,\ m=1$ in the gauge theory language.
\subsection{Vanishing blowup equations}\label{sec:vanish}
%%%%%%%%%%%%%%%%%%%%%%%%%%%%%%%%%%%%%%%%%%%%%%%%%%%%%%%%%%%%
The vanishing blowup equations for general local Calabi-Yau were already written down in \cite{Gu:2017ccq}, which is
\be\label{eq:blv}
	\sum_{\md n \in \mb Z^{g}}  (-1)^{|\md n|} \exp\(\widehat{F}_{\rm ref}\( \md t+ \epsilon_1\bR, \epsilon_1, \epsilon_2 - \epsilon_1 \) + \widehat{F}_{\rm ref}\( \md t+ \epsilon_2\bR, \epsilon_1 - \epsilon_2, \epsilon_2 \) \)=0,
\ee
where $\bR=\md C\cdot \md n + \mathbf{r}/2$. We call the integral vectors $\md r$ making the above equation hold as the \emph{vanishing $\md r$ fields}. The prerequisite such fields should satisfy is the $\md B$ field condition
\be\md r\equiv\md B\ (\mathrm{mod}\ (2\IZ)^b).
\ee
It is obvious that two different vectors $\md r$, $\md r'$ are equivalent for the vanishing blowup equation if
	\begin{equation}\label{eq:r-equivalence}
	\md r' = \md r + 2\md C \cdot \md n,\quad\quad \md n\in \IZ^g.
	\ee
We denote the number of non-equivalent vanishing $\md r$ fields as $w_{\rm v}$ and the set of non-equivalent vanishing $\md r$ fields as $\mathcal{S}_{\rm vanish}$. It was conjectured in \cite{Sun:2016obh} that
\be
g\le w_{\rm v}<\infty,
\ee
where $g$ is genus of the mirror curve. We will give a physical explanation for this conjecture later in Section \ref{ch:solver}.

It is easy to see the NS limit of the vanishing blowup equations give exactly the compatibility formulae (\ref{conjecture}), as mentioned before. Indeed,
\be
\lim_{\ve_1=\ri \hbar,\ \ve_2\to 0}\widehat{F}_{\rm ref}\( \md t+ \epsilon_1\(\md C\cdot \md n + \mathbf{r}/2\), \epsilon_1, \epsilon_2 - \epsilon_1 \)=F_{\text{unref}}\left(\mathbf{t}+ \ri\hbar \mathbf{n}\cdot\mathbf{C}+\frac{1}{2}\ri\hbar\mathbf{r},\hbar\right),
\ee
and
\be
\lim_{\ve_1=\ri \hbar,\ \ve_2\to 0}\widehat{F}_{\rm ref}\( \md t+ \epsilon_2\(\md C\cdot \md n + \mathbf{r}/2\), \epsilon_1 - \epsilon_2, \epsilon_2 \) =-\ri n_jC_{ji}\frac{\partial }{\partial t_i}F_{\text{NS}}\left(\mathbf{t},\hbar\right).
\ee

Now let us have a close look at the structure of the vanishing blowup equations. It easy to show that after dropping some irrelevant factors, equation (\ref{eq:blv}) can be written as
\be
\sum_{\md n \in \mb Z^{g}}  (-1)^{|\md n|} \exp\(G^{\rm v}(\md t,\md R,\eq,\et)\)=0,
\ee
where
\be
G^{\rm v}=G^{\rm v}_{\rm pert}+G^{\rm v}_{\rm inst},
\ee
\be\label{eq:Gvpert}
G^{\rm v}_{\rm pert}=-\half a_{ijk}t_iR_jR_k+\(\eq+\et\)\(-\frac{1}{6}\aijk R_iR_jR_k+\bi R_i-\bins R_i\),
\ee
and
\be\label{eq:Gvinst}
\begin{split}
G^{\rm v}_{\rm inst}&=F^{\rm inst}_{\rm ref}\( \md t+\ri\pi\md B+ \epsilon_1\bR, \epsilon_1, \epsilon_2 - \epsilon_1 \) + F^{\rm inst}_{\rm ref}\( \md t+\ri\pi\md B+ \epsilon_2\bR, \epsilon_1 - \epsilon_2, \epsilon_2 \)\\
&= \sum_{g,n} (-1)^{\md d\cdot \md B} n^{\md d}_{g,n}\( (\epsilon_1(\epsilon_2 - \epsilon_1))^{g-1}\epsilon_2^{2n} \re^{-\epsilon_1\md d\cdot \md R} + (\epsilon_2(\epsilon_1 - \epsilon_2))^{g-1} \epsilon_1^{2n} \re^{-\epsilon_2 \md d\cdot \md R} \)\re^{-\md d\cdot \bt}.
\end{split}
\ee
Here $n^{\md d}_{g,n}$ is the refined Gromov-Witten invariants defined by
\begin{equation}\label{eq:F-inst-exp}
	F^{\rm inst}_{\rm ref}(\md t,\epsilon_1,\epsilon_2) = \sum_{g,n,\md d}^\infty (\epsilon_1\epsilon_2)^{g-1}(\epsilon_1+\epsilon_2)^{2n} n^{\md d}_{g,n} \re^{-\md d \cdot \md t}.
	\end{equation}
It is important that $G^{\rm v}_{\rm pert}$ is a linear function of $\md t$ and the coefficients of $t_i$ are quadratic for $\md R$, as we will see later. Apparently, the vanishing blowup equations can be expanded with respect to $\md Q=e^{-\md t}$ and the vanishment of their coefficients at each degree gives some constraints among the refined BPS invariants.

The vanishing blowup equations (\ref{eq:blv}) can also be expanded with respect to $\eq$ and $\et$:
\be
\sum_{\md n \in \mb Z^{g}}  (-1)^{|\md n|} \exp\(\sum_{r=0}^{\infty}\sum_{s=0}^{\infty}G^{\rm v}_{(r,s)}(\md t,\md R,\eq,\et)\)=\sum_{r=0}^{\infty}\sum_{s=0}^{\infty}I^{\rm v}_{(r,s)}(\md t,\br)\epsilon_1^r\epsilon_2^s=0.
\ee
Since $\eq$ and $\et$ are arbitrary, we have 
\be
I^{\rm v}_{(r,s)}(\md t,\br)=0,\quad\quad\forall\, r \ge0,\,s\ge 0.
\ee 
We call these equations as the \emph{component equations of the vanishing blowup equations}. Dropping an irrelevant factor, the leading equations of vanishing blowup equations are 
\be
I^{\rm v}_{(0,0)}=\sum_{\md n \in \mb Z^{g}}  (-1)^{|\md n|} \exp \left(-\frac{1}{2} R^2 F_{(0,0)}''\right)=0,
\ee
where
\be
R^2 F_{(0,0)}''=\sum_{i,j}R_iR_j\frac{\d^2 F_{(0,0)}}{\d t_i\d t_j}.
\ee
In the following, we also use the abbreviation
\be
R^mF^{(m)}_{(n,g)}=\sum_{i_1,\dots,i_m} R_{i_1}R_{i_2}\cdots R_{i_{m}}\frac{\d}{\d t_{i_1}}\frac{\d}{\d t_{i_2}}\cdots\frac{\d}{\d t_{i_{m}}}F_{(n,g)}.
\ee
The expression of $I^{\rm v}_{(0,0)}$ is quite like the definition of Riemann theta function with characteristic, therefore we also use the notation
\be
\Theta_{\rm v}(\br)=(-1)^{|\md n|} \exp \left(-\frac{1}{2} R^2 F_{(0,0)}''\right).
\ee
Then the leading order of vanishing blowup equations can be written as
\be\label{eq:thetav}
\sum_{\md n \in \mb Z^{g}}\Theta_{\rm v}=0.
\ee

In the spirit that the classical information can uniquely determine the local Calabi-Yau and therefore determine the full instanton part of the refined partition function, we conjecture that all the $\md r$ fields satisfying the above equation give exactly the set $\mathcal{S}_{\rm vanish}$. This means that as long as an $\md r$ field make the leading order of vanishing blowup equation hold, it will make the whole vanishing blowup equation holds. This is a very strong conjecture and can be used to determine $\mathcal{S}_{\rm vanish}$ since the formulae for $F_{(0,0)},\ F_{(1,0)}$ and $F_{(0,1)}$ of local Calabi-Yau are well-known. 
%We will give evidences on this conjecture later.

The subleading ($r+s=1$) equations are
\be
I^{\rm v}_{(1,0)}=I^{\rm v}_{(0,1)}=\sum_{\md n \in \mb Z^{g}}\left(-\frac{1}{6} R^3 F_{(0,0)}^{(3)}+R\left(F_{(0,1)}'+F_{(1,0)}'\right)\right)\Theta_{\rm v}(\br)=0.
\ee
The sub-subleading ($r+s=2$) equations are
\be
\begin{split}
I^{\rm v}_{(2,0)}&=I^{\rm v}_{(0,2)}=\sum_{\md n \in \mb Z^{g}} \Bigg(\frac{1}{72} R^6 \left(F_{(0,0)}^{(3)}\right)^2+\frac{1}{24} R^4 \left(-4
   F_{(0,0)}^{(3)} F_{(0,1)}'-4 F_{(0,0)}^{(3)}
   F_{(1,0)}'-F_{(0,0)}^{(4)}\right)\\
   &+\frac{1}{2} R^2
   \left(\left(F_{(0,1)}'+F_{(1,0)}' \right)^2+F_{(0,1)}''\right)-F_{(0,2)}-3
   F_{(2,0)}
\Bigg)\Theta_{\rm v}(\br)=0,
\end{split}
\ee
and
\be
\begin{split}
I^{\rm v}_{(1,1)}&=\sum_{\md n \in \mb Z^{g}} \Bigg(\frac{1}{36} R^6 \(F_{(0,0)}^{(3)}\)^2+\frac{1}{24} R^4 \left(-8
   F_{(0,0)}^{(3)} F_{(0,1)}'-8 F_{(0,0)}^{(3)}
   F_{(1,0)}'-F_{(0,0)}^{(4)}\right)\\
   &+R^2 \left(\left(F_{(0,1)}'+F_{(1,0)}' \right)^2+\frac{1}{2}
   F_{(1,0)}''\right)+F_{(0,2)}-2 F_{(1,1)}-5 F_{(2,0)}\Bigg)\Theta_{\rm v}(\br)=0.
\end{split}
\ee
Due to leading order equation (\ref{eq:thetav}), obviously $F_{(0,2)}, F_{(1,1)},F_{(2,0)}$ in the above two equations can be dropped out. This fact will be used in section \ref{sec:counting} for the counting of independent component equations.

The order ($r+s=3$) equations are
\be
\begin{split}
I^{\rm v}_{(3,0)}&=I^{\rm v}_{(0,3)}=\sum_{\md n \in \mb Z^{g}}\Bigg(-\frac{1}{1296}R^9 \left(F_{(0,0)}^{(3)}\right)^3+\frac{1}{144} R^7
   F_{(0,0)}^{(3)} \left(2 F_{(0,0)}^{(3)}
   \left(F_{(0,1)}'+F_{(1,0)}'\right)+F_{(0,0)}^{(4)}\right)\\
   &+\frac{1}{120
   } R^5 \left(-10 F_{(0,0)}^{(3)}
   \left(\left(F_{(0,1)}'+F_{(1,0)}'\right)^2+F_{(0,1)}''\right)-5
   F_{(0,0)}^{(4)}
   \left(F_{(0,1)}'+F_{(1,0)}'\right)-F_{(0,0)}^{(5)}\right)\\
   &+\frac{1}{6}
   R^3 \left(3 \left(F_{(0,1)}'+F_{(1,0)}'\right)
   F_{(0,1)}''+\left(F_{(0,1)}'+F_{(1,0)}'\right)^3+(F_{(0,2)}+3
   F_{(2,0)}) F_{(0,0)}^{(3)}+F_{(0,1)}^{(3)}\right)\\
   &+R
   \left(-F_{(0,2)}'-(F_{(0,2)}+3 F_{(2,0)})
   \left(F_{(0,1)}'+F_{(1,0)}'\right)+F_{(2,0)}'\right)\Bigg)\Theta_{\rm v}(\br)=0,
\end{split}
\ee
and
\be
\begin{split}
I^{\rm v}_{(2,1)}&=I^{\rm v}_{(1,2)}=\sum_{\md n \in \mb Z^{g}}\Bigg(-\frac{1}{432} R^9 \left(F_{(0,0)}^{(3)}\right)^3+\frac{1}{72} R^7
   F_{(0,0)}^{(3)} \left(3 F_{(0,0)}^{(3)}
   \left(F_{(0,1)}'+F_{(1,0)}'\right)+F_{(0,0)}^{(4)}\right)\\
   &+\frac{1}{120
   } R^5 \left(-10 F_{(0,0)}^{(3)} \left(3
   \left(F_{(0,1)}'+F_{(1,0)}'\right)^2+F_{(0,1)}''+F_{(1,0)}''\right)-10
   F_{(0,0)}^{(4)}
   \left(F_{(0,1)}'+F_{(1,0)}'\right)-F_{(0,0)}^{(5)}\right)\\
&+\frac{1}{6}R^3 \left(3 \left(F_{(0,1)}'+F_{(1,0)}'\right)\left(\left(F_{(0,1)}'+F_{(1,0)}'\right)^2+F_{(0,1)}''+F_{(1,0)}''\right)+2 (F_{(1,1)}+4 F_{(2,0)}) F_{(0,0)}^{(3)}\right)\\
&+R\left(F_{(0,2)}'-2 (F_{(1,1)}+4 F_{(2,0)})
   \left(F_{(0,1)}'+F_{(1,0)}'\right)+F_{(1,1)}'+F_{(2,0)}'\right)
\Bigg)\Theta_{\rm v}(\br)=0.
\end{split}
\ee

Let us look at the general form of the component equations. In the refined topological string theory, it is convenient to introduce the concept of \emph{total genus} which is $g_{\rm t}=g+n$. Then the general form of order $r+s=2g_{\rm t}-2$ component equations can be written as
\be
I^{\rm v}_{(r,s)}=\sum_{\md n \in \mb Z^{g}}\Bigg(\sum_{l=0}^{3g_{\rm t}-3}R^{6g_{\rm t}-6-2l}\sum_{6g_{\rm t}-6-2l=\sum_ih_{ir}}c^{\rm v}_{rsg_{\rm t}l\star}\prod_{n_{i}+g_{i}\le g_{\rm t}} F_{n_{i},g_{i}}^{(h_{i})}\Bigg)\Theta_{\rm v}(\br)=0,
\ee
where $c_{rg_{\rm t}l\star}$ are some rational coefficients for each possible product of $F_{n_{i},g_{i}}^{(h_{i})}$. While the general form of order $r+s=2g_{\rm t}-1$ component equations can be written as
\be
I^{\rm v}_{(r,s)}=\sum_{\md n \in \mb Z^{g}}\Bigg(\sum_{l=0}^{3g_{\rm t}-2}R^{6g_{\rm t}-3-2l}\sum_{6g_{\rm t}-3-2l=\sum_ih_{ir}}c^{\rm v}_{rsg_{\rm t}l\star}\prod_{n_{i}+g_{i}\le g_{\rm t}} F_{n_{i},g_{i}}^{(h_{i})}\Bigg)\Theta_{\rm v}(\br)=0.
\ee
These structures will be used in the counting of independent component equations in section \ref{sec:counting}.

Now we consider how the blowup equations and the $\md r$ fields behave under the reduction of local Calabi-Yau. Here the reduction means to set some of the mass parameters to zero while the genus of mirror curve does not change. Such procedure is quite common. For example, local $\IP^2$ can be reduced from local $\IF_1$ and resolved $\mathbb{C}^3/\mathbb{Z}_5$ orbifold can be reduced from $SU(3)$ geometry $X_{3,2}$.

Since blowup equations can be expanded with respect to $\mathbf{Q}=\re^{-\mathbf{t}}$, thus under the reduction, one can simply set some $Q_m=e^{-t_m}$ to be zero in the blowup equations. Obviously, all the original vanishing $\br$ fields will result in the vanishing $\br $ fields of the reduced local Calabi-Yau. Note this is \emph{not} true for unity $\br$ fields, where some of the original unity $\br$ fields could turn to vanishing $\br$ fields after the reduction.
%%%%%%%%%%%%%%%%%%%%%%%%%%%%%%%%%%%%%%%%%%%%%%%%%%%%%%%%%%%%%%%%%%%%%%%%%%%%%%%%%%%%%%%%%%%%%%%%%%%%%%%%%%%%%%%
\subsection{Unity blowup equations}\label{sec:unity}
We propose the unity blowup equations for general local Calabi-Yau as
\be\label{eq:blu}
\sum_{\md n \in \mb Z^{g}}  (-1)^{|\md n|} \exp\(\widehat{F}_{\rm ref}\( \md t+ \epsilon_1\bR, \epsilon_1, \epsilon_2 - \epsilon_1 \) + \widehat{F}_{\rm ref}\( \md t+ \epsilon_2\bR, \epsilon_1 - \epsilon_2, \epsilon_2 \)- \widehat{F}_{\rm ref}\( \md t, \epsilon_1, \epsilon_2\)\)=\Lambda(\md m,\eq,\et,\br).
\ee
Clearly, for arbitrary vector $\br$ field, the left side of the equation can always be represented as some function $\Lambda(\md t,\eq,\et,\br)$. However, only when $\md r\in\mathcal{S}_{\mathrm{unity}}$, the factor $\Lambda(\md t,\eq,\et,\br)$ will be significantly simplified, in particular, independent from the true moduli of Calabi-Yau. That is why we write it as $\Lambda(\md m,\eq,\et,\br)$ where $\md m$ denote the mass parameters.

Separate the perturbative and instanton part of the refined partition function, the unity blowup equations (\ref{eq:blu}) can be written as
\be
\sum_{\md n \in \mb Z^{g}}  (-1)^{|\md n|} \exp\(G^{\rm u}(\md t,\md R,\eq,\et)\)=0,
\ee
where
\be
G^{\rm u}=G^{\rm u}_{\rm pert}+G^{\rm u}_{\rm inst},
\ee
\be\label{eq:Gupert}
G^{\rm u}_{\rm pert}=\(b_i+\bins\)t_i-\half a_{ijk}t_iR_jR_k+\(\eq+\et\)\(-\frac{1}{6}\aijk R_iR_jR_k+\bi R_i-\bins R_i\),
\ee
and
\be\label{eq:Guinst}
G^{\rm u}_{\rm inst}=F^{\rm inst}_{\rm ref}\( \md t+\ri\pi\md B+ \epsilon_1\bR, \epsilon_1, \epsilon_2 - \epsilon_1 \) + F^{\rm inst}_{\rm ref}\( \md t+\ri\pi\md B+ \epsilon_2\bR, \epsilon_1 - \epsilon_2, \epsilon_2 \)-F^{\rm inst}_{\rm ref}\( \md t+\ri\pi\md B, \epsilon_1, \epsilon_2 \).
\ee
Like the vanishing case, these equations can be expanded with respect to $\md Q=e^{-\md t}$. Once the exact form of $\Lambda$ factor is found, the equations of the expansion coefficients at each degree gives some constraints among the refined BPS invariants.

The unity blowup equations (\ref{eq:blu}) can also be expanded with respect to $\eq$ and $\et$ as
\be
\sum_{r=0}^{\infty}\sum_{s=0}^{\infty}I^{\rm u}_{(r,s)}(\md t,\br)\epsilon_1^r\epsilon_2^s=\Lambda_{(r,s)}(\md t,\br)\epsilon_1^r\epsilon_2^s.
\ee
Since $\eq$ and $\et$ are arbitrary, all $I^{\rm u}_{(r,s)}(\md t,\br)=\Lambda_{(r,s)}(\md t,\br)$. The leading order of the unity blowup equations are 
\be\label{eq:u00}
I^{\rm u}_{(0,0)}=\sum_R   (-1)^{|\md n|} \exp \left(-\frac{1}{2} R^2 F_{(0,0)}''+F_{(0,1)}-F_{(1,0)}\right)=\Lambda_{(0,0)}.
\ee
We also denote the summand in the above expression as $\Theta_{\rm u}(\br)$. Therefore the leading equations of the unity blowup equations can be written as
\be\label{first}
\sum_R\Theta_{\rm u}(\br)=\Lambda_{(0,0)}.
\ee
We call these equations as well as the leading order of vanishing blowup equations (\ref{eq:thetav}) as \emph{generalized contact term equations}. Unlike the vanishing case where leading order equations usually trivially vanish, the leading order equations for the unity case give genuine constrain on the free energy, as the well-known contact term equation on Seiberg-Witten prepotential. 

The other component equations are
\be\label{second}
I^{\rm u}_{(1,0)}=I^{\rm u}_{(0,1)}=\sum_{\md n \in \mb Z^{g}} \left(-\frac{1}{6} R^3 F_{(0,0)}^{(3)}+R\left(F_{(0,1)}'+F_{(1,0)}'\right)\right)\Theta_{\rm u}(\br)=\Lambda_{(1,0)}=\Lambda_{(0,1)}.
\ee
\be
\begin{split}
I^{\rm u}_{(2,0)}&=I^{\rm u}_{(0,2)}=\sum_{\md n \in \mb Z^{g}}\Bigg(\frac{1}{72} R^6 \left(F_{(0,0)}^{(3)}\right)^2+\frac{1}{24} R^4 \left(-4
   F_{(0,0)}^{(3)} F_{(0,1)}'-4 F_{(0,0)}^{(3)}
   F_{(1,0)}'-F_{(0,0)}^{(4)}\right)\\
   &+\frac{1}{2} R^2
   \left(\left(F_{(0,1)}'+F_{(1,0)}' \right)^2+F_{(0,1)}''\right)-F_{(0,2)}-3
   F_{(2,0)}
\Bigg)\Theta_{\rm u}(\br)=\Lambda_{(2,0)}=\Lambda_{(0,2)},
\end{split}
\ee
\be
\begin{split}
I^{\rm v}_{(1,1)}&=\sum_{\md n \in \mb Z^{g}}\Bigg(\frac{1}{36} R^6 \(F_{(0,0)}^{(3)}\)^2+\frac{1}{24} R^4 \left(-8
   F_{(0,0)}^{(3)} F_{(0,1)}'-8 F_{(0,0)}^{(3)}
   F_{(1,0)}'-F_{(0,0)}^{(4)}\right)\\
   &+R^2 \left(\left(F_{(0,1)}'+F_{(1,0)}' \right)^2+\frac{1}{2}
   F_{(1,0)}''\right)+F_{(0,2)}-2 F_{(1,1)}-5 F_{(2,0)}\Bigg)\Theta_{\rm u}(\br)=\Lambda_{(1,1)},\dots
\end{split}
\ee
The only difference between the l.h.s. of unity and vanishing blowup equations lies in $\Theta_{\rm u}(\br)$ and $\Theta_{\rm v}(\br)$. Note that unlike the vanishing case, the l.h.s of unity blowup equations could not be multiplied by a free factor. In some sense, the unity blowup equations are mote delicate and restrictive than the vanishing blowup equations. 

The general form of order $r+s=2g_{\rm t}-2$ component equations can be written as
\be\label{eq:urscomp1}
I^{\rm u}_{(r,s)}=\sum_{\md n \in \mb Z^{g}}\Bigg(\sum_{l=0}^{3g_{\rm t}-3}R^{6g_{\rm t}-6-2l}\sum_{6g_{\rm t}-6-2l=\sum_ih_{ir}}c^{\rm u}_{rsg_{\rm t}l\star}\prod_{n_{i}+g_{i}\le g_{\rm t}} F_{n_{i},g_{i}}^{(h_{i})}\Bigg)\Theta_{\rm u}(\br)=\Lambda^{\rm u}_{(r,s)},
\ee
where $c_{rg_{\rm t}l\star}$ are some rational coefficients for each possible product of $F_{n_{i},g_{i}}^{(h_{i})}$. The general form of order $r+s=2g_{\rm t}-1$ component equations can be written as
\be\label{eq:urscomp2}
I^{\rm u}_{(r,s)}=\sum_{\md n \in \mb Z^{g}}\Bigg(\sum_{l=0}^{3g_{\rm t}-2}R^{6g_{\rm t}-3-2l}\sum_{6g_{\rm t}-3-2l=\sum_ih_{ir}}c^{\rm u}_{rsg_{\rm t}l\star}\prod_{n_{i}+g_{i}\le g_{\rm t}} F_{n_{i},g_{i}}^{(h_{i})}\Bigg)\Theta_{\rm u}(\br)=\Lambda^{\rm u}_{(r,s)}.
\ee

A crucial fact about the factor $\Lambda(\mathbf{m},\ep_1,\ep_2,\mathbf{r})$ is that it only contains the mass parameters! For those geometries without mass parameter, like local $\IP^2$, the factor will be even simpler as $\Lambda(\ep_1,\ep_2,\mathbf{r})$. A good explanation perhaps can be the requirement of modular invariance of $\Lambda$, which we will elaborate in Section \ref{subsection:modular}. Besides, it is possible that under certain reduction of local Calabi-Yau, some $\Lambda(\mathbf{m},\ep_1,\ep_2,\mathbf{r})$ will become zero. In such cases, the unity $\br$ fields will become vanishing $\br$ fields for the reduced Calabi-Yau.
%%%%%%%%%%%%%%%%%%%%%%%%%%%%%%%%%%%%%%%%%%%%%%%%%%%%%%%%%%%
\subsection{Solving the $\mathbf{r}$ fields}\label{ch:solver}
In this section, we show the constraints on $\mathbf{r}$ fields from the leading expansion with respect to $\mathbf{Q}=\re^{-\mathbf{t}}$. Surprisingly, for all geometries we studied in this paper and our previous paper, these constraint result in all correct $\mathbf{r}$ fields, which perfectly agree with those obtained by scanning case by case.
\subsubsection{Unity $\mathbf{r}$ fields}
Motivated by \cite{Aganagic:2006wq}, it will be elaborated in section \ref{subsection:modular} that the left hand side of unity blowup equation (\ref{eq:blu}) is a quasi-modular form with weight 0. If blowup equations hold, then $\Lambda(\mathbf{t},\ep_1,\ep_2,\mathbf{r})$ is also a quasi-modular form of weight 0. In principle, for arbitrary $\mathbf{r}$ fields, $\Lambda(\mathbf{t},\ep_1,\ep_2,\mathbf{r})$ is an infinite series of $e^{-\mathbf{t}}$. But the interesting thing is that there exist a set of $\mathbf{r}$ fields that make $\Lambda(\mathbf{t},\ep_1,\ep_2,\mathbf{r})$ finite series. However, a finite series of $e^{-\mathbf{t}}$ normally could not be invariant under the modular transformation, so the only possible case is that $\Lambda(\mathbf{t},\ep_1,\ep_2,\mathbf{r})$ do not depend on the true K\''{a}hler parameters $\mathbf{t}$ or true moduli, but only depend on the mass parameters $\mathbf{m}$, and deformation parameters $\ep_1,\ep_2$ and $\mathbf{r}$ fields. The discuss
 ion above shows that if we expand the left hand side of (\ref{eq:blu}) with respect to $e^{-\mathbf{t}}$, the K\"ahler parameter related to the compact cycles of the mirror curve, then the zeroth order of the expansion should be the only remaining term, which could be a function $\Lambda(\mathbf{m},\ep_1,\ep_2,\mathbf{r})$ or simply 0. 

In section \ref{sec:np}, it will be shown that the blowup equations still hold after we complete them with Lockhart-Vafa non-perturbative partition function:
\be
\Lambda\(\md t,\tau_1,\tau_2,\md r\)=\sum_{\md n \in \mb Z^{g}}  (-1)^{|\md n|}\frac{Z_{\rm ref}^{(\np)}(\bt+2\pi\ri \tau_1\bR,\tau_1,\tau_2-\tau_1)Z_{\rm ref}^{(\np)}(\bt+2\pi\ri \tau_2\bR,\tau_1-\tau_2,\tau_2)}{Z_{\rm ref}^{(\np)}(\bt,\tau_1,\tau_2)}.
\ee
Make the transformation $\mathbf{t} \rightarrow \frac{2\pi\mathbf{t} }{\hbar},\tau_2\rightarrow \frac{2\pi}{\hbar} $ and take the limit 
\be\label{limit}
\tau_1,e^{\frac{2\pi{-\md t}}{\hbar}} \rightarrow 0,
\ee 
we have
\be\label{definer}
\Lambda\(\frac{2\pi{\md m}}{\hbar},0,\frac{2\pi}{\hbar},\md r\)=\lambda\(\frac{2\pi}{\hbar},\md r\)\sum_{\mathbf{n}\in \mathcal{I}}(-1)^{|\mathbf{n}|} e^{-\mathrm{i}\frac{2\pi^2}{3\hbar}a_{ijk}R_iR_jR_k-\frac{\pi}{3\hbar}a_{ijk}R_iR_jt_k},
\ee
where $\lambda$ is an overall term which does not depend on $\mathbf{r}$, and  $  \mathcal{I}=\{\mathbf{n}\in \mathbb{Z}^g|\forall k,f^k(\mathbf{n})\equiv\sum_{i,j}a_{ijk}R_iR_j=\text{min} \{f^k(\mathbf{n})|\mathbf{n}\in \mathbb{Z}^g\}\}$.

Let us explain equation (\ref{definer}) more. When we take the limit  $e^{-\frac{2\pi{\md t}}{\hbar}} \rightarrow 0$, the only possible contributions come from the lowest order of  $e^{-\frac{2\pi{\md t}}{\hbar}} $, because in physics, $a_{ijk}$ come from three point correlation function which indicate the lowest bounds always exist. Since we do the summation over all integers, we can always assume that the lowest bounds happen at $\mathbf{n}=0$. Then we can simplify the set $  \mathcal{I}$ as 
\be\label{Iset}
 \mathcal{I}=\{\mathbf{n}\in \mathbb{Z}^g|\forall k,f^k(\mathbf{n})\equiv\sum_{i,j}a_{ijk}R_iR_j=f^k(\mathbf{0})\}.
\ee

There is one more thing we should take into consideration. If $t_m$ is a pure mass parameter, we should exclude this $t_m$ in (\ref{Iset}). And for all other $t_k$ related to true K\"ahler parameter, the minima should happen simultaneously and must all be zero. Otherwise, $\Lambda$ will depend on the true moduli, which is prohibited by the modular invariance. 
 
We should also emphasize that the minima happen simultaneously is a very strong constraint that for incorrect $\mathbf{r}$ fields, in which cases $ \mathcal{I}$ is usually an empty set. Surprisingly, when we try to solve $\mathbf{r}$ fields which make (\ref{Iset}) not empty and select the solutions satisfying $\mathbf{r}=\mathbf{B}\mod\md 2$, we obtain exactly \textit{all} the correct $\mathbf{r}$ fields.

Once we have the correct unity $\br$ fields, the corresponding $\Lambda$ factor is just determined by the set $\mathcal{I}$ and the perturbative part of the refined free energy:
\be
\Lambda\(\md t,\tau_1,\tau_2,\md r\)=\sum_{\md n \in \mathcal{I}}  (-1)^{|\md n|}F_{\rm pert}(\bt+2\pi\ri \tau_1\bR,\tau_1,\tau_2-\tau_1)+F_{\rm pert}(\bt+2\pi\ri \tau_2\bR,\tau_1-\tau_2,\tau_2)-F_{\rm pert}(\bt,\tau_1,\tau_2).
\ee
This is easily computable and normally has simple expression.
%%%%%%%%%%%%%%%%%%%%%%%%%%%%%%%%%%%%%%%%%%%%%
\subsubsection{Vanishing $\mathbf{r}$ fields}\label{sec:vanishingr}
We now give similar method to derive the $\mathbf{r}$ fields for the vanishing blowup equation. The vanishing blowup equations in principle have no difference with the unity blowup equations, but only with $\Lambda=0$, which is trivially a  modular form of weight zero. Similarly there should exist some $\mathbf{r}$ vectors such that $f^k(0)$ are the minima for all $\mathbf{n}$. Different from the unity case, here it is not necessary $f^k(0)=0$, but the coefficients of $e^{-\mathbf{t}}$ should be cancelled. The simplest solution of this cancellation is that there is some sets $\mathcal{I}^a$ labeled by $a$ which only contain two elements.
\be
\mathcal{I}^a=\{\mathbf{0},n_j={\delta_{aj}}\}, \ a=1,\cdots,g.
\ee
Such choice of $\mathcal{I}^a$ has some interpretation in the quantization of mirror curves \cite{Grassi:2014zfa}. The limit we have chosen in (\ref{limit}) is actually the perturbative limit in quantum mechanics. In this limit, the grand potential becomes
\begin{equation}
J(\mathbf{t}+\ri\pi\mathbf{r},\hbar) \sim -\mathrm{i}\frac{2\pi^2}{3\hbar}a_{ijk}R_iR_jR_k-\frac{\pi}{3\hbar}a_{ijk}R_iR_jt_k+\mathrm{i} n_j \text{Vol}^{p}_{j}/\hbar,
 \end{equation}
 and the quantization conditions become
   \begin{equation}\label{qcsdnew}
\sum_{\mathbf{n}\in \mathbb{Z}^g}\re^{J(\mathbf{t}+\ri\pi\mathbf{r}+2\pi i \mathbf{n}\cdot\mathbf{C},\hbar) }\sim \sum_{\mathbf{n}\in \mathcal{I}}\re^{\mathrm{i} n_j \text{Vol}^{p}_{j}/\hbar}=0, 
 \end{equation}
If we choose $\mathcal{I}^a=\{\mathbf{0},n_j={\delta_{aj}}\}, \ a=1,\cdots,g$, then for each $r^a$, we have a divisor 
\be
 \text{Vol}^{p}_{a}=2\pi \hbar(n+1/2), a=1,\cdots,g.
\ee
The intersection of these divisors means to solve all these equations at the same time, which gives the NS quantization conditions. Here, we can also see that the number of vanishing $\mathbf{r}$ fields, and the divisors could not be smaller than the genus $g$ of the mirror curve. For higher genus cases, we may have more minima besides $\mathcal{I}^a=\{\mathbf{0},n_j={\delta_{aj}}\}$. For example, for $g=2$, we may (or may not, depend on models) have $\mathcal{I}^a=\{(0,0),(0,1),(1,0),(1,1)\}$ which may all make $\Lambda$ to be zero. This situation actually happens quite often. 
%we could check that, this kind of situation derive $\mathbf{r}$ field in the inner point of diamond graph for $SU(N)$ geometry.
 
To illustrate this method, the simplest non-trivial model is the resolved $\mathbb{C}^3/\mathbb{Z}_5$ orbifold. in this model, after some rescaling and shifting, the $f^k(\mathbf{n})$ functions are
\be
\begin{split}
&f^1(\mathbf{n})=-r_1 n_1 + 3 n_1^2 - r_2 n_2 - 2 n_1 n_2 + 2 n_2^2,\\
&f^2(\mathbf{n})=-r_1 n_2 - 3 r_2 n_2 + 5 n_2^2.
\end{split}
\ee
We can see that when $\mathcal{I}=\{(0,0),(1,0)\}$, the solution of $r_i$ is $(3,0),(3,-1),(3,-2)$. And when $\mathcal{I}=\{(0,0),(0,1)\}$, the solution of $r_i$ is $(-1,2)$. With the $\mathbf{B}$ field condition, $(3,-1)$ is ruled out. Then we obtain all the non-equivalent vanishing $\md r$ fields, $\mathbf{r}=(3,0),(3,-2),(-1,2)$. We have checked for all the models we have studied in the current paper and \cite{Sun:2016obh}, the $\mathbf{r}$ fields can always be obtained in this way.
%%%%%%%%%%%%%%%%%%%%%%%%%%%%%%%%%%%%%%%%%%%%%%%%%%%%%%
\subsection{Reflective property of the $\md r$ fields}
In this section, we will prove an important property of the $\md r$ fields, which we call the \emph{reflective property}, i.e. if $\md r$ makes the vanishing (unity) blowup equation hold, then $-\md r$ makes the vanishing (unity) blowup equation hold as well.

It is convenient to write the vanishing and unity blowup equations together:
\begin{equation}\label{eq:bltoge}
	\begin{aligned}
	\Lambda&\(\md t,\eq,\et,\md r\)=\sum_{\md n \in \mb Z^{g}}  (-1)^{|\md n|} \widehat{Z}_{\rm ref}\( \md t+ \epsilon_1\bR, \epsilon_1, \epsilon_2 - \epsilon_1 \) \widehat{Z}_{\rm ref}\( \md t+ \epsilon_2\bR, \epsilon_1 - \epsilon_2, \epsilon_2 \)/ \widehat{Z}_{\rm ref}\( \md t, \epsilon_1, \epsilon_2\)\\
	=&\sum_{\md n \in \mb Z^{g}}  (-1)^{|\md n|} \exp\(\widehat{F}_{\rm ref}\( \md t+ \epsilon_1\bR, \epsilon_1, \epsilon_2 - \epsilon_1 \) + \widehat{F}_{\rm ref}\( \md t+ \epsilon_2\bR, \epsilon_1 - \epsilon_2, \epsilon_2 \) -\widehat{F}_{\rm ref}\( \md t, \epsilon_1, \epsilon_2\)\)\\
	\end{aligned}
	\end{equation}
Here for $\md r=\mathbf{r}_{\rm v}$, $\Lambda=0$.

Let us take a substitution $\epsilon_{1,2}\to -\epsilon_{1,2}$ in equation (\ref{eq:bltoge}) and use the property of refined free energy (\ref{eq:con1}), we have
\be
\ba
&\Lambda\(\md t,-\eq,-\et,\md r\)\\
=&\sum_{\md n \in \mb Z^{g}}  (-1)^{|\md n|} \widehat{Z}_{\rm ref}\( \md t-\epsilon_1\bR, -\epsilon_1, -\epsilon_2 + \epsilon_1 \) \widehat{Z}_{\rm ref}\( \md t- \epsilon_2\bR, -\epsilon_1 + \epsilon_2, -\epsilon_2 \)/ \widehat{Z}_{\rm ref}\( \md t, -\epsilon_1, -\epsilon_2\)\\
=&\sum_{\md n \in \mb Z^{g}}  (-1)^{|\md n|} \widehat{Z}_{\rm ref}\( \md t- \epsilon_1\bR, \epsilon_1, \epsilon_2 - \epsilon_1 \) \widehat{Z}_{\rm ref}\( \md t- \epsilon_2\bR, \epsilon_1 - \epsilon_2, \epsilon_2 \)/ \widehat{Z}_{\rm ref}\( \md t, \epsilon_1, \epsilon_2\)
\ea
\ee
Consider the blowup equations for $-\md r$ and use the invariance of the summation under $\md n\to -\md n$,
\be
\Lambda\(\md t,\eq,\et,-\md r\)
=\sum_{\md n \in \mb Z^{g}}  (-1)^{|\md n|} \widehat{Z}_{\rm ref}\( \md t- \epsilon_1\bR, \epsilon_1, \epsilon_2 - \epsilon_1 \) \widehat{Z}_{\rm ref}\( \md t- \epsilon_2\bR, \epsilon_1 - \epsilon_2, \epsilon_2 \)/ \widehat{Z}_{\rm ref}\( \md t, \epsilon_1, \epsilon_2\)
\ee
Clearly, if one $\md r$ field makes the unity (vanishing) blowup equation hold, once we require
\be
\Lambda\(\md t,\eq,\et,-\md r\)=\Lambda\(\md t,-\eq,-\et,\md r\),
\ee
then $-\md r$ field makes the unity (vanishing) blowup equation hold as well.
%%%%%%%%%%%%%%%%%%%%%%%%%%%%%%%%%%%%%%%%%%%%%%%%%%%%%%%%%%%%%%%%%%%%%%%%%%%%%%%%%%%%%%%%%%%%%%%%%%%%%%%%%%%%%%%%%%%%%%

\subsection{Constrains on refined BPS invariants}\label{sec:constrain}
In this section, we will show the $\md B$ field condition of the refined BPS invariants $\N$ can actually be derived from the blowup equations. The ${\bf B}$ field condition is the key of the pole cancellation in both exact NS quantization conditions \cite{Wang:2015wdy} and HMO mechanism \cite{Hatsuda:2012dt}\cite{Hatsuda:2013gj}. At first, this condition was found in \cite{Hatsuda:2013oxa} for local del Pezzo CY threefolds. In our previous paper \cite{Sun:2016obh}, we gave a physical explanation on the existence of ${\bf B}$ field and an effective way to calculate ${\bf B}$ field for arbitrary toric Calabi-Yau threefold. Here, we show that the condition (\ref{eq:rcondition}) of the $\md r$ fields is the result of blowup equations. This implies the existence of the $\md B$ field and all $\md r$ fields are the representatives of the $\md B$ field.

In \cite{Sun:2016obh}, it was shown that to keep the form of quantum mirror curve unchanged when the Planck constant $\hbar$ is shifted to $\hbar+2\pi\ri$, the complex moduli must have the following transformation:
\be
z_i\rightarrow(-1)^{B_i}z_i,\quad\quad i=1,2,\dots,b.
\ee
Accordingly the K\"{a}hler moduli are shifted:
\be
\bt\to\bt+\pi\ri\md B.
\ee
Although the theory of refined mirror curve of local Calabi-Yau is not well developed yet, there has been much knowledge on the gauge theory side, called the double quantization of Seiberg-Witten geometry, see for example \cite{Kimura:2016ebq}\cite{Kimura:2017auj}\cite{Bourgine:2017jsi}. One of their basic observation is that in the refined case, it is $\ep_1+\ep_2$ that plays the role of quantum parameter. Following the same argument as in \cite{Sun:2016obh}, we find that when one shifts the deformation parameters in the following way:
\be\label{eq:echange}
\ep_1\to\ep_1+2\pi\ri m,\quad \ep_2\to\ep_2+2\pi\ri n,\quad m,n\in \IZ,
\ee
to keep the refined mirror curve unchanged, the complex moduli must have certain phase change:
\be\label{eq:zchange}
z_i\rightarrow(-1)^{B_i(m+n)}z_i,\quad\quad i=1,2,\dots,b.
\ee
For now, we only know such $\bf B$ field must exists but do not know any of its property. Accordingly the K\"{a}hler moduli are shifted:
\be\label{eq:tchange}
\bt\to\bt+\pi\ri(m+n){\bf B}.
\ee

Now let us look at how the blowup equations (\ref{eq:bltoge}) change under the simultaneous transformation (\ref{eq:echange}) and (\ref{eq:tchange}). We are not interested in the polynomial part here, because the polynomial contribution (\ref{eq:Gupert}) from the three refined free energies is linear with respect to $\bt$ and $\ep_1+\ep_2$ and its shift
\be
\pi\ri(m+n)\(b_i B_i+\bins B_i-\half a_{ijk}B_iR_jR_k-\frac{1}{6}\aijk R_iR_jR_k+\bi R_i-\bins R_i\)
\ee
should be absorbed into the phase change of the factor $\Lambda(\md t,\eq,\et,\br)$. Let us focus on the instanton part of the refined free energy. Using the fact that $\bf C$ is a integral matrix, we find that under the simultaneous transformation (\ref{eq:echange}) and (\ref{eq:tchange}), every summand in refined BPS formulation of $\widehat{F}_{\rm ref}\( \md t+ \epsilon_1\bR, \epsilon_1, \epsilon_2 - \epsilon_1 \)$ obtains a phase change:
\be\label{eq:factor1}
(-1)^{nw(2j_L+2j_R-1-{\bf B}\cdot{\bf d})+mw\({\bf B}+{\bf r}\)\cdot{\bf d}}.
\ee
Every summand in $\widehat{F}_{\rm ref}\( \md t+ \epsilon_2\bR, \epsilon_1 - \epsilon_2, \epsilon_2 \)$ obtains a phase change:
\be\label{eq:factor2}
(-1)^{mw(2j_L+2j_R-1-{\bf B}\cdot{\bf d})+nw\({\bf B}+{\bf r}\)\cdot{\bf d}}.
\ee
Every summand in $\widehat{F}_{\rm ref}\( \md t, \epsilon_1, \epsilon_2\)$ obtains a phase change:
\be\label{eq:factor3}
(-1)^{(m+n)w(2j_L+2j_R-1-{\bf B}\cdot{\bf d})}.
\ee
We know that the B model of local Calabi-Yau is determined by the mirror curve and the refined free energy is determined by the refined mirror curve. Since the refined mirror curve remains the same under the simultaneous transformation (\ref{eq:echange}) and (\ref{eq:zchange}), the blowup equations must still hold under the simultaneous transformation (\ref{eq:echange}) and (\ref{eq:tchange}). It is obvious that the only way to achieve that is to require all three factors in (\ref{eq:factor1}, \ref{eq:factor2}, \ref{eq:factor3}) to be identical to one. First, to require (\ref{eq:factor3}) to be one for arbitrary $m,n,w$, it means for all non-vanishing refined BPS invariants $\N$, there are constraints:
\be
2j_L+2j_R-1-{\bf B}\cdot{\bf d}\equiv 0\ (\mathrm{Mod}\ 2).
\ee
This is exactly the $\bf B$ field condition we introduced in the first place! Now we see it can actually be derived from the blowup equations. Substitute this definition into the (\ref{eq:factor1}) and (\ref{eq:factor2}), it is easy to see that to require them to be one for arbitrary $m,n,w$, there must be constraints:
\be
{\bf r}\equiv {\bf B}\ (\mathrm{Mod}\ {\bf 2}).
\ee
This means all $\md r$ fields are the representatives of the $\md B$ field, which is the prerequisite of $\md r$ fields we introduced in the first place.
%%%%%%%%%%%%%%%%%%%%%%%%%%%%%%%%%%%%%%%%%%%%%%%%%
\subsection{Blowup equations and (Siegel) modular forms}\label{subsection:modular}
Topological strings are closely related to modular forms. Such relation was first systematically studied in \cite{Aganagic:2006wq} and soon was used to solve the (refined) holomorphic anomaly equations for local Calabi-Yau in a series of papers \cite{Huang:2006si}\cite{Grimm:2007tm}\cite{Haghighat:2008gw}\cite{Huang:2010kf}\cite{Huang:2013yta}\cite{Huang:2014nwa}. All geometries studied in those paper have mirror curve of genus one and the corresponding free energies of topological string can be expressed by Eisenstein series, Dedekind eta function and Jacobi theta functions. For the geometries with mirror curve of higher genus, the free energies are related to Siegel modular forms \cite{Klemm:2015iya}, which are the high dimensional analogy of classical modular forms.

For the B model on a local Calabi-Yau manifold, there exists a discrete symmetry group $\Gamma$, which is generated by the monodromies of the periods. For example, for local $\IP^2$, the symmetry group is $\Gamma_3$, which is a subgroup of classical modular group $\Gamma_0=SL(2,\IZ)$. The main statement of \cite{Aganagic:2006wq} is that the genus $g$ topological string free energy, depending on the polarization, is either a holomorphic quasi-modular form or an almost holomorphic modular form of \emph{weight zero} under $\Gamma$. This fact can be directly generalized to refined topological string, which means every refined free energy $F_{(n,g)}$ is certain modular form of \emph{weight zero} under certain discrete group $\Gamma$.\footnote{Here the modular parameters come from the period matrix $\tau_{ij}$ which is connected to the prepotential $F_0$. There is fundamental difference between these general cases and the local Calabi-Yau with elliptic fibration, where the refined free ene
 rgy $F_{(n,g)}$ can be written as the modular form of elliptic fiber moduli $\tau$ and the weights are typically non-zero and related to $n$ and $g$.} 

The basic idea of $F_{(n,g)}$ is a (quasi-)modular form comes from Witten's observation that BCOV holomorphic anomaly equation
\be
\bar{\partial}_{\bar{i}} \mathcal{F}_g=\frac{1}{2}\bar{C}_{\bar{i}}^{jk}(D_jD_k \mathcal{F}_{g-1}+\sum_{r=1}^{g-1}D_j \mathcal{F}_r D_k\mathcal{F}_{g-r})
\ee
can be derived as quantization condition of quantizing Hilbert space parameterized by $x^I=t_i, p_I=\frac{\partial F^{(0,0)}}{\partial t_i}$. Here $W(t_i,\bar{t}_{\bar{i}})=e^{\sum_{g=0}^{\infty}g_s^{2g-2}\mathcal{F}_g(t_i,\bar{t}_{\bar{i}})}$ becomes quantized wave function, which should be invariant under $Sp(2n,\IZ)$ transformation $M$
\be
\begin{split}
\tilde{p}_I&={A_I}^J p_J+B_{IJ}x^J,\\
\tilde{x}^I&=C^{IJ} p_J+{D^I}_{J}x^J,
\end{split}
\ee
where
\be
M=\left(\begin{array}{cc}A& B \\C & D\end{array}\right) \in Sp(2n,\mathbb{Z}). 
\ee
More precisely, the invariance means a state $\left|Z\right\rangle$ is invariant under $Sp(2n,\IZ)$, but after choosing polarization $x$, the wave function $\left\langle x |Z\right\rangle $ indeed have changed.

In the holomorphic polarization, the refined free energy $\mathcal{F}_{(n,g)}(t,\bar{t})$ is invariant under $\Gamma$, which means they are modular forms of $\Gamma$ of weight zero. Besides, they are almost holomorphic, which means their anti-holomorphic dependence can be summarized in a finite power series in $(\tau-\bar{\tau})^{-1}$. While in the real polarization, $F_{(n,g)}(t)$ is holomorphic but not quasi-modular which means they are the constant part of the series expansion of $\mathcal{F}_{(n,g)}(t,\bar{t})$ in $(\tau-\bar{\tau})^{-1}$. It is convenient to introduce a holomorphic quasi-modular form of $E_{IJ}(\tau)$ of $\Gamma$ transform as \cite{Aganagic:2006wq}
\be
E^{IJ}(\tau) \rightarrow
{(C\tau +D)^{I}}_{K}\;{(C\tau+D)^{J}}_L \;E^{KL}(\tau)+C^{IL}{(C\tau+D)^{J}}_{L},
\ee
such that
\be
{\hat{E}}^{IJ}(\tau,\bar{\tau}) = E^{IJ}(\tau)+ \left(( \tau-{\bar{\tau}})^{-1}\right)^{IJ}
\ee
is a modular form and transforms as
\be
{\hat{E}}^{IJ}(\tau, \bar{\tau}) \rightarrow
{(C\tau +D)^{I}}_{K}\;{(C\tau+D)^{J}}_L \;{\hat{E}}^{KL}(\tau,\bar{\tau}),
\ee
under the modular transformation
\be
\tau\rightarrow (A\tau+B)(C\tau+D)^{-1}
\ee
where
\be
\begin{pmatrix}
A & B \\
C & D 
\end{pmatrix} 
\in\Gamma\subset Sp(2n, \IZ).
\ee
Here $E^{IJ}$ and ${\hat E}^{IJ}$ are just $\Gamma$ analogues of the second Eisenstein series $E_2(\tau)$ of $SL(2,
\IZ)$, and its modular but non-holomorphic counterpart $E_2(\tau,{\bar{\tau}})$. For the explicit construction for $E^{IJ}$ at genus two, see \cite{Klemm:2015iya}. Then the refined free energy $\mathcal{F}_{(n,g)}(t,\bar{t})$ in holomorphic polarization can be written as
\be
\mathcal{F}_{(n,g)}(t,\bar{t}) = h_{(n,g)}^{(0)}(\tau) +
({h_{(n,g)}^{(1)}})_{IJ}\, \hat{E}^{IJ}(\tau,\bar{\tau})+\ldots
+ ({h_{(n,g)}^{(3(n+g)-3)}})_{I_1\ldots I_{6(n+g)-6}}\,
\hat{E}^{I_1I_2}(\tau,\bar{\tau})\ldots
\hat{E}^{I_{6(n+g)-7}I_{6(n+g)-6}}(\tau,\bar{\tau}),
\ee
where $h_{(n,g)}^{(k)}(\tau)$ are holomorphic modular forms of $\Gamma$. This property is actually a direct consequence of the refined holomorphic anomaly equations. Sending $\bar{\tau}$ to infinity, 
\be
{{F}}_{(n,g)}(\tau) = \lim_{\bar{\tau} \rightarrow \infty}
\mathcal{F}_{(n,g)}(\tau,\bar{\tau})
\ee
one obtains the modular expansion of refined free energy in real polarization:
\be
{F}_{(n,g)}(t) = h_{(n,g)}^{(0)}(\tau) +
({h_{(n,g)}^{(1)}})_{IJ}\, {E}^{IJ}(\tau)+\ldots
+ ({h_{(n,g)}^{(3(n+g)-3)}})_{I_1\ldots I_{6(n+g)-6}}\,
{E}^{I_1I_2}(\tau)\ldots
{E}^{I_{6(n+g)-7}I_{6(n+g)-6}}(\tau).
\ee
These formulae also show that certain combinations of $F_{(n,g)}$ and their derivatives can be both modular and holomorphic.

Note that in our convention, there is $\mathbf{B}$ field adding on K\"ahler parameter, while it doesn't appear in the original paper \cite{Aganagic:2006wq}. We should explain why these two results match. As pointed out in \cite{Sun:2016obh}, $\mathbf{r}$ fields can be obtained by shifting the complex parameter $z_i$ with a phase $(-1)^{r_i}$:
\be\label{rfieldgenerat}
z_i\rightarrow (-1)^{r_i} z_i.
\ee
%we can indeed have $\mathbf{B}$ field in \cite{Aganagic:2006wq} by (\ref{rfieldgenerat}). 
This is actually only a change of variable, thus the free energies $F^{(n,g)}(\tau)$ do not change even though the period matrix $\tau(z)$ may be different. The only thing we need to care about is that the genus 0,1 parts indeed have changed. For genus 1 part, only a constant phase emerges. For genus 0 part,  we found the genus 0 free energy of local $\mathbb{P}^2$ becomes the same as our computation. We assume this happens quite general and will not mention the difference especially in B model.

In the previous sections, we write down the blowup equations in the real polarization. Our main assertion here is that \emph{the unity blowup equations (\ref{eq:bltoge}) are holomorphic modular forms of $\Gamma$ of weight $0$}.\footnote{The vanishing blowup equations can be regarded as the special cases of unity blowup equations.} This claim contains two parts: the first is that the $\Lambda(\bt,\br,\ep_1,\ep_2)$ factor is modular invariant, which is equivalent to say they are independent of the true K\"{a}hler moduli, since the mass parameters and $\epsilon_i$ do not change under the modular transformation. This is why we write the factor as $\Lambda(\mathbf{m},\br,\ep_1,\ep_2)$. This fact is extremely important in that it gives significant constraints on the unity $\br$ fields, which make it possible to solve all unity $\br$ fields merely from the polynomial part of the refined topological string. The second part is even more nontrivial, which claims the summation
\be
\sum_{\md n \in \mb Z^{g}}  (-1)^{|\md n|} \exp\(\widehat{F}_{\rm ref}\( \md t+ \epsilon_1\bR, \epsilon_1, \epsilon_2 - \epsilon_1 \) + \widehat{F}_{\rm ref}\( \md t+ \epsilon_2\bR, \epsilon_1 - \epsilon_2, \epsilon_2 \) -\widehat{F}_{\rm ref}\( \md t, \epsilon_1, \epsilon_2\)\)
\ee
is holomorphic modular forms of $\Gamma$ of weight $0$. Equivalently, that is to say all the expand coefficients $I^{\rm u}_{(r,s)}(\md t,\br)$ with respect to $\ep_1,\ep_2$ are holomorphic modular forms of $\Gamma$ of weight $0$. Let us look at the leading order of the unity blowup equations (\ref{eq:u00}). We first argue the infinite summation
\be\label{eq:sumtheta}
\sum_R   (-1)^{|\md n|} \exp \left(-\frac{1}{2} R^2 F_{(0,0)}''\right)
\ee
is a modular form of weight $1/2$. Since the period matrix
\be
\tau=\tau_{ij}=-C_{ik}C_{jl}\frac{\partial^2{F_{0,0}}}{\partial{t_k}\partial{t_l}},\quad\quad i,j=0,1,\dots,g,\quad k,l=0,1,\dots,b,
\ee
and $R_k=C_{ik}n_i+r_k/2$, we have
\be
\sum_R   (-1)^{|\md n|} \exp \left(-\frac{1}{2} R^2 F_{(0,0)}''\right)=\vartheta \left[\begin{array}{cc} \boldsymbol{\alpha}\\ \boldsymbol{\beta}
 \end{array}\right] \left(\tau_{ij},{\boldsymbol{0}}\right),
\ee
where $\vartheta$ is the Riemann theta function with rational characteristic $\boldsymbol{\alpha}$ and $\boldsymbol{\beta}$. Very much similar to the cases of Jacobi theta function, such theta functions at special value should be Siegel modular form of certain modular group $\Gamma$ of weight $1/2$. Although to our knowledge no mathematical theorem was established for the general cases, there is indeed a theorem for the genus one case: \textit{Every modular form of weight 1/2 is a linear combination of unary theta series.} This is called Serre-Stark theorem. We expect certain theorem can be established to the general cases (\ref{eq:sumtheta}). 

On the other hand, a general formula for $F_{(0,1)}$ was given in \cite{Klemm:2015iya}:
\be
\ba
\mathcal{F}_{(0,1)}(t,\bar{t})=&\frac{1}{2}\log\((\tau-\bar{\tau})^{-1}\)+\frac{1}{2}\log\(|\(\bar{G}^{-1}\)^{\bar{j}}_{\bar{i}}|\)+\frac{1}{2}\log\(\Delta^a\prod_iu_i^{a_i}m_j^{b_j}|\(G^{-1}\)^{j}_{i}|\)\\
=&\frac{1}{2}\log\((\tau-\bar{\tau})^{-1}\)+\frac{1}{2}\log\(|\(\bar{G}^{-1}\)^{\bar{j}}_{\bar{i}}|\)+F_{(0,1)}(t),
\ea
\ee
where
\be
G_i^j=\frac{\partial{t_i}}{\partial{z_j}}.
\ee
%We should emphasize that this form is only true for local Calabi-Yau. 
Since $(\tau-\bar{\tau})$ can be regarded as a almost-holomorphic modular form of weight $-2$, we have that $G_i^j$ is of weight $1$. Besides, $\exp{F_{(0,1)}(t)}$ becomes a quasi-modular form of weight $-1/2$, and ${F_{(1,0)}(t)}$ is modular invariant. Therefore, we observe that the weight of
\be
\sum_R   (-1)^{|\md n|} \exp \left(-\frac{1}{2} R^2 F_{(0,0)}''+F_{(0,1)}-F_{(1,0)}\right)
\ee
is zero. In fact, this expression should be both holomorphic and modular invariant.

%Since $e^{F_{(1,0)}-F_{(0,1)}}$ is always a weight $1/2$ form, the first also may be regarding as a method to find this theta series. It also seems a weight $1/2$ form will also corresponding to multi-variable theta series.

The higher order $I_{(r,s)}$ consist of many terms with form 
\be\label{eq:summand}
\sum_{\md n \in \mb Z^{g}}\Bigg(R^{m}\prod_{m=\sum_ih_{i}} F_{n_{i},g_{i}}^{(h_{i})}\Bigg)\Theta_{\rm u}(\br).
\ee
In order to study the modular property of $I_{(r,s)}$, we first have a review on Siegel modular forms. Siegel modular form of weight $\rho$ is defined as a holomorphic function $f$ on 
\be
\mathcal{H}_g=\{\tau\in M_{g\times g}(\mathbb{C})|\tau^T=\tau,\text{Im}(\tau)\text{ positive definite}\},
\ee
after modular transformation 
$\gamma=\begin{pmatrix}
A & B \\
C & D 
\end{pmatrix} \in \Gamma$,
\be
f(\gamma \tau)=\rho(C\tau+D) f(\tau),
\ee
where $\rho$ is a rational representation
\be
\rho:GL_g(\mathbf{C})\rightarrow GL(V),
\ee
 where $V$ is a finite-dimensional complex vector space. We may also study $n$th root of a Siegel modular form, which may have a constant term whose $n$th power is 1. And will also study the division of modular forms, which will not be holomorphic, but with some poles. In the following, we always refer (Siegel) modular form with such cases.

Some typical Siegel modular forms are the Riemann theta function with $z=0$. A Riemann theta function is defined as
\be
\Theta\left[\begin{array}{c}\bm{\alpha} \\ \bm{\beta}\end{array}\right](\bm{\tau},\bm{z})=\sum_{\bm{n}\in \mathbb{Z}^g}e^{\frac{1}{2} (\bm{n}+\bm{\alpha})\cdot\bm{\tau}\cdot(\bm{n}+\bm{\alpha})+(\bm{n}+\bm{\alpha})\cdot(\bm{z}+\bm{\beta})}.
\ee
Under modular transformation $\bm{\Gamma}\in Sp(2g,\mathbb{Z})$
\be
\bm{\Gamma}=\left(\bm{\begin{array}{cc} A& B \\C & D\end{array}}\right),
\ee
it has the transformation rule:
\bdm\label{modularrules}
\Theta\left[\begin{array}{c}\bm{D\alpha-C\beta}+\frac{1}{2}\text{diag}(\bm{CD}^T)\\ \bm{-B\alpha+A\beta}+\frac{1}{2}\text{diag}(\bm{AB}^T)\end{array}\right](\bm{(A\tau+B)(C\tau+D)^{-1}},((\bm{C\tau+D})^{-1})^T\bm{z})=\bm{\kappa(\alpha,\beta,\Gamma)}\sqrt{\det(\bm{C\tau+D})}e^{\pi i\bm{z}\cdot(\bm{(C\tau+D)^{-1}C})\cdot\bm{z}}\Theta\left[\begin{array}{c}\bm{\alpha} \\ \bm{\beta}\end{array}\right](\bm{\tau},\bm{z}),
\edm
where $\bm{\kappa(\alpha,\beta,\Gamma)}$ is a complex number depend on $\bm{\alpha,\beta,\Gamma}$. It is easy to see $\Theta\left[\begin{array}{c}\bm{\alpha} \\ \bm{\beta}\end{array}\right](\bm{\tau},\bm{0})$ is a weight $1/2$ Siegel modular form, with scalar valued weight 
\be
\sqrt{\det(\bm{C\tau+D})}.
\ee

In order to show 
\be\label{eq:sumtheta2}
\sum_R   (-1)^{|\md n|}R^m \exp \left(-\frac{1}{2} R^2 F_{(0,0)}''\right),
\ee
is a (quasi-)modular form, note that every $R_iR_j$ appearing there can be written as the $\tau_{ij}$ derivatives of 
\be\label{theta1}
\sum_R   (-1)^{|\md n|} \exp \left(-\frac{1}{2} R^2 F_{(0,0)}''\right),
\ee
or
\be\label{theta2}
\sum_R   (-1)^{|\md n|}R_I \exp \left(-\frac{1}{2} R^2 F_{(0,0)}''\right).
\ee
Obviously, (\ref{theta1}) is a special case of $\Theta\left[\begin{array}{c}\bm{\alpha} \\ \bm{\beta}\end{array}\right](\bm{\tau},0)$, and is a modular form. While (\ref{theta2}) is a special case of $\left.\frac{\partial}{\partial z_i}\Theta\left[\begin{array}{c}\bm{\alpha} \\ \bm{\beta}\end{array}\right](\bm{\tau},\bm{z})\right|_{\bm{z}=0}$ transform as 
\be
\left.\frac{\partial}{\partial z_I}\Theta\left[\begin{array}{c}\bm{\alpha} \\ \bm{\beta}\end{array}\right](\bm{\tau},\bm{z})\right|_{\bm{z}=0} \longrightarrow {(C\tau+D)^I }_J\det(\bm{C\tau+D})^{\frac{1}{2}} \frac{\partial}{\partial z_J}\left.\Theta\left[\begin{array}{c}\bm{\alpha} \\ \bm{\beta}\end{array}\right](\bm{\tau},\bm{z})\right|_{\bm{z}=0}.
\ee

We now show that $\tau_{IJ}$ derivative of a (quasi-)modular form is a quasi-modular form. Suppose $f^{I_1,\cdots,I_k}(\tau)$ symmetric in ${I_1,\cdots,I_k}$ is a modular form and transforms as 
\be
f^{I_1,\cdots,I_k}(\tilde\tau)= {(C\tau+D)^{I_1} }_{J_1}\cdots{(C\tau+D)^{I_k} }_{J_k}f^{J_1,\cdots,J_k}(\tau),
\ee
it is easy to show that
\be
\begin{split}
\left(\frac{\partial}{\partial \tilde\tau_{IJ}}-k E^{IJ}(\tilde \tau)\right)f^{I_1,\cdots,I_k}(\tilde\tau)=& {(C\tau+D)^{I_1} }_{J_1}\cdots{(C\tau+D)^{I_k} }_{J_k}\\
& \times {(C\tau+D)^{I} }_{L} {(C\tau+D)^{J} }_{M}\left({\partial \tau_{LM}}-kE^{LM}(\tau)\right)f^{J_1,\cdots,J_k}(\tau),
\end{split}
\ee
then we can see that the $\tau_{IJ}$ derivative of a modular form $f^{J_1,\cdots,J_k}(\tau)$ is indeed a quasi-modular form.
With the same method, we can show that the $\tau_{KL}$ derivative of $E^{IJ}$ is also a polynomial of $E^{IJ}$, and is also a quasi-modular form. And it is easy to understand any other $\tau_{KL}$ derivative of a (quasi-)modular is also a quasi-modular form. We conclude that (\ref{eq:sumtheta2}) is a quasi-modular form. Also because $\frac{\partial}{\partial t^I}=C_{IJK}\frac{\partial}{\partial \tau_{JK}}$, we know $\prod_{m=\sum_ih_{i}} F_{n_{i},g_{i}}^{(h_{i})}$ in (\ref{eq:summand}) is also a quasi-modular form. Now we conclude that (\ref{eq:summand}) is a quasi modular form. 

Let us now study the weight. Since $R_i \partial_{t_i}$ always appear together, and $R_i$ can be written as a $z_i$ derivative of Riemann theta function, their weights cancel with each other and contribute to zero weight (or $\rho(\cdot)=1$). Then we conclude that (\ref{eq:summand}) and $I_{(r,s)}$ is a quasi-modular form of weight zero for arbitrary $\mathbf{r}$ fields.

It is obvious that for general choice of $\mathbf{r}$ field, $\Lambda$ is a infinite series of $e^{-\md t}$, and $\Lambda$ is a quasi-modular form of weight zero. However, for our interest, we may expect that there exist some special $\mathbf{r}$ fields, which make $\Lambda$ a finite series of $e^{-\md t}$. It is known that a single $e^{-t_i}$ is not a modular form if $t_i$ is a true moduli, and finite sum of them should not be either. Then the only possible case is that $\Lambda$ is $t_i$ independent. Then $\Lambda$ could only be a function of $\mathbf{m},\ep_1,\ep_2$, and of weight zero. This will be the key point for us to derive all $\mathbf{r}$ fields and write down the blowup equations at conifold point and orbifold point.

%%%%%%%%%%%%%%%%%%%%%%%%%%%%%%%%%%%%%%%%%%%%%%%%%%%%%%%%%%

%%%%%%%%%%%%%%%%%%%%%%%%%%%%%%%%%%%%%%%%%%%%%%%%%%%%%%%%%%%%%%%%%%%%%%%%%%%%%%%%%%%%%%%%%%%%%%%%%%%%%%%%%%%%%%%%%%%%%%
\subsection{Relation with the GNY K-theoretic blowup equations}\label{sec:relationgny}
In this section, we show the relation between the G\"{o}ttsche-Nakajima-Yoshioka K-theoretic blowup equations and the blowup equations for the refined topological string on $X_{N,m}$ geometries. The part of vanishing blowup equations has been studied in \cite{Grassi:2016nnt}. We follow their calculation to study the unity blowup equations. 

The $X_{N,m}$ geometries are some local toric Calabi-Yau threefolds which engineer the 5D $SU(N)$ gauge theories with Chern-Simons level $m$. They describe the fibration of ALE singularity of $A_{N-1}$ type or $\IP^1$. Here $m$ is an integer satisfying $0\le m\le N$. As such geometries have been studied extensively, we do not want repeat the construction here. See for example \cite{Grassi:2016nnt} for the details about their toric fans, divisors and $\md C$ matrices. There are two typical choices for the K\''{a}hler moduli. One is ${t_1,\dots,t_{N-1},t_B}$, which directly correspond to the toric construction. The other is ${t_1,\dots,t_{N-1},t_N}$ where $t_N$ is purely a mass parameter. The two basis are related by
\begin{equation}\label{eq:tN-tB}
		t_N = t_B -  \sum_{i=1}^{\lfloor \tfrac{N+m-1}{2}\rfloor} (i-i\,  m/N) \, t_i - \sum_{i=\lfloor \tfrac{N+m+1}{2} \rfloor}^{N-1} (N+m-i - i\, m/N) \, t_i   \ .
	\end{equation}
In the following we will use the basis ${t_1,\dots,t_{N-1},t_N}$ to separate the mass moduli $t_N$. This is convenient because we want to show the $\Lambda$ factor in the unity blowup equations only depend on $t_N$ but not on $t_1,\dots,t_{N-1}$.

The $\dB$-field for the $X_{N,m}$ geometries with K\"{a}hler moduli $\{t_1,\ldots,t_{N-1},t_B\}$ is known to be
	\begin{equation}\label{eq:B-Nm}
		\dB \equiv (0, \ldots, 0, N+m) \mod (2\mbb Z)^{N-1} \ .
	\end{equation}
For $m=0$, the polynomial part of the refined topological string free energy is known to be
\begin{equation}\label{fpertSN}
		F^{\rm pert}_{\rm ref}(\bt; \epsilon_1,\epsilon_2) = \frac{1}{\epsilon_1\epsilon_2}\sum_{1\leq i< j\leq N}\left(\frac{t_{ij}^3 }{6} + \frac{t_N}{2N} t_{ij}^2 \right)+ \frac{\epsilon_1^2+\epsilon_2^2 + 3\epsilon_1\epsilon_2 - 4\pi^2}{\epsilon_1\epsilon_2}\sum_{i=1}^{N-1} \frac{i(N-i)}{12}t_i  \ ,
	\end{equation}
where
	\begin{equation}
		t_{ij} = \sum_{k=i}^{j-1} t_k \ .
	\end{equation}
For $m\neq 0$, the polynomial part does not have a general formula so far from the Calabi-Yau side. But one can use the gauge theory partition function to derive it.
	
Without loss of generality, we can assume $\beta=1$ in the K-theoretic partition function. Then the K-theoretic Nekrasov partition function in (\ref{eq:Zm}) with generic $m$ can be written as \cite{Grassi:2016nnt}
	\begin{equation}\label{eq:Zm-split}
		\log Z_m(\epsilon_1,\epsilon_2,\vec{a}; \fq) = F_m(\epsilon_1,\epsilon_2, \vec{a};\fq) + F_{\rm aux}(\epsilon_1,\epsilon_2,\vec{a};\fq) \ ,
	\end{equation}
	where
\begin{align}
		F_m(\epsilon_1,\epsilon_2,\vec{a};\fq) =& \;\frac{1}{\epsilon_1 \epsilon_2} \sum_{1\leq i< j \leq N}\(\frac{a_{ij}^3}{6} - \frac{\log(e^{-N \pi\ri}\fq)}{2N}a_{ij}^2 - 4\pi^2\frac{a_{ij}}{12} \) \nn
		& + 
		\frac{\epsilon_1^2 + \epsilon_2^2+3\epsilon_1\epsilon_2}{\epsilon_1\epsilon_2}\sum_{1\leq i < j \leq N}\frac{a_{ij}}{12} - \frac{m}{6\epsilon_1\epsilon_2} \sum_{\alpha=1}^{N} a^3_{\alpha} \nn
		&+\sum_{1\leq i < j \leq N} F^{\rm (1-loop)}(\epsilon_1,\epsilon_2,a_{ij})+ \log Z_m^{\rm inst}(\epsilon_1,\epsilon_2,\vec{a};\fq)  \ , \label{eq:Fm-pert}\\
		F_{\rm aux}(\epsilon_1,\epsilon_2,\vec{a};\fq) = &\sum_{1\leq i < j \leq N} 
		\( \frac{2\zeta(3)}{\epsilon_1\epsilon_2} - \frac{\epsilon_1+\epsilon_2}{\epsilon_1\epsilon_2} \pi \ri \frac{a_{ij}}{2} \right. \nn
		&\left. - \frac{\epsilon_1^2 + \epsilon_2^2 + 3\epsilon_1\epsilon_2}{24N\epsilon_1 \epsilon_2}\log(e^{-N \pi\ri }\fq) + \frac{\epsilon_1^3 + \epsilon_1^2\epsilon_2 + \epsilon_1\epsilon_2^2 + \epsilon_2^3}{48\epsilon_1\epsilon_2}\),
	\end{align}
and
	\be \label{eq:F-1loop}
		F^{\rm (1-loop)} (\epsilon_1, \epsilon_2, x) = \sum_{n=1}^\infty \frac{1}{n} \frac{e^{(\epsilon_1+\epsilon_2)n/2}+{e^{-(\epsilon_1+\epsilon_2)n/2}}}
		{(e^{\epsilon_1 n/2} -e^{-\epsilon_1 n/2})(e^{\epsilon_2 n/2} -e^{-\epsilon_2 n/2})} e^{-n x} \ .
	\ee
Here we use the notation
	\begin{equation}
		a_{ij} = \sum_{k=i}^{j-1} a_k \ .
	\end{equation}
Then under the substitution
	\begin{equation}\label{eq:dict}
	\left\{\begin{aligned}
		& a_i = t_i \ ,\\
		& \log \fq = -t_N + N\pi \ri \ ,
	\end{aligned}\right.
	\end{equation}
one can obtain the following identification between the Nekrasov partition function of gauge theory and the partition function of refined topological string theory on $X_{N,m}$ geometry:
	\begin{equation}\label{eq:Fm-Fref}
		F_m(\epsilon_1,\epsilon_2,\vec{a};\fq) = \widehat{F}_{\rm ref}(\dt;\epsilon_1,\epsilon_2) -ct_N.
	\end{equation}
Here $c$ is a constant depend on the convention. In fact, the linear terms corresponding to the mass parameters in the refined topological string partition function could not be fixed, nor do them matter. Different choices result in slightly different $\Lambda$ factor, but they do not affect the $\br$ fields. 

In the following, we will use (\ref{eq:blowup-Zm}), (\ref{eq:fullblm}) and (\ref{eq:Fm-Fref}) to obtain the blowup equations for the partition function of refined topological strings. The summation $\vec{k}$ in the blowup formula for the K-theoretic Nekrasov partition function (\ref{eq:blowup-Zm}) is subject to the constraint \eqref{kconst}. It is useful to write $\vec{k}$ as
	\begin{equation}\label{eq:k-prm}
		\vec{k} = \sum_{i=1}^{N-1} k_i \, \vec{\alpha}_i^\vee = \sum_{i=1}^{N-1} \tilde{k}_i \, \vec{\omega}_i \ ,
	\end{equation}
	where $\vec{\alpha}_i^\vee$ are the coroots of $su(N)$ and $\vec{\omega}_i$ are the fundamental weights of $su(N)$. Then we have
	\begin{equation}\label{eq:k-n}
		k_i = n_i + \frac{i k}{N} \ ,\quad n_i \in \IZ \ ,
	\end{equation}
	and
	\begin{equation}\label{eq:tk-n}
		\tilde{k}_i = \sum_{j=1}^{N-1} C_{ij}n_j + k \delta_{i,N-1} \ .
	\end{equation}
	Here $C_{ij}$ is the Cartan matrix of $su(N)$. 
	
	Now we can replace the Nekrasov partition functions in the blowup equation by the twisted topological string free energy $\widehat{F}_{\rm ref}(\dt;\epsilon_1,\epsilon_2)$ as well as the auxiliary function $F_{\rm aux}(\epsilon_1,\epsilon_2,\vec{a};\fq)$ using identifications \eqref{eq:Zm-split}, \eqref{eq:Fm-Fref}. In the unity blowup equations, the contribution of the three auxiliary functions is
	\begin{equation}\label{eq:compt-1}
	\begin{aligned}
	 \sum_{\{\vec{k}\}=-k/N} &\exp\left[F_{\rm aux}(\epsilon_1,\epsilon_2-\epsilon_1,\vec{a} + \epsilon_1 \vec{k};\fq)+F_{\rm aux}(\epsilon_1-\epsilon_2,\epsilon_2,\vec{a} + \epsilon_2 \vec{k};\fq)-F_{\rm aux}(\epsilon_1,\epsilon_2,\vec{a};\fq)\right] \\
	=& \exp \(\sum_{1\leq i <j \leq N} \(-{\pi\ri \over 2}\sum_{\ell =i}^{j-1}\tilde k_{\ell}+\frac{(N+m-2d)(\eq+\et)}{24 N}-\frac{\eq+\et}{24}\)\)\\ 
=	& \exp\( -\pi\ri \sum_{i=1}^{N-1} n_i\)\exp\(\frac{(m-2d)(N-1)(\eq+\et)}{48}\) \ .
	\end{aligned}
	\end{equation}
Considering blowup equations (\ref{eq:blowup-Zm}) and (\ref{eq:fullblm}), we have 
\begin{equation}
	\begin{aligned}
&\sum_{\{\vec{k}\}=-k/N} Z_m\left(\epsilon_1,\epsilon_2-\epsilon_1,\vec{a}+ \epsilon_1 \vec{k} ; \exp\left( \epsilon_1 (d+m(-\tfrac{1}{2}+\tfrac{k}{N}) - \tfrac{N}{2})\right) \fq \right) \\
		&\times Z_m\left(\epsilon_1-\epsilon_2,\epsilon_2,\vec{a}+ \epsilon_2 \vec{k} ; \exp\left( \epsilon_2 (d+m(-\tfrac{1}{2}+\tfrac{k}{N}) - \tfrac{N}{2})\right) \fq \right)/Z_m(\ve_1,\ve_2,\vec{a};\fq) \\
		= &g(m,k,d,N,\ve_1,\ve_2,\fq) \exp\left[\left( \frac{(4(d+m(-\tfrac{1}{2}+\tfrac{k}{N}))-N)(N-1)}{48}  -\frac{k^3 m}{6N^2} \right) (\epsilon_1+\epsilon_2)  \right].
	\end{aligned}	
	\end{equation}
On the other hand, using identifications \eqref{eq:Zm-split}, \eqref{eq:Fm-Fref} we have
\be
\begin{aligned}
&\sum_{\{\vec{k}\}=-k/N} Z_m\left(\epsilon_1,\epsilon_2-\epsilon_1,\vec{a}+ \epsilon_1 \vec{k} ; \exp\left( \epsilon_1 (d+m(-\tfrac{1}{2}+\tfrac{k}{N}) - \tfrac{N}{2})\right) \fq \right) \\
		&\times Z_m\left(\epsilon_1-\epsilon_2,\epsilon_2,\vec{a}+ \epsilon_2 \vec{k} ; \exp\left( \epsilon_2 (d+m(-\tfrac{1}{2}+\tfrac{k}{N}) - \tfrac{N}{2})\right) \fq \right)/Z_m(\ve_1,\ve_2,\vec{a};\fq) \\
		=&\sum_{\md n \in \mb Z^{g}}  (-1)^{|\md n|} \exp\(\widehat{F}_{\rm ref}\( \md t+ \epsilon_1\bR, \epsilon_1, \epsilon_2 - \epsilon_1 \) + \widehat{F}_{\rm ref}\( \md t+ \epsilon_2\bR, \epsilon_1 - \epsilon_2, \epsilon_2 \)- \widehat{F}_{\rm ref}\( \md t, \epsilon_1, \epsilon_2\)\)\times e^{-ct_N}\\
		&\times\sum_{\{\vec{k}\}=-k/N}\exp\left[F_{\rm aux}(\epsilon_1,\epsilon_2-\epsilon_1,\vec{a} + \epsilon_1 \vec{k};\fq)+F_{\rm aux}(\epsilon_1-\epsilon_2,\epsilon_2,\vec{a} + \epsilon_2 \vec{k};\fq)-F_{\rm aux}(\epsilon_1,\epsilon_2,\vec{a};\fq)\right]\\
=&\sum_{\md n \in \mb Z^{g}} \exp\(\widehat{F}_{\rm ref}\( \md t+ \epsilon_1\bR, \epsilon_1, \epsilon_2 - \epsilon_1 \) + \widehat{F}_{\rm ref}\( \md t+ \epsilon_2\bR, \epsilon_1 - \epsilon_2, \epsilon_2 \)- \widehat{F}_{\rm ref}\( \md t, \epsilon_1, \epsilon_2\)\)\\
		&\times e^{-ct_N} \exp\( -\pi\ri \sum_{i=1}^{N-1} n_i\)\exp\(\frac{(m-2d)(N-1)(\eq+\et)}{48}\) ,
\end{aligned}
\ee
where $\bR=\md C\cdot \md n + \mathbf{r}/2$ and the vectors $\mbf r = (r_1,\ldots,r_{N-1},r_N)$ are
	\begin{equation}\label{eq:B-N}
	r_i = \begin{cases}
		0 \ , & i\leq N-2 \\
		2k \ ,& i=N-1 \\
		N-2d- 2m\(-\tfrac{1}{2} + \tfrac{k}{N}\)  \ , & i=N \; (t_N)\ .
	\end{cases}
	\end{equation}
	with $(k,d)\in\mathcal{S}_{\rm unity}$. Finally we arrive at the unity blowup equations for $X_{N,m}$ geometries:
\be
\begin{split}
&\sum_{\md n \in \mb Z^{g}}  (-1)^{|\md n|} \exp\(\widehat{F}_{\rm ref}\( \md t+ \epsilon_1\bR, \epsilon_1, \epsilon_2 - \epsilon_1 \) + \widehat{F}_{\rm ref}\( \md t+ \epsilon_2\bR, \epsilon_1 - \epsilon_2, \epsilon_2 \)- \widehat{F}_{\rm ref}\( \md t, \epsilon_1, \epsilon_2\)\)\\
=&g(m,k,d,N,\ve_1,\ve_2,(-1)^Ne^{-t_N})\exp\(\(\frac{\(6d-3m+\frac{4km}{N}-N\)(N-1)}{48}-\frac{k^3m}{6N^2}\)(\ve_1+\ve_2)+ct_N\).
\end{split}
\ee
One can see that the $\Lambda$ factor in the second indeed does not depend on the true moduli $t_1,\dots,t_{N-1}$. Note that the above equations only hold for the $\md r$ fields satisfying (\ref{eq:B-N}). It is possible $r_N$ is not integer. This is not a problem at all, because $t_N$ may not be an integral flat coordinate for general $N$ and $m$. Once we switch to the basis $t_1,\dots,t_{N-1},t_{B}$, the $\md r$ fields will always be integral. 
	%%%%%%%%%%%%%%%%%%%%%%%%%%%%%%%%%%%%%%%%%%%%%%%%%%%%%%%%%%%%%%%%%%%%%%%%%%%%%%%%%%%%%%%%%%%%%%%%%%%%%%%%%%%%%%%%%%%%%%%%%%%%%%%%%%%%%%%%%%%%%%%%%%%%%%%%%%%%%%%%%%%%%%%%%%%%%%%%%%%%%%%%%%%%%%%%%%%%%5

%number of r fields
%%%%%%%%%%%%%%%%%%%%%%%%%%%%%%%%%%%%%%%%%%%%%%%%%%%%%%%%%%%%%%%
\subsection{Interpretation from M-theory}\label{sec:brane}
In this section, we would like to give a highly speculative interpretation on the blowup equations from M-theory. Before going into M-theory, let us first go back to Nekrasov's Master Formula for the partition function of $\mathcal{N}=2$ gauge theories on general toric four dimensional manifolds \cite{Nekrasov:2003}. It was shown by Nekrasov that any toric four-manifold $M$ admits a natural $\eq,\et$ deformation and $\mathcal{N}=2$ gauge theories can be well-defined on them. Using the equivariant version of Atiyah-Singer index theorem, the partition function of $U(N)$ $\mathcal{N}=2$ gauge theories on $X$ can be expressed via the original Nekrasov partition function on $\IC^2$:
\be
Z_{\vec{w}}(\vec{a},\vec{\ep})=\sum_{\vec{k}_a\in \IZ^N,\{\vec{k}_a\}=w_a}\prod_v Z(\vec{a}+\sum_ak_a\phi_a^{(v)}(\ep),w_{v_1},w_{v_2}),
\ee
in which $H^2(M)=\IZ^d$. In the presence of Higgs vev, the $U(N)$ gauge bundle is reduced to $U(1)^N$ bundle and the $d$ vectors $\vec{k}_a=(k_{a,l}),\ a=1,\dots,d$ classify all equivalence classes of $U(1)^N$ bundle. One need to sum over all equivalence classes and fix the traces : $\{\vec{k}_a\}=w_a=\sum_{l=1}^Nk_{a,l}$. In fact, in \cite{Nekrasov:2003} Nekrasov calculated the more general cases which include the presence of 2-observables, but here we only focus on the partition function.

The simplest example beyond $\IC^2$ is just its one-point blowup $\hat{\IC}^2$. For this toric complex surface, the Master Formula reduces to:
\be
Z_{\hat{\IC}^2}(\vec{a},\ep_1,\ep_2,\ep_3)=\sum_{\vec{k}\in \IZ^N,\{\vec{k}\}=w}Z(\vec{a}+\vec{k}\ep_1,\ep_1+\ep_3,\ep_2-\ep_1)Z(\vec{a}+\vec{k}\ep_2,\ep_1-\ep_2,\ep_2+\ep_3),
\ee
where $\ep_3$ controls the size of the exceptional divisor $\IP^1$. For $\ep_3=0$, the above formula coincides with the G\"{o}ttsche-Nakajima-Yoshioka K-theoretic formulae.

To understand the blowup equations for general local Calabi-Yau takes two steps. First we want to argue that the situations for $\hat{\IC}^2$ with the exceptional divisor $\IP^1$ of vanishing size is very similar to those for $\IC^2$. Of course $\hat{\IC}^2$ with the exceptional divisor $\IP^1$ of vanishing size itself is almost the same as $\IC^2$ except for the singular origin. It is well-known that partition function of M-theory compactified on local Calabi-Yau $X$ and five dimensional Omega background is equivalent to the partition function of refined topological string on $X$ \cite{Dijkgraaf:2006um}:
\be\label{eq:Mc2}
Z_{\mathrm{M-theory}}(X\times S^1\ltimes \IC_{\ep_1,\ep_2})=Z_{\rm ref}(X,\eq,\et)
\ee
In physics, the refined BPS invariants encoded in refined topological string partition function count the refined BPS states on 5D Omega background which comes from M2-branes wrapping the 11th dimensional circle $S^1$ and the holomorphic curves in $X$. Let us further consider M-theory compactified on local Calabi-Yau $X$ and Omega deformed $\hat{\IC}^2$, see Figure \ref{fig:Mbackground}. In this case, the M2-branes can either warp $S^1$ and the holomorphic curves in $X$ or the exceptional divisor $\IP^1$ or the both. In the first circumstance, the refined BPS counting should be exactly the same with $\IC^2$ case. While in the second and third circumstance, it will contribute to the M-theory partition function with terms relevant to the size of the divisor $\IP^1$. However, when we shrink the size of blowup divisor to be zero, it can be expected that the second and third circumstances only contribute to the M-theory partition function an overall factor:
\be\label{eq:Mc2blowup}
Z_{\mathrm{M-theory}}(X\times S^1\ltimes \widehat{\IC}_{\ep_1,\ep_2})\propto Z_{\rm ref}(X,\eq,\et).
\ee
We expect such factor explains the existence of the $\Lambda$ factor in the blowup equations (\ref{eq:bltoge}). 
\begin{figure}[htbp]
\centering\includegraphics[width=5in]{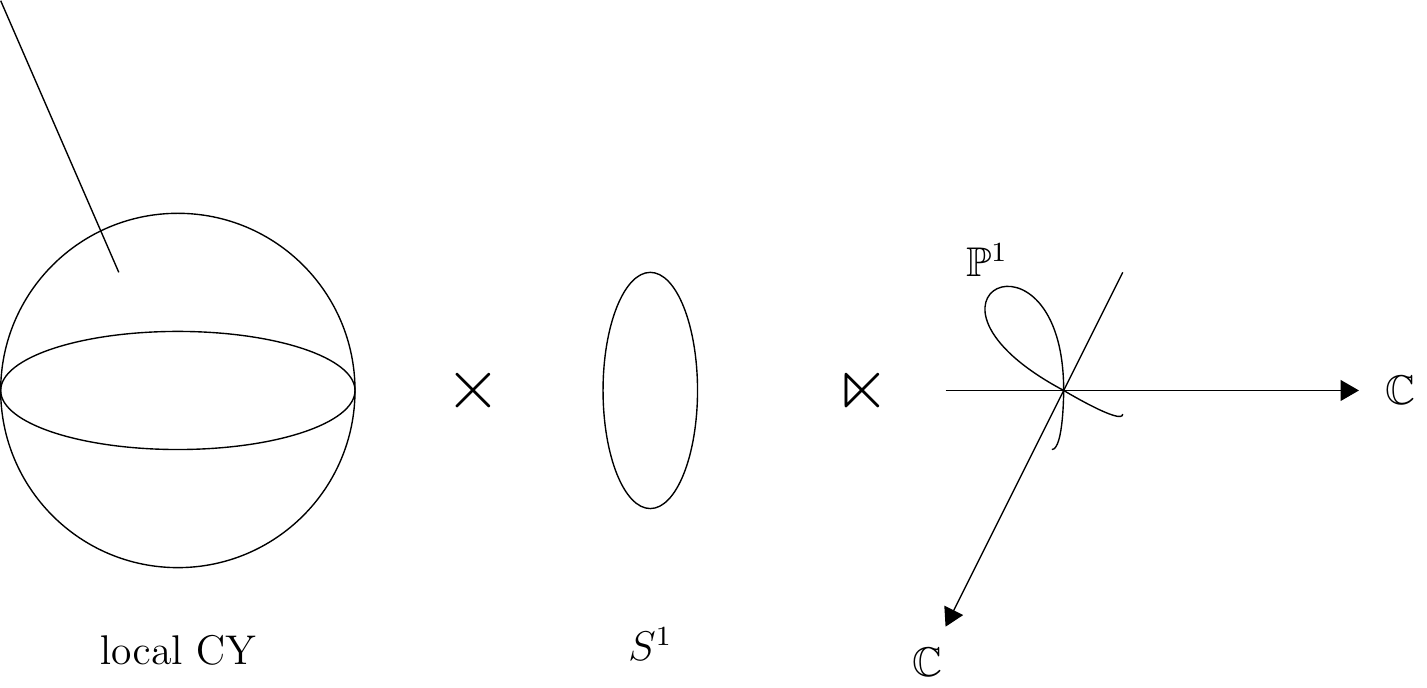}
\caption{The background of M-theory.}\label{fig:Mbackground}
\end{figure}

Now it leaves the question how to actually compute the $Z_{\mathrm{M-theory}}(X\times S^1\ltimes \widehat{\IC}_{\ep_1,\ep_2})$. This relies much on the inspiration from supersymmetric gauge theories. Let us have a close look at $\hat{\IC}^2$, see the toric diagram of $\hat{\IC}^2$ in Figure \ref{fig:c2blowup}. The length of the slash controls the size of the blowup divisor $\IP^1$. There are two fixed points of torus $\mathbb{T}^2_{\ep_1,\ep_2}$ action. Remembering the $\mathbb{T}^2_{\ep_1,\ep_2}$ acts on $\IC^2$ as
\be
(z_1,z_2)\ \sim\  (z_1\re^{\beta\ep_1},z_2\re^{\beta\ep_2})
\ee
Since the homogeneous coordinates near the fixed points are respectively $(z_1,z_2/z_1)$ and $(z_1/z_2,z_2)$, thus the $\mathbb{T}^2$ weight are $(\ep_1,\ep_2-\ep_1)$ and $(\ep_1-\ep_2,\ep_2)$ respectively on the two patches. This actually explains the behavior of $\ep$ in the blowup equations.
\begin{figure}[htbp]
\centering\includegraphics[width=5in]{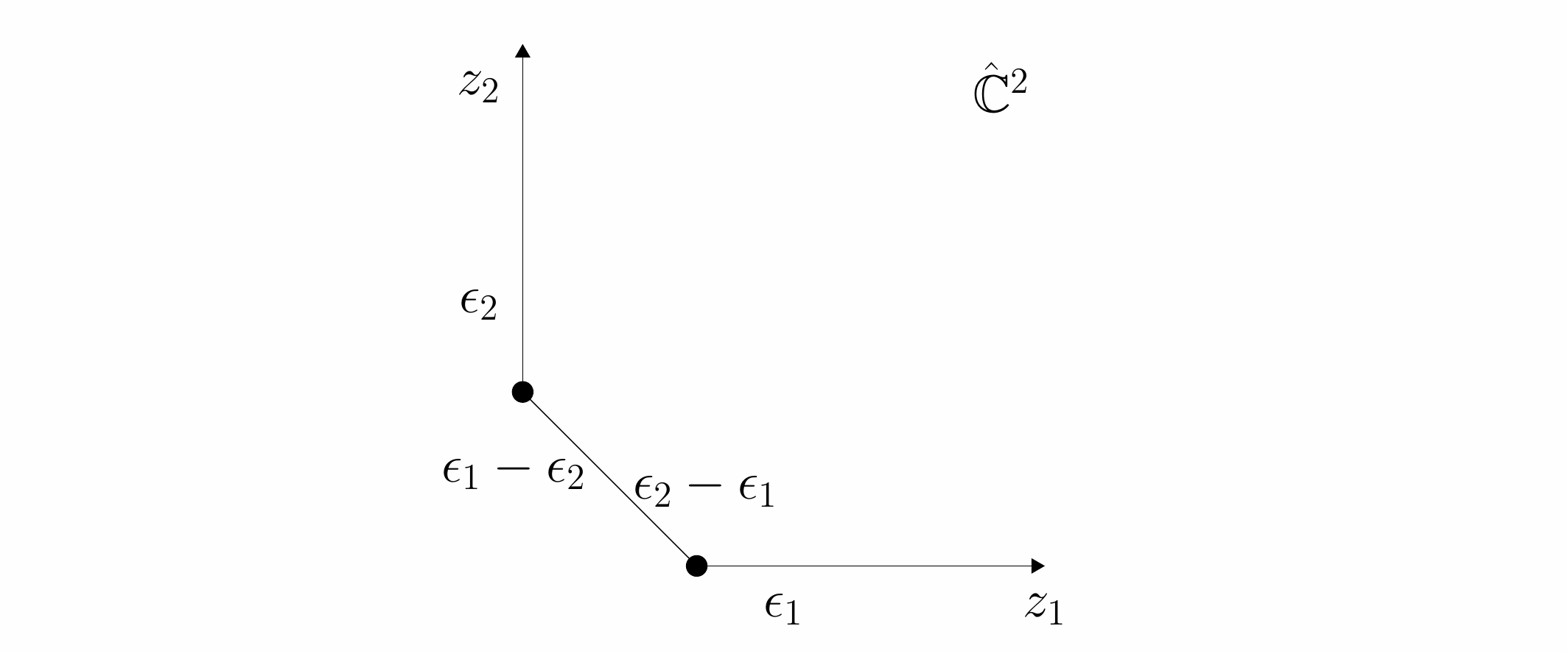}
\caption{The toric diagram of $\hat{\IC}^2$.}\label{fig:c2blowup}
\end{figure}

To express the partition function on $\hat{\IC}^2$ via the partition function on $\IC^2$, we need to calculate the partition function on the two patches near the two fixed points. In the blowup circumstance, certain background field emerges and has nontrivial flux through the exceptional divisor $\IP^1$. The K\"{a}hler moduli in the partition function must receive certain shifts proportional to the flux. This is very much like the complexified K\"{a}hler parameters $r^{\alpha}=\int C^{\alpha}J+\ri B$ where $J$ denotes the K\"{a}hler class and $B$ is the Kalb-Ramond field. The flux is quantized, independent of the size of divisor $\IP^1$ and can only take some special values. The quantization is reflected in the summation over $\bf n$ in $\bR=\md C\cdot \md n + \mathbf{r}/2$ and the $\bf r$ fields characterize the zero-point energy. Although not in refined topological string, similar structure already appeared in the context of traditional topological string theory when I-branes or NS
 5-branes are in presence, see \cite{Dijkgraaf:2002ac} and the chapter four of \cite{Hollands:2009ar}. It should be stressed that even when the size of divisor $\IP^1$ is shrinker to zero, the flux and the two fixed points still exist. In summary, the partition function on $\hat{\IC}^2$ with vanishing size of exceptional divisor should be the product of the partition function on two patches with the K\"{a}hler moduli shifted by the background field flux and summed over all possible flux:
\be
Z_{\mathrm{M-theory}}(X\times S^1\ltimes \widehat{\IC}_{\ep_1,\ep_2})\sim\sum_{\bR}Z_{\mathrm{M-theory}}(X_{\bt+\ep_1\bR}\times S^1\ltimes\IC_{\ep_1,\ep_2-\ep_1})Z_{\mathrm{M-theory}}(X_{\bt+\ep_2\bR}\times S^1\ltimes\IC_{\ep_1-\ep_2,\ep_2}).
\ee
Together with (\ref{eq:Mc2}) and (\ref{eq:Mc2blowup}), we can see why the blowup equations for $Z_{\mathrm{ref}}$ exist.

This is of course a very rough picture. We did not even include the crucial $(-1)^{|\bf n|}$ factor before the double product and distinguish the refined partition function $Z_{\mathrm{ref}}$ from the twisted version $\widehat{Z}_{\mathrm{ref}}$. Nevertheless, we can see from this picture why such structure could exist for general local Calabi-Yau.
%%%%%%%%%%%%%%%%%%%%%%%%%%%%%%%%%%%%%%%%%%%%%%%%%%%%%%%%%%%%%%%%%%
\subsection{Non-perturbative formulation}\label{sec:np}\label{sec:LV}
In this section, we introduce the non-perturbative formulation of blowup equations. It is obvious that if $\eq$ or $\et$ equal to $2\pi\ri p/q$, the refined free energy (\ref{eq:F-ref}) is the divergent. This means one needs to add non-perturbative contributions in the correspondence to the quantization of mirror curve. This idea was first proposed from the study of ABJM theory in \cite{Hatsuda:2012dt}. We define the non-perturbative completion for the refined partition function as
\be\label{eq:nppar}
Z_{\rm ref}^{(\np)}(\bt,\tau_1,\tau_2)=\frac{Z_{\rm ref}(\bt,\tau_1+1,\tau_2)Z_{\rm ref}(\frac{\bt}{\tau_1},\frac{1}{\tau_1},\frac{\tau_1}{\tau_2}+1)}{Z_{\rm ref}(-\frac{\bt}{\tau_2},-\frac{1}{\tau_2},-\frac{\tau_1}{\tau_2}-1)}.
\ee
Equivalently,
\be\label{eq:npfree}
F_{\rm ref}^{(\np)}(\bt,\tau_1,\tau_2)=F_{\rm ref}(\bt,\tau_1+1,\tau_2)+F_{\rm ref}(\frac{\bt}{\tau_1},\frac{1}{\tau_1},\frac{\tau_1}{\tau_2}+1)-F_{\rm ref}(-\frac{\bt}{\tau_2},-\frac{1}{\tau_2},-\frac{\tau_1}{\tau_2}-1).
\ee
Here $2\pi\ri \tau_{1,2}=\epsilon_{1,2}$, and the polynomial part is not included. Our non-perturbative completion is slightly different from the Lockhart-Vafa partition function \cite{Lockhart:2012vp} or the one in \cite{Hatsuda:2013oxa}. However, it still satisfies the requirement that in the limit $\tau_1\to\hbar/2\pi,\tau_2\to 0$ it gives the exact Nekrasov-Shatashvili quantization conditions and in the limit $\tau_1+\tau_2\to 0$, it gives the Grassi-Hatsuda-Mari\~no quantization condition. Besides, this non-perturbative completion is the direct result of the Mellin-Barnes representation in our previous paper \cite{Sun:2016obh}.

We propose the non-perturbative blowup equations as
\be\label{eq:npbl}
\ba
\Lambda&\(\md t,\tau_1,\tau_2,\md r\)=\sum_{\md n \in \mb Z^{g}}  (-1)^{|\md n|}\frac{Z_{\rm ref}^{(\np)}(\bt+2\pi\ri \tau_1\bR,\tau_1,\tau_2-\tau_1)Z_{\rm ref}^{(\np)}(\bt+2\pi\ri \tau_2\bR,\tau_1-\tau_2,\tau_2)}{Z_{\rm ref}^{(\np)}(\bt,\tau_1,\tau_2)}\\
=&\sum_{\md n \in \mb Z^{g}}  (-1)^{|\md n|}\exp{\(F_{\rm ref}^{(\np)}(\bt+2\pi\ri \tau_1\bR,\tau_1,\tau_2-\tau_1)+F_{\rm ref}^{(\np)}(\bt+2\pi\ri \tau_2\bR,\tau_1-\tau_2,\tau_2)-F_{\rm ref}^{(\np)}(\bt,\tau_1,\tau_2)\)},
\ea
\ee
where $\Lambda$ is the same with the one in the original perturbative blowup equations (\ref{eq:bltoge}). Let us prove the non-perturbative blowup equations (\ref{eq:npbl}) from the perturbative one. First, notice that
\be
\widehat{F}_{\rm ref}(\md t, \epsilon_1, \epsilon_2)=F_{\rm ref}(\bt,\tau_1+1,\tau_2).
\ee
This means the first term in the non-perturbative refined free energy (\ref{eq:npfree}) is exactly the twisted perturbative refined free energy. The other two terms in (\ref{eq:nppar}) are the non-perturbative contributions. Since the perturbative blowup equations hold, all we need to prove is the non-perturbative contributions from all three refined free energies in the left side of (\ref{eq:npbl}) cancel with each others. Let us consider the non-perturbative contributions from $F_{\rm ref}^{(\np)}(\bt+2\pi\ri \tau_1\bR,\tau_1,\tau_2-\tau_1)+F_{\rm ref}^{(\np)}(\bt+2\pi\ri \tau_2\bR,\tau_1-\tau_2,\tau_2)$, they are
\be\label{eq:npcomp}
\ba
&F_{\rm ref}(\frac{\bt+2\pi\ri \tau_1\bR}{\tau_1},\frac{1}{\tau_1},\frac{\tau_2-\tau_1}{\tau_1}+1)-F_{\rm ref}(-\frac{\bt+2\pi\ri \tau_1\bR}{\tau_2-\tau_1},-\frac{1}{\tau_2-\tau_1},-\frac{\tau_1}{\tau_2-\tau_1}-1)\\
&+F_{\rm ref}(\frac{\bt+2\pi\ri \tau_2\bR}{\tau_1-\tau_2},\frac{1}{\tau_1-\tau_2},\frac{\tau_2}{\tau_1-\tau_2}+1)-F_{\rm ref}(-\frac{\bt+2\pi\ri \tau_2\bR}{\tau_2},-\frac{1}{\tau_2},-\frac{\tau_1-\tau_2}{\tau_2}-1)\\
=&F_{\rm ref}(\frac{\bt}{\tau_1}+2\pi\ri \bR,\frac{1}{\tau_1},\frac{\tau_2}{\tau_1})-F_{\rm ref}(\frac{\bt+2\pi\ri \tau_2\bR}{\tau_1-\tau_2}+2\pi\ri\bR,\frac{1}{\tau_1-\tau_2},\frac{\tau_2}{\tau_1-\tau_2})\\
&+F_{\rm ref}(\frac{\bt+2\pi\ri \tau_2\bR}{\tau_1-\tau_2},\frac{1}{\tau_1-\tau_2},\frac{\tau_1}{\tau_1-\tau_2})-F_{\rm ref}(-\frac{\bt}{\tau_2}-2\pi\ri \bR,-\frac{1}{\tau_2},-\frac{\tau_1}{\tau_2})\\
\ea
\ee
Since $\bR=\md C\cdot \md n + \mathbf{r}/2$, then 
\be
F_{\rm ref}(\bt+2\pi\ri \bR,\tau_1,\tau_2)=F_{\rm ref}(\bt+\pi\ri\md B,\tau_1,\tau_2)=F_{\rm ref}(\bt,\tau_1,\tau_2+1).
\ee
Using this relation to (\ref{eq:npcomp}), we have
\be
\ba
&F_{\rm ref}(\frac{\bt}{\tau_1},\frac{1}{\tau_1},\frac{\tau_2}{\tau_1}+1)-F_{\rm ref}(\frac{\bt+2\pi\ri \tau_2\bR}{\tau_1-\tau_2},\frac{1}{\tau_1-\tau_2},\frac{\tau_2}{\tau_1-\tau_2}+1)\\
&+F_{\rm ref}(\frac{\bt+2\pi\ri \tau_2\bR}{\tau_1-\tau_2},\frac{1}{\tau_1-\tau_2},\frac{\tau_1}{\tau_1-\tau_2})-F_{\rm ref}(-\frac{\bt}{\tau_2},-\frac{1}{\tau_2},-\frac{\tau_1}{\tau_2}-1)\\
=&F_{\rm ref}(\frac{\bt}{\tau_1},\frac{1}{\tau_1},\frac{\tau_2}{\tau_1}+1)-F_{\rm ref}(-\frac{\bt}{\tau_2},-\frac{1}{\tau_2},-\frac{\tau_1}{\tau_2}-1)
\ea
\ee
Obviously, this is just the non-perturbative contributions from $F_{\rm ref}^{(\np)}(\bt,\tau_1,\tau_2)$. Now we complete the proof that all non-perturbative contributions in (\ref{eq:npbl}) cancel with each other.

%%%%%%%%%%%%%%%%%%%%%%%%%%%%%%%%%%%%%%%%%%%%%%%%%%%%

%%%%%%%%%%%%%%%%%%%%%%%%%%%%%%%%%%%%%%%%%%%%%%%%%%%%%%%%%%%%%
%%%%%%%%%%%%%%%%%%%%%%%%%%%%%%%%%%%%%%%%%%%%%%%%%%%%%%%%%%%%%%%%%%%
\section{Examples}\label{sec:ex}
In this section, we test the blowup equations with various local toric geometries. Many simple local toric Calabi-Yau like local $\mathbb{P}^2$, $\mathbb{F}_n$ and resolved $\mathbb{C}^3/\mathbb{Z}_5$ orbifold can be realized as $X_{N,m}$ geometries or their reduction. For such cases, the blowup equations can be derived from the G\"{o}ttsche-Nakajima-Yoshioka K-theoretic blowup equations. We will also check local $\mathfrak{B}_3$, the three-point blowup of local $\IP^2$, which via geometric engineering gives gauge theory with two fundamental matters. For blowup equations with matters, certain primary results have been found in \cite{Nakajima:2009qjc}.
%%%%%%%%%%%%%%%%%%%%%%%%%%%%%%%%%%%%%%%%%%%%%%%%%%%%%%%%%%%%%%%%%%%%%%%%%%%%%%%%%%%%%%
\subsection{Genus zero examples}\label{sec:zero1}
The models with genus-zero mirror curves are quite special and not entirely trivial. Various examples has been studied in \cite{Aganagic:2011mi}.  The typical example is the well known resolved conifold.
\subsubsection{$\CO(-1)\oplus\CO(-1)\mapsto\IP^1$}\label{sec:zero11}
The resolved conifold is a non-compact Calabi-Yau threefold described by the constraint equation
\be
xy-zw=0,
\ee
where the singularity is resolved by a two-sphere $x=\rho z$, $w=\rho y$. Thus the resulting space can be characterized as $\CO(-1)\oplus\CO(-1)\mapsto\IP^1$. There is a single K\"{a}hler parameter $t$ measuring the size of base $\IP^1$. It is well known the only non-vanishing Gopakumar-Vafa invariant of the resolved conifold is $n_0^1=1$, and the only non-vanishing refined BPS invariant is $n_{0,0}^1=1$. Thus the B field is $1$.

The resolved conifold involves a lot of interesting physics. For example, the large-$N$ duality, or later known as the open/closed duality originated from the observation that the closed topological string theory on the resolved conifold is exactly dual to the $U(N)$ Chern-Simons theory on $S^3$ \cite{Gopakumar:1998ki}. In geometric engineering, the compactification of M-theory on resolved conifold gives rise to $U(1)$ supersymmetric gauge theory \cite{Katz:1996fh}. The resolved conifold has the simplest toric diagram, and its refined partition function was computed with the refined topological vertex in \cite{Iqbal:2007ii} as
\be
Z(q,t,Q)=\exp\l-\sum_{n=1}^{\infty}\frac{Q^n}{n(q^{\frac{n}{2}}-q^{-\frac{n}{2}})(t^{\frac{n}{2}}-t^{-\frac{n}{2}})}\r,
\ee
where $q=e^{\eq}$, $t=e^{-\et}$ and $Q=e^{-t}$.

It is easy to check that
\be\label{eq:blowupconi}
Z(q,qt,\frac{1}{\sqrt{q}}Q)Z(qt,t,\sqrt{t}Q)=Z(q,t,Q),
\ee
which means the unity blowup equation holds for $r=1$. This equation is the result of nothing but a simple identity:
\be\label{eq:id1}
\frac{\frac{1}{x}}{(x-\frac{1}{x})(\frac{y}{x}-\frac{x}{y})}+\frac{\frac{1}{y}}{(\frac{x}{y}-\frac{y}{x})(y-\frac{1}{y})}=\frac{1}{(x-\frac{1}{x})(y-\frac{1}{y})}.
\ee
Here we do not have to make the twist of $t+\ri \pi$ in partition function since there is no perturbative part as comparison. Similarly, we can check $r=-1$ is also an unity $r$ field:
\be
Z(q,qt,{\sqrt{q}}Q)Z(qt,t,\frac{1}{\sqrt{t}}Q)=Z(q,t,Q),
\ee 
which is the result of identity
\be
\frac{{x}}{(x-\frac{1}{x})(\frac{y}{x}-\frac{x}{y})}+\frac{{y}}{(\frac{x}{y}-\frac{y}{x})(y-\frac{1}{y})}=\frac{1}{(x-\frac{1}{x})(y-\frac{1}{y})}.
\ee
It is easy to prove that there is no other unity $r$ fields. 

It is worthwhile to point out that these two $r$ fields $\pm 1$ are non-equivalent, since there is no $\Gamma_{\md C}$ symmetry for genus zero models. However, they are in fact related by the reflective property of the $r$ fields. In section \ref{sec:proofconi}, we will further prove that a local Calabi-Yau satisfying the blowup equation (\ref{eq:blowupconi}) can only be the resolved conifold.

It is also worthwhile to point out that geometries with genus-zero mirror curve do not have vanishing blowup equations. This is not very surprising since there is no traditional quantization condition for genus-zero curves. 

\subsubsection{$\CO(0)\oplus\CO(-2)\mapsto\IP^1$}\label{sec:zero2}
This geometry is the resolution of $\IC\times\IC^2/\IZ_2$ and can be obtained from local $\IP^1\times\IP^1$ by taking the size of one of the $\IP^1$ very large. The refined partition function was computed with the refined topological vertex in \cite{Iqbal:2007ii} as
\be
Z(q,t,Q)=\exp\l-\sum_{n=1}^{\infty}\frac{Q^n\(\frac{q}{t}\)^{\frac{n}{2}}}{n(q^{\frac{n}{2}}-q^{-\frac{n}{2}})(t^{\frac{n}{2}}-t^{-\frac{n}{2}})}\r.
\ee
It is easy to check the unity $r$ field of this geometry is $r=0$:
\be
Z(q,qt,Q)Z(qt,t,Q)=Z(q,t,Q),
\ee
which is the result of identity
\be
\frac{{y}}{(x-\frac{1}{x})(\frac{y}{x}-\frac{x}{y})}+\frac{{x}}{(\frac{x}{y}-\frac{y}{x})(y-\frac{1}{y})}=\frac{xy}{(x-\frac{1}{x})(y-\frac{1}{y})}.
\ee

\subsection{Local $\mathbb{P}^2$}\label{sec:p2}
In this section, we study the unity and vanishing blowup equations for local $\IP^2$. This geometry is the simplest local toric Calabi-Yau with genus-one mirror curve and compact four-cycle. We will not only check its blowup equations to high degree of $Q$, but also give a rigorous proof for the fist two equations in the $\epsilon$ expansion of both vanishing and unity equations. Interestingly, as we will see later, the leading order of the unity blowup equation of local $\IP^2$ is just the pentagonal number theorem, originally due to Euler.

Local $\mathbb{P}^2$ is a geometry of line bundle $\mathcal{O}(-3)\rightarrow\mathbb{P}^2$. The toric data are 
\begin{equation}\label{P2toric}
\begin{array}{c|crrr|rl|}
    \multicolumn{5}{c}{v_i }    &Q_i&    \\
    D_u    &&     1&     0&   0&         -3&        \\
    D_1    &&     1&     1&   0&         1&         \\
    D_2    &&     1&     0&   1&         1&          \\
    D_3    &&     1&    -1&   -1&         1&          \\ 
  \end{array}
\end{equation}
The moduli space of local $\mathbb{P}^2$ in B-model is described globally by complex structure parameter $z$. The moduli space contains three singular points: large radius point, conifold point and orbifold point with $z\sim 0, \  z\sim \frac{1}{27},\  \frac{1}{z}\sim 0$ respectively.
We can write down the mirror curve in B-model as
\be\label{remannsurface}
1+x+y+\frac{z}{x y}=0,
\ee
which is an elliptic curve, with meromorphic 1-form $\lambda=\log y \frac{dx}{x}.$ The periods are defined as 
\be
t=\int_{\alpha} \lambda,\ \ p=\int_{\beta} \lambda=\frac{\partial F_0}{\partial t},
\ee
where $F_0$ is the genus 0 free energy of topological string. In A-model, the moduli space is described by K\"{a}hler parameter $t$. We can compute $t(z)$ in B-model via Picard-Fuchs equation
\be
(\theta^3+3z (3\theta-2)(3\theta-1)\theta)\Pi=0,
\ee
where $\theta =t\frac{\partial}{\partial t}.$ There are three solutions $\Pi_0=1,\Pi_1=t,\Pi_2=\frac{\partial F_0}{\partial t}$ to this equation. At large radius point, solving from Picard-Fuchs equation, also as is computed in \cite{Aganagic:2006wq}, we have
\be
F_0=-\frac{1}{18}t^3+\frac{1}{12}t^2+\frac{1}{12}t+3 Q-\frac{45}{4} Q^2+\cdots,
\ee
where $Q=e^t$. Define modular parameter $2\pi i \tau=3\frac{\partial^2}{\partial t^2}F_0$ of Riemann surface (\ref{remannsurface}), the modular group of local $\mathbb{P}^2$ is $\Gamma(3)\in SL(2,\mathbb{Z})$. It has generators
\be
a:=\theta^3\left[{1\over 6}\atop {1\over 6}\right], \quad
b:=\theta^3\left[{1\over 6}\atop {1 \over 2} \right], \quad
c:=\theta^3\left[{1\over 6}\atop {5\over 6}\right], \quad
d:=\theta^3\left[{1 \over 2} \atop {1\over 6}\right],
\ee
all have weight $3/2$. The Dedekind $\eta$ function satisfies the identity $\eta^{12}=\frac{i}{3^{3/2}}abcd$. As in \cite{Aganagic:2006wq}, the genus one free energy can be compute from holomorphic anomaly equation:
\be
 F^{(0,1)}=-\frac{1}{6}\log(d\eta^3),\ \  F^{(1,0)}=\frac{1}{6}\log(\eta^3/d).
\ee
Besides, it is obvious from the toric data that for local $\IP^2$, $C=3$, and from the curve that $B=1$.

Let us first consider the unity blowup equations. It is easy to find the non-equivalent unity $r$ field for local $\IP^2$ are $r=\pm1$. They are reflexive, so we can merely look at $r=1$. In this case, $R=3n+1/2$ and the leading order of unity blowup equation (\ref{first}) gives
\be\label{first:P2}
\sum_{n=-\infty}^{\infty} (-1)^ne^{\frac{1}{2}(n+1/6)^2 3\cdot 2\pi i \tau}=\eta(\tau),
\ee
where the right side comes from
\be
F^{(0,1)}-F^{(1,0)}=\log(\eta(\tau)).
\ee
This is exactly the Euler identity, or the Pentagonal number theorem! We can see both sides of the equation are weight $1/2$ modular forms of $\Gamma(3)$. For higher order of the blowup equation, we obtain more such identities. For example, the subleading order requires the following identity:
\be\label{theta3}
\sum R\Theta\equiv\sum_{n=-\infty}^{\infty}(-1)^n \frac{3n+1/2}{2}e^{\frac{1}{2}(n+1/6)^2 3\cdot2\pi i \tau}=\frac{b}{2i}+\frac{d}{2\sqrt{3}}.
\ee
Substitute (\ref{theta3}) into (\ref{second}), we can prove subleading unity equation in the following form
\be
d \partial_\tau(\sum R\Theta)-(\sum R\Theta)\partial_\tau d=-\frac{\sqrt{3}}{2}\eta^{10}.
\ee
One can in principle prove the component equations order by order. 
%However, a simple analysis shows that all of this identities are weight zero modular forms, then we can perform the modular transformations to generalize the blowup equations to other points in the moduli space. 
For the higher order identities, only $F^{(n,g)}$ and their derivatives with respect to $t$ and $\sum_R R^k (-1)^k e^{\frac{1}{2}(n+1/6)^2 3\cdot2\pi i\tau}$ appear. The free energies $F^{(n,g)}$ are quasi-modular, and the replacement of partial differentiation
\be
\partial_t=C_{ttt} \partial_{\tau}
\ee
does not break the modular property.\footnote{$\partial_{\tau}$ of a quasi-modular form is also a quasi-modular form, add the weight by 2. $C_{ttt}=C_{zzz} (\frac{\partial z}{\partial t})^3$ is a weight $-3$ form.} $\sum R^n (-1)^k e^{\frac{1}{2}(n+1/6)^2 3\cdot2\pi i \tau}$ can be understood as the $k$th $z$-derivative of a theta function and is also a modular form. Using these facts, we find that the unity blowup equation at all orders are weight zero quasi-modular form, if we only leave the $\Lambda$ factor on one side.
\begin{figure}[htbp]
\centering\includegraphics[width=5in]{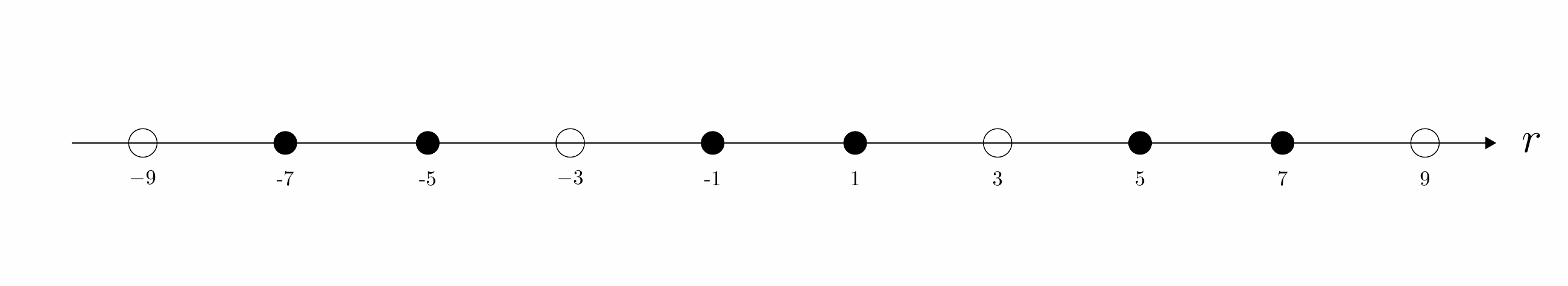}
\caption{The $r$ lattice of local $\IP^2$.}\label{fig:p2r}
\end{figure}

Now we turn to the vanishing blowup equation. The sole vanishing $r$ field of local $\IP^2$ is $r=3$. Then we have $R=3(n+\frac{1}{2})$. Because of the symmetry under $n\rightarrow -n$, it is easy to see that half of the component equations including the leading order of vanishing blowup equation vanish trivially. The first nontrivial identity is the subleading order equation: 
\be\label{P2:vanishing_2}
\sum_{n=-\infty}^{\infty}(-1)^n \left((n+1/2)^3 F'''_{(0,0)}+ 6(n+1/2)\left(F'_{(0,1)}+F'_{(1,0)}\right)\right)e^{\frac{1}{2}(n+1/2)^2 3\cdot2\pi i \tau}=0.
\ee
Integrate the above equation and fix the integration constant, we obtain the following identity of level three modular forms:
\be
\sum_{n=-\infty}^{\infty}(-1)^n(n+1/2)e^{\frac{1}{2}(n+1/2)^2 \cdot 6\pi i\tau}=\frac{d}{3\sqrt{3}}.
\ee

All the unity and vanishing $r$ fields can be gathered into a lattice, which we call the \emph{$r$ lattice}, as is shown in Figure \ref{fig:p2r}. The white dots represent the vanishing $r$ fields, while the black dots represent the unity $r$ fields.
%%%%%%%%%%%%%%%%%%%%%%%%%%%%%%%%%%%%%%%%%%%%%%%%%%%%%%%%
\subsection{Local Hirzebruch surfaces}\label{sec:fn}
Local Hirzebruch surfaces $F_n$ are typical local toric Calabi-Yau threefolds with genus-one mirror curve. They have two charges $Q^1=(-2,1,0,1,0), \ Q^2=(-2+n,0,1,n,1)$, and complex structure parameters $z_1,z_2$ which are defined by the homogeneous coordinates $x_i$ of the toric geometry with $z_a=\prod x_i ^{Q^{ai}},a=1,2$. For detailed notations, see \cite{Huang:2013yta}. 
\subsubsection{Local $\mathbb{F}_0$}\label{sec:f0}
In the following, we study local $\mathbb{F}_0$ in details. We can write down the mirror curve $\Sigma(z_1,z_2)$ as
\be\label{F0_mirrorcurve}
H(x,y)=y^2-x^3-(1-4z_1-4z_2) x^2-16z_1z_2 x,
\ee
and Picard-Fuchs operators as
\be
\begin{split}
\mathcal{D}_1 &=\theta_1^2-2z_1(\theta_1+\theta_2)(1+2\theta_1+2\theta_2),\\
\mathcal{D}_2 &=\theta_2^2-2z_2(\theta_1+\theta_2)(1+2\theta_1+2\theta_2).\\
\end{split}
\ee
We can compute the periods from Picard-Fuchs operators, and then determine the genus zero free energy. We can also integrate the mirror curve, which is an elliptic curve, and obtain all information at genus zero. Before doing that, we introduce the following convention:
\be
z_1=z,\ z_2= z m.
\ee
where we separate the "true" modulus $z$ and mass parameter $m$. Then the mirror map related to mass parameter is just $t_m=\log m$, which is invariant under modular transformation. And $t$ is related to the mirror map of $z$. It is convenient since we only need to deal with only one K\"ahler parameter under modular transformation. Therefore we can only consider the massless case $m=1$ in the following discussion.

The discriminant and $j$-invariant of $\Sigma(z_1,z_2)$ are
\be 
\Delta=1-8(z_1 +z_2)+16(z_1 -z_2)^2,
\ee
\be\label{p1p1_j}
j(\tau)=\frac{\left(16 z_1^2-8 \left(2 z_2+1\right) z_1+\left(1-4
   z_2\right){}^2\right){}^3}{1728 z_1^2 z_2^2 \left(16 z_1^2-8 \left(4
   z_2+1\right) z_1+\left(1-4 z_2\right){}^2\right)},
\ee
where $2\pi i \tau=2\frac{\partial^2 F_0}{\partial t^2}$.

From holomorphic anomaly equation, we can fix the genus 1 free energy as
\be
F^{(1,0)}=\frac{1}{24}\log(\Delta z^{-2}m^{-1})=-\frac{1}{6}\log(\frac{\theta_2^2}{\theta_3 \theta_4}),
\ee
and
\be
F^{(0,1)}=-\frac{1}{2}\log(\frac{\partial t}{\partial z}\frac{\partial t_m}{\partial m}-\frac{\partial t}{\partial m}\frac{\partial t_m}{\partial z})-\frac{1}{12}\log(\Delta z^{7}m^{13/2})=-\log(\eta (\tau)).
\ee

The unique vanishing $\md r$ field is $(2,0)$. And all non-equivalent unity $\md r$ fields and the corresponding $\Lambda$ factors are listed in table \ref{defaulttable}. The $\md r$ lattice is shown in Figure \ref{fig:f0r}.

For the unity $\md r$ $(0,0)$, the leading order of unity blowup equation gives the following identity:
\be\label{p1p1:first}
2^{1/3}e^{F^{(1,0)}-F^{(0,1)}}=\sum_{n=-\infty}^{\infty} e^{\frac{1}{2}(n)^2*2\tau+i\pi n}.
\ee
It is also easy to check the higher order results and the other $\mathbf{r}$ fields with mass parameter. Since we have checked it via BPS invariants formalism, we do not list the details here. 
\begin{table}
\begin{center}
\begin{tabular}{|c|c|}
\hline
Unity $\mathbf{r}$ fields & $\Lambda$ \\ \hline
 $(0,0)$  & $1$ \\ \cline{1-2}
 $(0,2)$  & $1$ \\ \cline{1-2}
 $(0,-2)$  & $1$ \\ \cline{1-2}
 $(0,4)$  & $1-e^{ \ep_1+\ep_2+t_m}$ \\ \cline{1-2}
 $(0,-4)$  & $1-e^{- \ep_1-\ep_2+t_m}$ \\ \cline{1-2}
 $(2,-2)$  & $e^{-\frac{1}{4}  \left(- t_m+\epsilon _1+\epsilon _2\right)}$ \\ \cline{1-2}
 $(2,2)$  & $-e^{\frac{1}{4}  \left( t_m+\epsilon _1+\epsilon _2\right)}$ \\ \cline{1-2}
\hline

\end{tabular}
\end{center}
\caption{The non-equivalent unity $\mathbf{r}$ fields and $\Lambda$ of local $\mathbb{F}_0$.}
\label{defaulttable}
\end{table}

\begin{figure}[htbp]
\centering\includegraphics[width=5in]{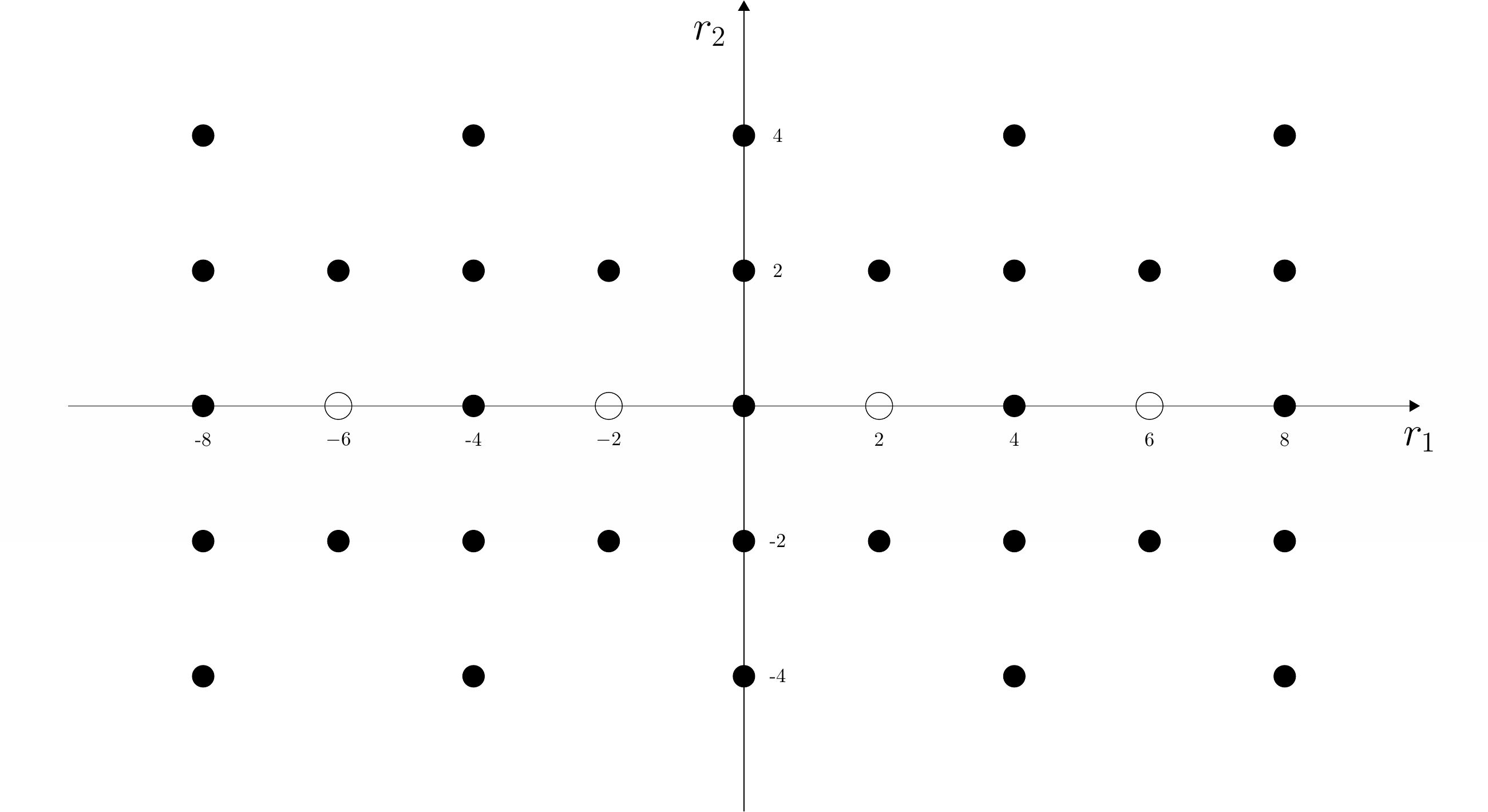}
\caption{The $\bf r$ lattice of local $\mathbb{F}_0$.}\label{fig:f0r}
\end{figure}
%%%%%%%%%%%%%%%%%%%%%%%%%%%%%%%%%%%%%%%%%%%%%%%%%%%%
\subsubsection{Local $\mathbb{F}_1$}\label{sec:f1}
In the following, we will consider local $\mathbb{F}_1$ in detail. One can also following the above method to give genus 0,1 free energy, and from refined topological vertex to get BPS contents. But from the quantization of the mirror curve, we can have the information of this information directly. 
The mirror curve of local $\mathbb{F}_1$ is parameterized by two complex parameters
$$
z_1=\frac{m}{u^2},\ \ z_2= \frac{1}{mu},
$$
where we re-express it with respect to the mass parameter $m$ and "true" parameter $u$. Then it is obvious that
$$
\log(t_1)-2\log(t_2)=3\log(m),
$$
is a pure mass parameter do not depend on $\tau$.
The refined free energy of local $\mathbb{F}_1$ is 
\be
F=-\frac{\left(t_1+t_2\right) \left(\epsilon _1+\epsilon _2\right){}^2}{18 \epsilon _1
   \epsilon _2}+\frac{2 \pi ^2 \left(-t_1-t_2\right)}{9 \epsilon _1 \epsilon
   _2}+\frac{\frac{t_1^3}{3}-\frac{1}{2} t_2 t_1^2+\frac{1}{2} t_2^2 t_1}{\epsilon _1
   \epsilon _2}+\frac{1}{18} \left(-t_1-t_2\right)+\mathcal{O}(e^{t_i}),
\ee
where $\mathcal{O}(e^{t_i})$ is computed from the refined topological vertex \cite{Iqbal:2007ii}. We also have checked the unity blowup equation, and the results are listed in table \ref{default2}.
\begin{table}

\begin{center}
\begin{tabular}{|c|c|}
\hline
Unity $\mathbf{r}$ fields & $\Lambda$ \\ \hline
$(1,-2)$  & $e^{-\frac{5}{4}t_1+\frac{5}{8}t_2+\frac{1}{4}(\ep_1+\ep_2)}$ \\ \cline{1-2}
$(1,0)$  & $e^{-\frac{1}{4}t_1+\frac{1}{8}t_2-\frac{1}{8}(\ep_1+\ep_2)}$ \\ \cline{1-2}
$(3,0)$  & $e^{-\frac{9}{4}t_1+\frac{9}{8}t_2+\frac{5}{8}(\ep_1+\ep_2)}$ \\ \cline{1-2}
$(-1,2)$  & $e^{-\frac{5}{4}t_1+\frac{5}{8}t_2-\frac{1}{4}(\ep_1+\ep_2)}$ \\ \cline{1-2}
$(-1,0)$  & $e^{-\frac{1}{4}t_1+\frac{1}{8}t_2+\frac{1}{8}(\ep_1+\ep_2)}$ \\ \cline{1-2}
$(-3,0)$  & $e^{-\frac{9}{4}t_1+\frac{9}{8}t_2-\frac{5}{8}(\ep_1+\ep_2)}$ \\ \hline

\end{tabular}
\end{center}
\caption{The non-equivalent unity $\mathbf{r}$ fields and $\Lambda$ of local $\mathbb{F}_1$.}
\label{default2}
\end{table}%

\begin{figure}[htbp]
\centering\includegraphics[width=5in]{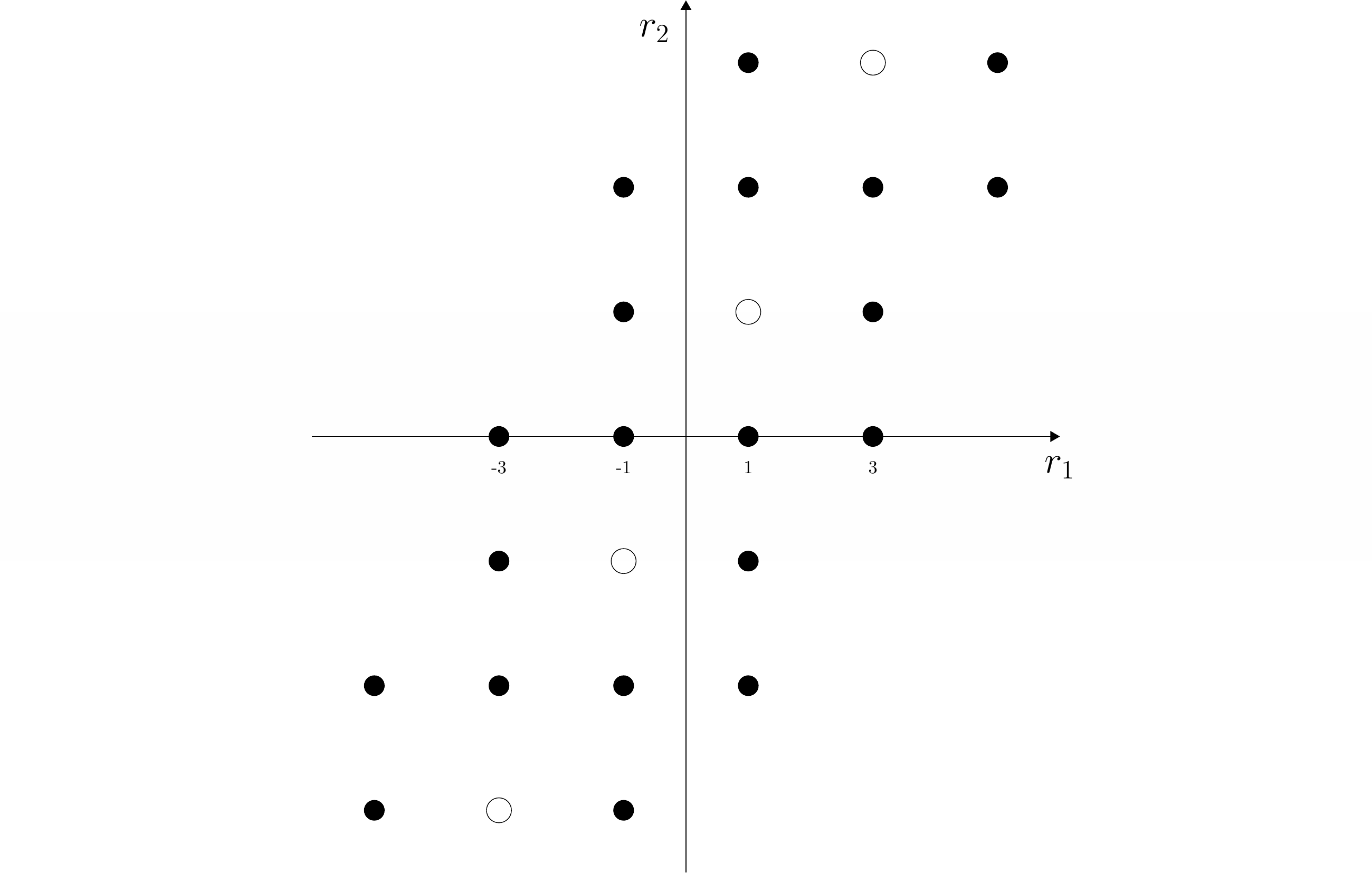}
\caption{The $\bf r$ lattice of local $\mathbb{F}_1$.}\label{fig:f1r}
\end{figure}
%%%%%%%%%%%%%%%%%%%%%%%%%%%%%%%%%%%%%%%%%%%%%%%%%%%%%%%%
\subsection{Local $\mathfrak{B}_3$}\label{sec:b3}
$\mathfrak{B}_3$ is obtained by blowing up $\IP^2$ at three generic points. The BPS invariants can be computed from the refined topological vertex, we use the notations and results in \cite{Hatsuda:2017zwn}. There are four independent K\"ahler parameters, writing as $Q,m_1,m_2,m_3$ in \cite{Hatsuda:2017zwn}. These bases are commonly used in mirror curve. From refined topological vertex, we can have another convenient base choice for K\"ahler parameters as 
\be
Q_1=\frac{Q}{m_1},\ Q_2=\frac{Q}{m_2},\ Q_3=\frac{Q}{m_3},\ Q_4=m_1m_2m_3.
\ee
These bases are more natural in the sense that the $\br$ fields will all be integral. In the base $Q_1, Q_2, Q_3, Q_4$, it is easy to obtain that the $\mathbf{B}$ field is $(1,1,1,0)$. We scanned all possible $\mathbf{r}$ fields, and find that there are 28 non-equivalent unity $\mathbf{r}$ fields and one unique non-equivalent vanishing $\mathbf{r}$ field $(1,1,1,0)$.\footnote{One can also choose bases $(H,e_1,e_2,e_3)$, where $H$ is the hyperplane class in $\IP^2$ and $e_i$ are blowup divisors. In this base, $\mathbf{C}=\br_{\rm v}=(3,1,1,1)$ and all unity $\br$ fields are also integers.} Having done that, we can transform the base back to the $Q,m_1,m_2,m_3$, and some $\br$ fields may become fraction numbers. Since all the three mass parameters are symmetric, the $\mathbf{r}$ field has a symmetry between this parameters. Note that the value of $\Lambda$ may depend on the coefficients of pure mass parameter terms $b_{\rm NS},b_{\rm GV}$, which we do not determine in this paper. 
We list some data of this geometry, these data are also computed in \cite{Hatsuda:2017zwn}.
\be
\ba
-t&=\log z+(m_1+m_2+m_3)z^2+2(1+m_1 m_2 m_3)z^3 \\
&\quad+\frac{3}{2}(m_1^2+m_2^2+m_3^2+4m_1 m_2+4m_2 m_3+4m_3 m_1)z^4+\cO(z^5),
\ea
\ee
where we fixed the integration constant so that $Q=\re^{-t}=z+\cO(z^2)$ as $z\to 0$.
Inverting this, one gets the mirror map
\be
\ba
z&=Q(1-(m_1+m_2+m_3)Q^2-2(1+m_1 m_2 m_3)Q^3\\
&\quad+(m_1^2+m_2^2+m_3^2-m_1m_2-m_2m_3-m_3m_1)Q^4+\cO(Q^5)).
\ea
\ee

The free energies are
\be
\ba
F_0(t)&=t^3-\frac{\log(m_1 m_2 m_3)}{2}t^2+C_0 t+F_0^{\text{inst}}(t), \\
F_0^{\text{inst}}(t)&=-\(\frac{1}{m_1}+\frac{1}{m_2}+\frac{1}{m_3}+m_1 m_2+m_2 m_3+m_3 m_1 \)\re^{-t}\\
&\quad -\frac{1}{8}\biggl( 16(m_1+m_2+m_3)-\frac{1}{m_1^2}-\frac{1}{m_2^2}-\frac{1}{m_3^2} \\
&\quad -m_1^2 m_2^2-m_2^2 m_3^2-m_3^2 m_1^2 \biggr)\re^{-2t}+\cO(\re^{-3t}), 
\ea
\label{eq:F0}
\ee
where
\be
C_0=-\pi^2-\frac{1}{2}(\log^2 m_1+\log^2 m_2+\log^2 m_3 ).
\label{eq:C0}
\ee
\be
\ba
F_{\text{NS}}^{\text{(1)}}(t)&=-\frac{t}{4}-\frac{\log{(m_1 m_2 m_3)}}{24}\\
&\quad-\frac{1}{24}\(\frac{1}{m_1}+\frac{1}{m_2}+\frac{1}{m_3}+m_1 m_2+m_2 m_3+m_3 m_1 \)\re^{-t}+\cdots.
\ea
\label{eq:F-NS-1-total}
\ee

With these data at hand, in order to find the unity $\mathbf{r}$ fields with the method in section \ref{ch:solver}, we obtain $f^k(n)$ defined there as
$$
f(n)=\frac{1}{24} \left(72 n^2+72 n r+12 n r_4+18 r^2-2 r_1^2-2 r_2^2-2 r_3^2-r_4^2+2 r_1
   r_2+2 r_1 r_3+2 r_2 r_3+6 r r_4+6\right),
$$
where $\mathbf{r}=(r,r_{m,1},r_{m,2},r_{m,3})$, and $r=\frac{1}{3}(r_1+r_2+r_3+r_4),\; r_{m,i}=r-r_i,i=1,2,3.$ Here, $\mathbf{r}'=(r_1,r_2,r_3,r_4)$ are in the base where all components are integers. Solving from $f(0) \leqslant f(n)$ within $\mathbf{r}'\in \mathbb{Z}^4$, we obtain all the unity $\mathbf{r}$ fields listed in table \ref{default66}. We also use computer to scan these $\mathbf{r}$ fields, and find exact agreement. We list the corresponding $\Lambda$ factor for each unity $\br$, where $t_{m_i}=\log m_i$. One can see that they all are indeed independent of the true modulus $t$.

\begin{table}

\begin{center}
\begin{tabular}{|c|c|}
\hline
  $\mathbf{r}$ fields & $\Lambda$ \\ \hline

$\left(-\frac{1}{3},-\frac{4}{3},-\frac{4}{3},-\frac{4}{3}\right)$& $\left(e^{\epsilon _1+\epsilon _2}-e^{t_{m_1}+t_{m_2}+t_{m_3}}\right) \exp \left(\frac{1}{54} \left(-3 t_{m_1}-3 t_{m_2}-3 t_{m_3}-43 \epsilon _1-43 \epsilon _2\right)\right)$\\ \cline{1-2}
$\left(\frac{1}{3},-\frac{2}{3},-\frac{2}{3},-\frac{2}{3}\right)$& $e^{\frac{1}{27} \left(3 t_{m_1}+3 t_{m_2}+3 t_{m_3}-\epsilon _1-\epsilon _2\right)}$\\ \cline{1-2}
$\left(-\frac{1}{3},\frac{2}{3},\frac{2}{3},\frac{2}{3}\right)$& $e^{\frac{1}{27} \left(3 \left(t_{m_1}+t_{m_2}+t_{m_3}\right)+\epsilon _1+\epsilon _2\right)}$\\ \cline{1-2}
$\left(\frac{1}{3},\frac{4}{3},\frac{4}{3},\frac{4}{3}\right)$& $\left(1-e^{t_{m_1}+t_{m_2}+t_{m_3}+\epsilon _1+\epsilon _2}\right) \exp \left(\frac{1}{54} \left(-3 t_{m_1}-3 t_{m_2}-3 t_{m_3}-11 \epsilon _1-11 \epsilon _2\right)\right)$\\ \cline{1-2}
$(1,0,0,-2)$& $e^{\frac{1}{6} \left(t_{m_1}+t_{m_2}+4 t_{m_3}-2 \epsilon _1-2 \epsilon _2\right)}$\\ \cline{1-2}
$\left(-\frac{1}{3},\frac{2}{3},\frac{2}{3},-\frac{4}{3}\right)$& $\exp \left(\frac{1}{54} \left(6 t_{m_1}+6 t_{m_2}-3 t_{m_3}+5 \epsilon _1+5 \epsilon _2\right)\right)$\\ \cline{1-2}
$\left(\frac{1}{3},\frac{4}{3},\frac{4}{3},-\frac{2}{3}\right)$& $\exp \left(\frac{1}{54} \left(-3 \left(t_{m_1}+t_{m_2}-2 t_{m_3}\right)-8 \epsilon _1-8 \epsilon _2\right)\right)$\\ \cline{1-2}
$(1,0,-2,0)$& $e^{\frac{1}{6} \left(t_{m_1}+4 t_{m_2}+t_{m_3}-2 \epsilon _1-2 \epsilon _2\right)}$\\ \cline{1-2}
$\left(-\frac{1}{3},\frac{2}{3},-\frac{4}{3},\frac{2}{3}\right)$& $\exp \left(\frac{1}{54} \left(6 t_{m_1}-3 t_{m_2}+6 t_{m_3}+5 \epsilon _1+5 \epsilon _2\right)\right)$\\ \cline{1-2}
$\left(\frac{1}{3},\frac{4}{3},-\frac{2}{3},\frac{4}{3}\right)$& $\exp \left(\frac{1}{54} \left(-3 \left(t_{m_1}-2 t_{m_2}+t_{m_3}\right)-8 \epsilon _1-8 \epsilon _2\right)\right)$\\ \cline{1-2}
$\left(-\frac{1}{3},\frac{2}{3},-\frac{4}{3},-\frac{4}{3}\right)$& $\exp \left(\frac{1}{54} \left(6 t_{m_1}-3 t_{m_2}-3 t_{m_3}+8 \epsilon _1+8 \epsilon _2\right)\right)$\\ \cline{1-2}
$\left(\frac{1}{3},\frac{4}{3},-\frac{2}{3},-\frac{2}{3}\right)$& $\exp \left(\frac{1}{54} \left(-3 t_{m_1}+6 t_{m_2}+6 t_{m_3}-5 \epsilon _1-5 \epsilon _2\right)\right)$\\ \cline{1-2}
$(1,2,0,0)$& $\exp \left(\frac{1}{6} \left(4 t_{m_1}+t_{m_2}+t_{m_3}+2 \epsilon _1+2 \epsilon _2-6 i \pi \right)\right)$\\ \cline{1-2}
$(1,2,0,-2)$& $\left(e^{t_{m_3}}-e^{t_{m_1}+\epsilon _1+\epsilon _2}\right) \exp \left(\frac{1}{6} \left(-2 t_{m_1}+t_{m_2}-2 t_{m_3}-3 \epsilon _1-3 \epsilon _2\right)\right)$\\ \cline{1-2}
$(1,2,-2,0)$& $\left(e^{t_{m_2}}-e^{t_{m_1}+\epsilon _1+\epsilon _2}\right) \exp \left(\frac{1}{6} \left(-2 t_{m_1}-2 t_{m_2}+t_{m_3}-3 \epsilon _1-3 \epsilon _2\right)\right)$\\ \cline{1-2}
$(1,-2,0,0)$& $e^{\frac{1}{6} \left(4 t_{m_1}+t_{m_2}+t_{m_3}-2 \epsilon _1-2 \epsilon _2\right)}$\\ \cline{1-2}
$\left(-\frac{1}{3},-\frac{4}{3},\frac{2}{3},\frac{2}{3}\right)$& $\exp \left(\frac{1}{54} \left(-3 t_{m_1}+6 t_{m_2}+6 t_{m_3}+5 \epsilon _1+5 \epsilon _2\right)\right)$\\ \cline{1-2}
$\left(\frac{1}{3},-\frac{2}{3},\frac{4}{3},\frac{4}{3}\right)$& $\exp \left(\frac{1}{54} \left(-3 \left(-2 t_{m_1}+t_{m_2}+t_{m_3}\right)-8 \epsilon _1-8 \epsilon _2\right)\right)$\\ \cline{1-2}
$\left(-\frac{1}{3},-\frac{4}{3},\frac{2}{3},-\frac{4}{3}\right)$& $\exp \left(\frac{1}{54} \left(-3 \left(t_{m_1}-2 t_{m_2}+t_{m_3}\right)+8 \epsilon _1+8 \epsilon _2\right)\right)$\\ \cline{1-2}
$\left(\frac{1}{3},-\frac{2}{3},\frac{4}{3},-\frac{2}{3}\right)$& $\exp \left(\frac{1}{54} \left(6 t_{m_1}-3 t_{m_2}+6 t_{m_3}-5 \epsilon _1-5 \epsilon _2\right)\right)$\\ \cline{1-2}
$(1,0,2,0)$& $\exp \left(\frac{1}{6} \left(t_{m_1}+4 t_{m_2}+t_{m_3}+2 \epsilon _1+2 \epsilon _2-6 i \pi \right)\right)$\\ \cline{1-2}
$(1,0,2,-2)$& $\left(e^{t_{m_3}}-e^{t_{m_2}+\epsilon _1+\epsilon _2}\right) \exp \left(\frac{1}{6} \left(t_{m_1}-2 t_{m_2}-2 t_{m_3}-3 \epsilon _1-3 \epsilon _2\right)\right)$\\ \cline{1-2}
$\left(-\frac{1}{3},-\frac{4}{3},-\frac{4}{3},\frac{2}{3}\right)$& $\exp \left(\frac{1}{54} \left(-3 \left(t_{m_1}+t_{m_2}-2 t_{m_3}\right)+8 \epsilon _1+8 \epsilon _2\right)\right)$\\ \cline{1-2}
$\left(\frac{1}{3},-\frac{2}{3},-\frac{2}{3},\frac{4}{3}\right)$& $\exp \left(\frac{1}{54} \left(6 t_{m_1}+6 t_{m_2}-3 t_{m_3}-5 \epsilon _1-5 \epsilon _2\right)\right)$\\ \cline{1-2}
$(1,0,0,2)$& $\exp \left(\frac{1}{6} \left(2 \left(2 t_{m_3}+\epsilon _1+\epsilon _2-3 i \pi \right)+t_{m_1}+t_{m_2}\right)\right)$\\ \cline{1-2}

$(1,0,-2,2)$& $\left(e^{t_{m_2}}-e^{t_{m_3}+\epsilon _1+\epsilon _2}\right) \exp \left(\frac{1}{6} \left(t_{m_1}-2 t_{m_2}-2 t_{m_3}-3 \epsilon _1-3 \epsilon _2\right)\right)$\\ \cline{1-2}
$(1,-2,2,0)$& $\left(e^{t_{m_1}}-e^{t_{m_2}+\epsilon _1+\epsilon _2}\right) \exp \left(\frac{1}{6} \left(-2 t_{m_1}-2 t_{m_2}+t_{m_3}-3 \epsilon _1-3 \epsilon _2\right)\right)$\\ \cline{1-2}
$(1,-2,0,2)$& $\left(e^{t_{m_1}}-e^{t_{m_3}+\epsilon _1+\epsilon _2}\right) \exp \left(\frac{1}{6} \left(-2 t_{m_1}+t_{m_2}-2 t_{m_3}-3 \epsilon _1-3 \epsilon _2\right)\right)$\\ \cline{1-2}

\hline

\end{tabular}
\end{center}
\caption{The non-equivalent unity $\mathbf{r}$ fields and $\Lambda$ of local $\mathfrak{B}_3$.}
\label{default66}
\end{table}%
%%%%%%%%%%%%%%%%%%%%%%%%%%%%%%%%%%%%%%%%%%%%%%%%%%%%%%%%%%%%%
\subsection{Resolved $\mathbb{C}^3/\mathbb{Z}_5$ orbifold}\label{sec:c3z5}
Resolved $\mathbb{C}^3/\mathbb{Z}_5$ orbifold is the simplest local toric Calabi-Yau with genus-two mirror curve.  It has two true complex moduli without mass parameter. This model has been extensively studied in \cite{Klemm:2015iya}\cite{Codesido:2015dia}\cite{Franco:2015rnr}. In this subsection, we amplify our theory with this example, determine all the $\mathbf{r}$ fields and check the identities (\ref{conjecture}). We find for resolved $\mathbb{C}^3/\mathbb{Z}_5$, there are three non-equivalent $\mathbf{r}$ fields.
\begin{equation}\label{c3z5charge}
\begin{array}{c|rrr|rr}
    \multicolumn{5}{c}{v_i}    $Q_1$&  Q_2 \\
    x_0    &     0&     0&  \ \ 1\,&         -3&  1  \\
    x_1    &     1&     0&   \ \ 1\, &         1&     -2\\
    x_2    &     2&     0&   \ \ 1\, &         0&     1  \\
    x_3    &     0&    1&   \ \ 1\, &         1&   0\\
    x_4    &     -1&    -1&   \ \ 1\, &         1&     0\\
  \end{array}
\end{equation}

\begin{figure}[htbp]
\centering\includegraphics[width=2.5in]{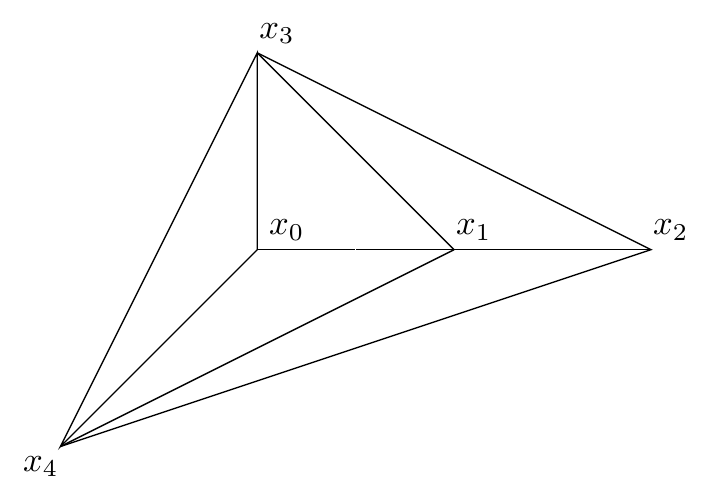}
\caption{Fan diagram of resolved $\cz$ model.}\label{fan:c3z5}
\end{figure}

Resolved $\mathbb{C}^3/\mathbb{Z}_5$ can be obtained by taking the limit $x_5=0$ in the $SU(3)$ geometry with $m=2$. The toric data of this model is listed in (\ref{c3z5charge})\footnote{One should not mix the charges $Q_i$ here and exponential of K\"{a}hler parameter $Q_i=e^{-t_i}$}. The fan diagram is illustrated in Figure \ref{fan:c3z5}. From the toric data, we can see there are two Batyrev coordinates,
\begin{equation}
z_1=\frac{x_1x_3x_4}{x_0^3}, \ \ \ z_2=\frac{x_0 x_2}{x_1^2}.
\end{equation}
The true moduli of this model are $x_0,x_1$ and the $C$ matrix is
\be
C=\left(\begin{array}{cc}-3 & 1 \\1 & -2\end{array}\right).
\ee
The genus zero free energy of this geometry is \cite{Klemm:2015iya}\footnote{In this and the following subsections, we use the notation $Q_i=\re^{t_i}$ for convenience.}
\be
F_0=\frac{1}{15}t_1^3+\frac{1}{10}t_1^2t_2+\frac{3}{10}t_1t_2^2+\frac{3}{10}t_2^3+3 Q_1-2 Q_2-{45 \over 8} Q_1^2 +4 Q_1 Q_2 -
{Q_2^2 \over 4}+\mathcal O(Q_i^3).
\ee
The genus one free energy in NS limit is
\be
F^{\rm NS}_1=-\frac{1}{12}t_1-\frac{1}{8}t_2-{7 Q_1 \over 8} +{Q_2 \over 6} + {129 Q_1^2 \over 16} -{5 Q_1 Q_2 \over 6}+{Q_2^2 \over 12}+ \mathcal O(Q_i^3),
\ee
and the genus one unrefined free energy is
\be
F^{\rm GV}_1=\frac{2}{15}t_1+\frac{3}{20}t_2+ {Q_1 \over 4} -{Q_2\over 6}-{3Q_1^2 \over 8} +{Q_1 Q_2 \over 3}-{Q_2^2 \over 12}+ \mathcal O(Q_i^3).
\ee
To check the identities (\ref{conjecture}), we need to find the correct $\mathbf{r}$ fields. The most direct method is that we could set some special value of $t_i,\hbar$ in (\ref{conjecture}), and scan all the integral vectors $\mathbf{r}(=\mathbf{B} \mod 2)$ in a region to see if the identity (\ref{conjecture}) holds. For the current case, we scan from $-5$ to $5$, and the following $\mathbf{r}$ fields make the identity (\ref{conjecture}) holds,
\be\label{rc3z5}
\begin{split}
&(-5,-4),(-5,0),(-5,4),(-3,-4),(-3,0),(-3,2),(-1,-4),(-1,-2),\\
&(-1,2),(1,-2),(1,2),(1,4),(3,-2),(3,0),(3,4),(5,-4),(5,0),(5,4).
\end{split}
\ee
Many of the above $\mathbf{r}$ fields may result in the same spectral determinant, which are defined to be equivalent. In fact, all equivalent $\mathbf{r}$ fields are generated by the shift,
\be\label{c3z5:symmetry}
\mathbf{r} \rightarrow \mathbf{r} +2\mathbf{n}\cdot C, \ \ \mathbf{n}\in \mathbb{Z}^{g}.
\ee
Under this symmetry, we obtain three classes of non-equivalent $\mathbf{r}$ fields. $\mathbf{r}=(-3,2)$ is equivalent to
$$
(-5,-4),(-3,2),(-1,-2),(1,4),(3,0),(5,-4),\dots.
$$
$\mathbf{r}=(-3,0)$ is equivalent to
$$
(-5,4),(-3,0),(-1,-4),(1,2),(3,-2),(5,4),\dots.
$$
$\mathbf{r}=(-1,2)$ is equivalent to
$$
(-5,0),(-3,-4),(-1,2),(1,-2),(3,4),(5,0),\dots.
$$
The list (\ref{rc3z5}) is exactly the union of these three classes of $\mathbf{r}$ fields, agree with the results compute in section \ref{sec:vanishingr}. The similar method can be used to find unity $\mathbf{r}$ fields, the results are listed in table \ref{fig:c3z5r}. The $\md r$ lattice is shown in Figure \ref{fig:c3z5r}, where black dots represent the unity $\md r$ fields and colored dots represent the vanishing $\md r$ fields.

\begin{table}
\begin{center}
\begin{tabular}{|c|c|}
\hline
$\mathbf{r}$ fields & $\Lambda$ \\ \hline
$\left(1,0\right)$  & $\exp{\left(-\frac{1}{10}(\ep_1+\ep_2)\right)}$\\ \cline{1-2}
$\left(-1,0\right)$ & $\exp{\left(\frac{1}{10}(\ep_1+\ep_2)\right)}$\\ \cline{1-2}
\hline
\end{tabular}
\end{center}
\caption{The non-equivalent unity $\mathbf{r}$ fields and $\Lambda$ of resolved $\mathbb{C}^3/\mathbb{Z}_5$ orbifold.}
\end{table}

\begin{figure}[htbp]
\centering\includegraphics[width=5in]{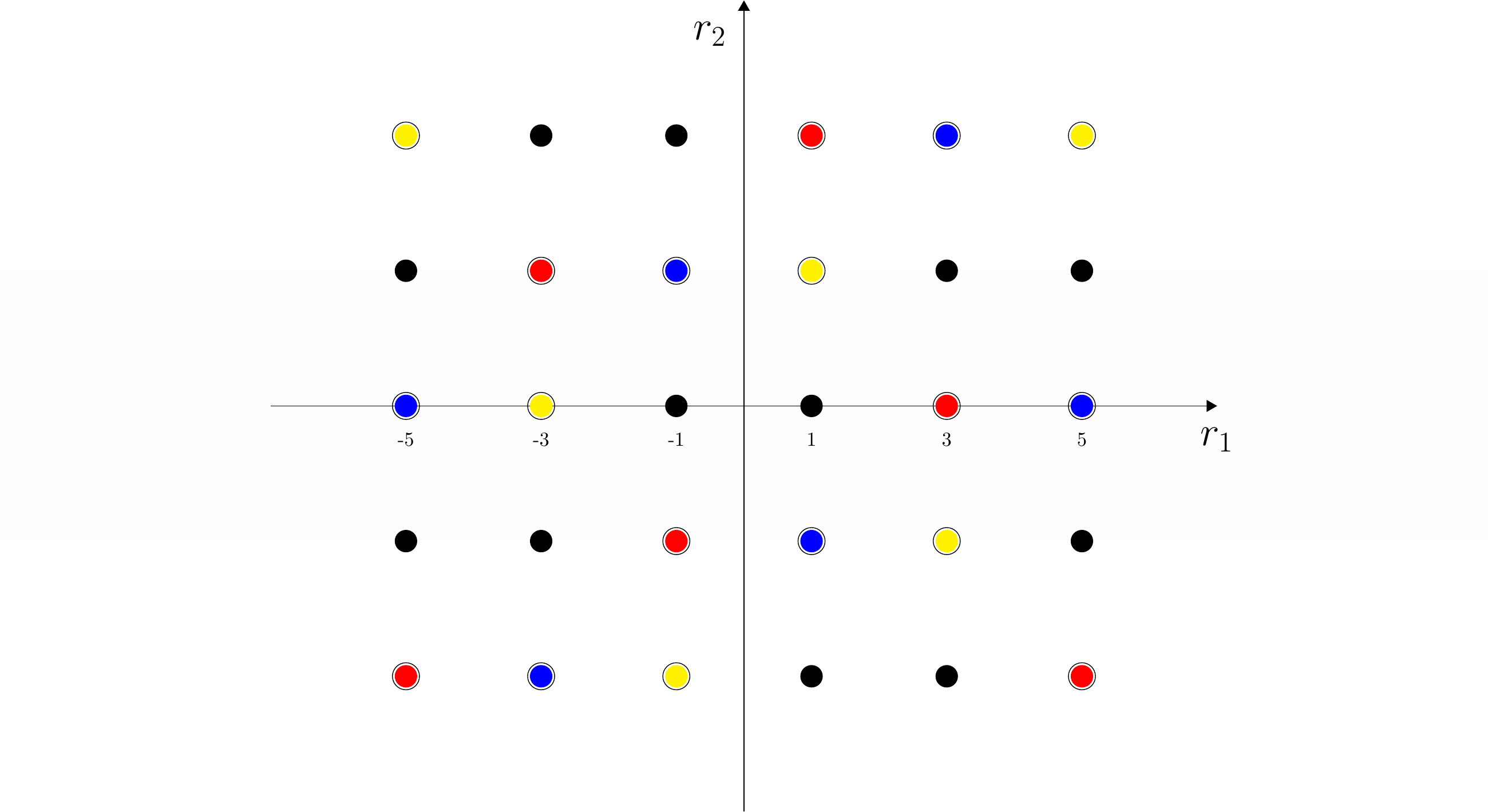}
\caption{The $\bf r$ lattice of resolved $\mathbb{C}^3/\mathbb{Z}_5$ orbifold.}\label{fig:c3z5r}
\end{figure}
%\subsection%%%%%%%%%%%%%%%%%%%%%%%%%%%%%%%%%%%%%%%%%%%%%%%%%%%%%%%%%%%%%%%%%%%%%%%%%%%%%%%%%%%%%%%%%%%%%%%%%%%%%%%%%
\section{A non-toric case: local half K3}\label{sec:nontoric}
\subsection{Local half K3 and E-string theory}\label{sec:halfk3}
All geometries we studied in the last section are toric. In this section, we want to test the blowup equations for non-toric geometries. A typical example of non-toric local Calabi-Yau is the local half K3 where the half K3 surface can be realized as nine-point blowup of $\IP^2$. It is well-known that the blowup surfaces $\mathfrak{B}_i(\IP^2)$ are non-toric for $i>3$ and not even del Pezzo for $i> 8$. Therefore it will be a strong support for the universality of the blowup equations if they can apply to local half K3.

Local half K3 Calabi-Yau can also be identified as the elliptic fibration over the total space of the bundle $\mathcal{O}(-1)\rightarrow \IP^1$. Such geometry is described by ten parameters, in which $t_b$ controls the size of base $\IP^1$ and $\tau$ controls the elliptic fiber and there are eight mass parameters $m_i,\ i=1,2,\dots,8$ which give a global $E_8$ symmetry. For details on the geometry of local half K3, see for example \cite{Gu:2017ccq}. In physics, the topological string theory on local half K3 corresponds to the E-string theory which is the simplest 6d $(1,0)$ SCFT \cite{Witten:1995gx}\cite{Ganor:1996mu}\cite{Seiberg:1996vs}. In the Ho\v rava-Witten picture of $E_8\times E_8$ heterotic string theory, E-strings can be realized by M2-branes stretched between a M5-brane and a M9-brane. The instanton partition function of refined topological string and the elliptic genus of $n$ E-strings are related by
\begin{equation}
Z_{\rm inst}(\epsilon_1,\epsilon_2,t_b,\tau,\vec{m}_{E_8}) = \sum_{n=0}^\infty Q_{b}^{n} Z^{\mathrm{E}}_{n}(\tau;\epsilon_1,\epsilon_2,\vec{m}_{E_8}),
\end{equation}
where $Q_b=e^{2\pi i t_b}$ with $t_b$ being the string tension.

There are many approaches to compute the above partition function. For example, on the topological string side, one can use the refined modular anomaly equations \cite{Huang:2013yta} to calculate the refined free energies $\mathcal{F}_{n,g,\ell}$ which are defined from
\begin{equation}
\log Z_{\rm inst}=\mathcal{F} = \sum_{n \geq 0}\sum_{g\geq 0}\sum_{l \geq 0} Q_b^{n} (\epsilon_1\epsilon_2)^{g-1}(\epsilon_1+\epsilon_2)^{2\ell}\mathcal{F}_{n,g,\ell}.
\end{equation}
The refined modular anomaly equations reads
\begin{eqnarray}
\partial_{E_2}\mathcal{F}_{n,g,\ell} &=& \frac{1}{24}\sum_{\nu = 1}^{n-1}\sum_{\gamma=0}^g \sum_{\lambda=0}^{\ell} \nu(n-\nu)\mathcal{F}_{\nu,\gamma,\lambda}\mathcal{F}_{n-\nu,g-\gamma,\ell-\lambda}\nonumber\\
&+& \frac{n(n+1)}{24}\mathcal{F}_{n,g-1,\ell}-\frac{n}{24}\mathcal{F}_{n,g,\ell-1}.\label{eq:modanomalyF}
\end{eqnarray}
However, due to the modular ambiguities, it is impossible to determine all $\mathcal{F}_{n,g,\ell}$ from anomaly equations. Other methods to determine the E-string elliptic genus include domain wall method \cite{Haghighat:2014pva}\cite{Cai:2014vka}, 2d $(0,4)$ gauge theories \cite{Kim:2014dza}, tau web \cite{Kim:2015jba} and so on. Recently another new ansatz exploiting the modular property and replying on Jacobi forms was proposed in \cite{Gu:2017ccq}. Some of the formulas for the elliptic genus of few E-strings have been proved identical. In the following, we will check the blowup equation using the formula from domain wall method.
%%%%%%%%%%%%%%%%%%%%%%%%%%%%%%%%%%%%%%%%%%%%%%%%%%%%%%%%%%%%%%%%%%%%%%%%%%%%%%%%%%%%%%%%%%%%%%%%%%%%%%%%%%%%%%%5
\subsection{Refined partition function of E-strings}\label{sec:Epart}
The polynomial part of E-string free energy has been computed  in \cite{Gu:2017ccq} for the massless case:
	\begin{equation}\label{eq:F0-pert}
	F_0(t_b, t_f, \mathbf{m}=0) = \frac{1}{2} t_b^2 t_f  + \frac{1}{2} t_b t_f^2 + \frac{1}{6}t_f^3 \ .
	\end{equation}
	\begin{equation}\label{eq:F1-pert}
	\begin{aligned}
	F_1^{\rm ST}(t_b,t_f,\mathbf{m}=0) &= -\frac{1}{2}t_b\ ,\\
	F_1^{\rm NS}(t_b,t_f,\mathbf{m}=0) &= -\frac{1}{2}t_b \ .
	\end{aligned}
	\end{equation}	
Here massless means all mass $m_i$ vanish. For our current purpose, the massless polynomial part is sufficient.

The elliptic genus for a single E-string is easy to compute: it is simply given by the torus partition function for eight bosons compactified on an internal $ E_8 $ lattice and four spacetime bosons:
\begin{equation}\label{eq:Z1E}
Z_1^{\textrm{E}} = -\left(\frac{A_1(\vec{m}_{E_8,L})}{\eta^8}\right)\frac{\eta^{2}}{\theta_1(\epsilon_1)\theta_1(\epsilon_2)},
\end{equation}
where $ A_1(\vec{m}_{E_8,L}) = \Theta_{E_8}(\tau;\vec{m}_{E_8,L}) $ is the $ E_8 $ theta function. The formulae for the $E_8$ Weyl invariant Jacobi form $A_i,\ i=1,\dots,5$ and $B_i,\ i=2,\dots,5$ can be found in for example the appendix of \cite{Huang:2013yta}.

From the domain wall method in \cite{Haghighat:2014pva}, the two E-string elliptic genus can be written as
\begin{equation}
Z_2^{\textrm{E}} = -\frac{N_{\ydiagram{2}}(\vec{m}_{E_8},\epsilon_1)/\eta^{16}}{\theta_1(\epsilon_1)\theta_1(\epsilon_2)
\theta_1(\epsilon_1-\epsilon_2)\theta_1(2\epsilon_1)\eta^{-4}}-\frac{N_{\ydiagram{1,1}}(\vec{m}_{E_8},
\epsilon_2)/\eta^{16}}{\theta_1(\epsilon_1)\theta_1(\epsilon_2)\theta_1(\epsilon_2-\epsilon_1)
\theta_1(2\epsilon_2)\eta^{-4}}.\label{eq:Z2Ansatz}
\end{equation}
Here $ N_{\ydiagram{2}} $ and $ N_{\ydiagram{1,1}}$ are some Jacobi form with weight 8 to guarantee the modular invariance of $Z_2^{\textrm{E}}$. Besides, they should also be written as linear combinations of the three level-two Weyl invariant $E_8$ Jacobi forms $ A_1^2, A_2,$ and $ B_2 $. By matching against the known free energy calculated from topological strings and exploiting the following structure theorem of weak Jacobi form: \cite{Zagierbook}\cite{Dabholkar:2012nd}\\\newline
{ \noindent\emph{The weak Jacobi forms with modular parameter $ \tau $ and elliptic parameter $ \epsilon $ of index $ k $ and even weight $ w $ form a polynomial ring which is generated by the four modular forms $ E_4(\tau), E_6(\tau), \phi_{0,1}(\epsilon,\tau), $ and $\phi_{-2,1}(\epsilon,\tau) $, where}
\[ \phi_{-2,1}(\epsilon,\tau) = -\frac{\theta_1(\epsilon;\tau)^2}{\eta^6(\tau)}\qquad \text{\emph{and}}\qquad
\phi_{0,1}(\epsilon,\tau) = 4\left[\frac{\theta_2(\epsilon;\tau)^2}{\theta_2(0;\tau)^2}+\frac{\theta_3(\epsilon;\tau)^2}{\theta_3(0;\tau)^2}+\frac{\theta_4(\epsilon;\tau)^2}{\theta_4(0;\tau)^2}\right]\]
\emph{are Jacobi forms of index 1, respectively of weight $ -2 $ and 0,}}\newline\\
the explicit expression of domain walls was determined as \cite{Haghighat:2014pva}
\begin{align}
&N_{\ydiagram{2}}(\vec{m}_{E_8,L},\epsilon_1)= \frac{1}{576}\bigg[4 A_1^2(\phi_{0,1}(\epsilon_1)^2- E_4\phi_{-2,1}(\epsilon_1)^2)\hspace{2in}\nonumber\\
&\hspace{.1in}+3A_2(E_4^2\phi_{-2,1}(\epsilon_1)^2-E_6\phi_{-2,1}(\epsilon_1)\phi_{0,1}(\epsilon_1))+5 B_2(E_6\phi_{-2,1}(\epsilon_1)^2-E_4 \phi_{-2,1}(\epsilon_1)\phi_{0,1}(\epsilon_1))\bigg],
\end{align}
The symmetry between $\ve_1$ and $\ve_2$ requires $N_{\ydiagram{1,1}}(\vec{m}_{E_8,L},\epsilon_1)=N_{\ydiagram{2}}(\vec{m}_{E_8,L},\epsilon_1)$. This gives an explicit formula for the two E-string elliptic genus.

For the elliptic genus of three E-strings, we have the following formula:
\begin{equation}
Z_3^{\textrm{E}} =  D_{\substack{\ydiagram{3}}}^{M9,L} D^{M5}_{\ydiagram{3}~\emptyset} +D_{\ydiagram{2,1}}^{M9,L}   D^{M5}_{\ydiagram{2,1}~\emptyset}
+D_{\ydiagram{1,1,1}}^{M9,L}   D^{M5}_{\ydiagram{1,1,1}~\emptyset},
\end{equation}
Based on the known results of M5 domain walls and the ansatz of M9 domain walls, this formula can be explicitly written as
\begin{equation}
\begin{split}
Z_3^{\mathrm{E}}=&\frac{N_{\ydiagram{3}}(\epsilon_1,\vec{m}_{E_8})}{\eta^{18}\theta_1(3\epsilon_1)\theta_1(2\epsilon_1-\epsilon_2)
\theta_1(2\epsilon_1)\theta_1(\epsilon_1-\epsilon_2)\theta_1(\epsilon_1)\theta_1(-\epsilon_2)}\\
+&\frac{N_{\ydiagram{2,1}}(\epsilon_1,\epsilon_2,\vec{m}_{E_8})}{\eta^{18}\theta_1(2\epsilon_1-\epsilon_2)
\theta_1(\epsilon_1-2\epsilon_2)\theta_1(\epsilon_1)^2\theta_1(-\epsilon_2)^2}\\
+&\frac{N_{\ydiagram{1,1,1}}(\epsilon_2,\vec{m}_{E_8})}{\eta^{18}\theta_1(\epsilon_1-2\epsilon_2)\theta_1(-3\epsilon_2)
\theta_1(\epsilon_1-\epsilon_2)\theta_1(-2\epsilon_2)\theta_1(\epsilon_1)\theta_1(-\epsilon_2)}.
\end{split}
\end{equation}
The formulae for $N_{\ydiagram{3}}$, $N_{\ydiagram{2,1}}$ and $N_{\ydiagram{1,1,1}}$ are extremely complicated. The explicit form can be found in \cite{Cai:2014vka}. For more three E-strings, the elliptic genus can be computed with the methods in \cite{Kim:2014dza} or \cite{Gu:2017ccq}, at least in principle. To our knowledge, there is no known explicit formula for the elliptic genus of arbitrary $n$ E-strings.
%%%%%%%%%%%%%%%%%%%%%%%%%%%%%%%%%%%%%%%%%%%%%%%%%%%%%%%%%%%%%%
\subsection{Vanishing blowup equation and Jacobi forms}\label{sec:Eblow}
The vanishing blowup equation of local half K3 has been checked in \cite{Gu:2017ccq}. Here we use a different method to study it. The $\md C$ matrix of local half K3 in the bases $(t_b,t_f,m_{i=1\dots,8})$ reads
	\begin{equation}
	\md C = (1,0,0,0,0,0,0,0,0,0 ) \ ,
	\end{equation}
and the $\md B$ field is
	\begin{equation}
	\md B \equiv (1,0,0,0,0,0,0,0,0,0)\ .
	\end{equation}
Substitute the E-string elliptic genus expansion with respect to $Q_b$ into the vanishing blowup equation, we obtain:
\be
\ba
0=&\sum_{N=-\infty}^{\infty}  (-1)^{N} \exp\(\widehat{F}_{\rm ref}\( \md t+ \epsilon_1\bR, \epsilon_1, \epsilon_2 - \epsilon_1 \) + \widehat{F}_{\rm ref}\( \md t+ \epsilon_2\bR, \epsilon_1 - \epsilon_2, \epsilon_2 \) \)\\
=&\sum_{N=-\infty}^{+\infty}(-1)^N q^{(N+\frac{1}{2})^2}\mathrm{exp}\(\sum_{n,g}^{\infty}\sum_{n_b=1}^{\infty}\(\ep_2^{2n}\delta_1^{g-1}q_1^{n_b(N+\frac{1}{2})}+\ep_1^{2n}\delta_2^{g-1}q_2^{n_b(N+\frac{1}{2})}\)(-1)^n_bQ_b^{n_b}F^{(n,g,n_b)}\)\\
&=\frac{1}{\ri}\sum_{n_b,n_b'}Z_{n_b}(-1)^{n_b+n_b'}\(\ep_1,\ep_2-\ep_1\)Z_{n_b'}\(\ep_1-\ep_2,\ep_2\)\theta_1\(n_b\ep_1+{n_b'}\ep_2\)Q_b^{n_b+n_b'}\\
&=\frac{1}{\ri}\sum_{n_b=1}^{\infty}\sum_{n=0}^{n_b}(-1)^{n_b}Z_{n_b-n}(\ep_1,\ep_2-\ep_1)Z_n(\ep_1-\ep_2,\ep_2)\theta_1((n_b-n)\ep_1+n\ep_2)Q_b^{n_b}.
\ea
\ee
Here we use the notation $q=\re^{t_f}$, $\delta_1=\ve_1(\ve_2-\ve_1)$, $\delta_2=\ve_2(\ve_1-\ve_2)$. The term $q^{(N+\frac{1}{2})^2}$ coming from the contribution of perturbative part.

A series vanishes means all its coefficients vanish. Therefore, we have
\be\label{eq:Evanish}
\sum_{n=0}^{n_b}Z_{n_b-n}(\ep_1,\ep_2-\ep_1)Z_n(\ep_1-\ep_2,\ep_2)\theta_1((n_b-n)\ep_1+n\ep_2)=0.
\ee
These formulae for arbitrary $n_b$ are the \emph{vanishing blowup equations for E-strings}.

For $n_b=1$, the vanishing blowup equation is
\be
Z_{1}(\ep_1,\ep_2-\ep_1)\theta_1(\ep_1)+Z_1(\ep_1-\ep_2,\ep_2)\theta_1(\ep_2)=0
\ee
From the formula for $Z_1$ in (\ref{eq:Z1E}), it is obvious the above equation holds.

For $n_b=2$, the vanishing blowup equation reads
\be
Z_2(\ep_1,\ep_2-\ep_1)\theta_1(2\ep_1)+Z_{1}(\ep_1,\ep_2-\ep_1)Z_1(\ep_1-\ep_2,\ep_2)\theta_1(\ep_1+\ep_2)+Z_2(\ep_1-\ep_2,\ep_2)\theta_1(2\ep_2)=0.
\ee
This formula is already highly nontrivial. We have checked this identity correct to high orders. Using the formula in \cite{Cai:2014vka}, we also checked the order $n_b=3$ vanishing blowup equation holds:
\be
Z_3(\ep_1,\ep_2-\ep_1)\theta_1(3\ep_1)+Z_{2}(\ep_1,\ep_2-\ep_1)Z_1(\ep_1-\ep_2,\ep_2)\theta_1(2\ep_1+\ep_2)+\Big(\ep_1\leftrightarrow\ep_2\Big)=0.
\ee

For arbitrary $n_b$, since there is no explicit formula for $Z_{n_b}$, we are unable to verify the vanishing blowup equation (\ref{eq:Evanish}) directly. Nevertheless, we can provide a nontrivial support here, which is all terms in the summation in (\ref{eq:Evanish}) shares the same index! The index of $Z_n$ was given in \cite{Haghighat:2014pva} as
\be
\frac{\d}{\d E_2}\log Z_n(\ep_1,\ep_2)=\frac{n}{24}\left[(\epsilon_1+\epsilon_2)^2-(n+1)\epsilon_1\epsilon_2-\left(\sum_{i=1}^8 (m^L_{E_8,i})^2\right)\right].
\ee
Then, for a fixed $n_b$ and arbitrary $n$, we have
\be
\ba
&\frac{\d}{\d E_2}\mathrm{log}{\Big(Z_{n_b-n}(\ep_1,\ep_2-\ep_1)Z_n(\ep_1-\ep_2,\ep_2)\theta_1((n_b-n)\ep_1+n\ep_2)\Big)}\\
=&\frac{1}{24}\left[n_b\epsilon_1^2+n_b\epsilon_2^2-(n_b^2+n_b)\epsilon_1\epsilon_2-2\left(\sum_{i=1}^8 (m^L_{E_8,i})^2\right)\right].
\ea
\ee
One can see that the second line of the above equation does not contain $n$. This is a strong support for the vanishment of the blowup equation (\ref{eq:Evanish}).

%We also give a conjecture on the vanishment of meromorphic Jacobi forms.
%Let $\phi:H\times\mathbb{C}^2\rightarrow\mathbb{C}$ be a Jacobi form with one modular parameter $\tau$ and two elliptic parameters $z_1,z_2$. If $\phi(\tau,z_1,z_2)$ with weight $k$ and positive-definite index quadratic form $az_1^2+bz_1z_2+cz_2^2$ satisfy the following condition
%\be
%\frac{\partial^{n_1+n_2}\phi}{\partial^{n_1}z_1\partial^{n_2}z_2}\bigg|_{z_1=z_2=0}=0,\quad\quad \forall n_1=0,1,\dots,2a,\quad n_2=0,1,\dots,2c,
%\ee
%then $\phi(z_1,z_2)\equiv 0$.

As for the unity blowup equations of E-string theory, we expect the $r$-field components on the fugacity $m_i$ will be non-zero. Thus unlike in the vanishing case, the $E_8$ symmetry will not manifest in unity blowup equations. Apparently, this makes the unity case more intricate. We would like to address this issue in a future paper \cite{GHSW}.
%%%%%%%%%%%%%%%%%%%%%%%%%%%%%%%%%%%%%%%%%%%%%%%%%%%%%%%%%%%%%%%%%%%%%%%%%%%%%%%%%
%\subsection{Remark on Jacobi forms of higher degree}
%In the recent study on refined topological string theory, one often encounters the Jacobi forms of higher degree, which have more than one elliptic parameters, see for example \cite{Haghighat:2014pva}\cite{Cai:2014vka}\cite{Gu:2017ccq}. Such objects are first systematically studied in \cite{Ziegler} at the most general setting. 
%%%%%%%%%%%%%%%%%%%%%%%%%%%%%%%%%%%%%%%%%%%%%%%%%%%%%%%%%%%%%%%%%%%%%%%%%%%%%%%%%%%%%%%%%%%%%%%%%%%%%%%%%%%%%%%%%%%%%%%%%%%%%%%%%%%
\section{Solving refined BPS invariants from blowup equations}\label{sec:solve}
In the previous sections, we check the validity of the blowup equations using the refined partition function computed from known techniques. In this section, we study the inverse problem which is to determine the refined partition function itself from blowup equations. It turns out the blowup equations are actually stronger than one may anticipate. In fact, we conjecture that assuming the knowledge of the classical information of a local Calabi-Yau, all vanishing and unity blowup equations combined together can uniquely determine the full free partition function of the refined topological string.

This conjecture is of course far beyond a strict proof. In the current paper, we would like to demonstrate some supports for the conjecture from three different aspects. First, we study the $\eq,\et$ expansion of refined free energy and the blowup equations. We count the numbers of independent unknown functions and constraint equations and find it normally to be an over-determined system. In section \ref{sec:proofconi}, we will give a strict proof that a local Calabi-Yau with genus-zero mirror curve and a single K\"{a}hler moduli whose refined free energy satisfies the unity blowup equation with $r=1$ must be the resolved conifold. In section \ref{sec:solvep2}, we go to genus-one model local $\IP^2$ and use the blowup equations to determine its refined BPS invariants to high degrees.

As we mentioned early, the blowup equations are directly related to the target physics and refined BPS formulation. They are most useful when the refined BPS invariants are the main concern. In general, to use the blowup equations to determine the refined BPS invariants, one need the following ingredients
\be
a_{ijk},b_i,\bins,C_{ij},j_L^{\rm max}(\bd),j_R^{\rm max}(\bd),
\ee
where $j_L^{\rm max},j_R^{\rm max}$ are the maximal value of $j_L,j_R$ for non-vanishing refined BPS invariants $\N$. $a_{ijk},b_i$ and $\bins$ are in the polynomial part of refined topological string, which we assume are known.  $C_{ij}$ and $j_L^{\rm max},j_R^{\rm max}$ can be obtained from the classical information of local Calabi-Yau. The exact formulae of $j_L^{\rm max},j_R^{\rm max}$ were given in \cite{Gu:2017ccq}. Suppose the base surface of a local Calabi-Yau is $S$ and the canonical class of $S$ is $K$. Given a smooth representative $C_{\kappa}$ of a class $\kappa\in H_2(S,\IZ)$, then
\be\label{eq:jjbound}
\ba
2j_L^{\rm max}=&\frac{C_{\kappa}^2+K\cdot C_{\kappa}}{2}+1,\\
2j_R^{\rm max}=&\frac{C_{\kappa}^2-K\cdot C_{\kappa}}{2}.
\ea
\ee
These formulae are very much like the Castelnuovo bound for the Gopakumar-Vafa invariants.

Once the concise formulae of $j_L^{\rm max}(\bd),j_R^{\rm max}(\bd)$ are obtained, one can begin with the smallest degree $\bd$, assume all $\N$ are unknown from $j_L=j_R=0$ to $j_L=j_L^{\rm max}(\bd),j_R=j_R^{\rm max}(\bd)$, then substitute them into the blowup equations and expand the equations with ${\bf Q}=e^{-\bt}$. The resulting constraints for the coefficients of $\bf Q$ are expected to solve all unknown $\N$. Then one move to second smallest degree, do the same and  so on. This is of course much more direct than to compute the refined free energy via topological vertex or holomorphic anomaly equations and convert it into the refined BPS formulation and then obtain the refined BPS invariants.
%%%%%%%%%%%%%%%%%%%%%%%%%%%%%%%%%%%%%%%%%%%%%%%%%%%%%%%%%%%%%%%%%%%%%%%%%%%%%%%%%%%%%%%%%%%%%%%%%%%%%%%%
\subsection{Counting the equations}\label{sec:counting}
In this section, we provide some general supports for our conjecture that all blowup equations combined together can uniquely determine the full free partition function of the refined topological string.

Obviously at total genus $g_{\rm t}=n+g$, there are $g_{\rm t}+1$ independent components of refined free energy $F_{(n,g)}$. From section \ref{sec:unity}, we see that for each unity $\md r$ field, the component equations of the unity blowup equation
\be
I_{r,s}^{\rm u}=\Lambda_{r,s}^{\rm u}
\ee
only contains those $F_{(n,g)}$ satisfying
\be
n+g\le g_{\rm t}\le [\frac{r+s}{2}]+1.
\ee
In fact, when $r+s=2g_{\rm t}-2$, the component equations can be written as
\be\label{eq:2g-2}
\sum_{g=0}^{g_{\rm t}}c_{gr} F_{(g_{\rm t}-g,g)}=f_r[F_{(n,g)}^{(i)}],
\ee
where $c_{gr}$ are some rational coefficients and $f_r$ are some functions of  $F_{(n,g)}$ and their derivatives with $n+g<g_{\rm t}$. Since there is symmetry between $r$ and $s$, we have $g_{\rm t}$ independent component equations. This gives some recursion relation to determine the $F_{(n,g)}$ of large total genus from those of small total genus. When $r+s=2g_{\rm t}-1$, the component equation can be written as
\be\label{eq:2g-1}
\sum_{g=0}^{g_{\rm t}}c'_{gr} d_{gr}F'_{(g_{\rm t}-g,g)}=f'_r[F_{(n,g)}^{(i)}],
\ee
where again $c'_{gr}$ are some rational coefficients, $d_{gr}$ are some functions computed from $\sum R_i \Theta_{\rm u}$ and $f'_r$ are some functions of  $F_{(n,g)}$ and their derivatives with $n+g<g_{\rm t}$. Here we have $g_{\rm t}$ new independent component equations for $F_{(n,g)}$. But unlike the equations in (\ref{eq:2g-2}), these are first-order differential equations. For the geometries with more than one K\"{a}hler moduli, the above differential equations can not be directly integrated.

Let us turn to vanishing blowup equations. For each vanishing $\md r$ field, the component equations of the vanishing blowup equation
\be
I_{r,s}^{\rm v}=0
\ee
only contains those $F_{(n,g)}$ satisfying
\be
n+g=g_{\rm t}\le [\frac{r+s}{2}]+1.
\ee
In fact, when $r+s=2g_{\rm t}-2$, the component equations do not result in new constraint equations for $F_{(n,g)}$ with $n+g=g_{\rm t}$, due to the leading order of vanishing blowup equations $\sum \Theta_{\rm v}=0$.  While for $r+s=2g_{\rm t}-1$, the component equation can again be written as certain first-order differential equations for $F_{(n,g)}$ with $n+g=g_{\rm t}$ like those in (\ref{eq:2g-1}). In principle, there are $g_{\rm t}$ of them for each vanishing $\bf r$ field.

Suppose the number of nonequivalent vanishing $\md r$ fields and unity $\md r$ fields are denoted as $w_{\rm v}$ and $w_{\rm u}$ respectively, then theoretically there are $g_{\rm t}w_{\rm u}$ independent algebraic equations and
\be
g_{\rm t}\(w_{\rm v}+w_{\rm u}\)
\ee
independent first-order differential equations for $F_{(n,g)}$ with $n+g=g_{\rm t}$. According to our previous conjecture, $w_{\rm v}\ge g_{\Sigma}$ and $w_{\rm u}\ge 2$. Note that the definition of non-equivalent $\bf r$ fields does not count the reflective property, therefore it is possible certain non-equivalent $\bf r$ fields may be reflective to each other. Thus effectively there may be just one unity $\bf r$ field. For the cases with more than one unity $\bf r$ fields which are not reflective to each other, obviously those algebraic equations are enough to determine all refined free energy recursively with the initial condition of $F_{(0,0)}$. For the cases with only one unity $\bf r$ field, there are $g_{\rm t}$ algebraic equations for $g_{\rm t}+1$ unknown $F_{(n,g)}$. One can only solve the algebraic equations up to an unfixed refined free energy, for example $F_{(0,g_{\rm t})}$. Then there are $g_{\rm t}\(w_{\rm v}+w_{\rm u}\)$ first-order differential equations for $F_{(0,
 g_{\rm t})}$, which are normally an overdetermined system. These $g_{\rm t}\(w_{\rm v}+w_{\rm u}\)$ first-order differential equations can also be regarded as $g_{\rm t}\(w_{\rm v}+w_{\rm u}\)$ algebraic equations for $\partial_i F_{(0,g_{\rm t})}$. It is an overdetermined system whenever $g_{\rm t}\(w_{\rm v}+w_{\rm u}\)$ is larger than the number of K\"{a}hler moduli $b$, which seems always to be true because $w_{\rm v}$ and $w_{\rm u}$ are normally much larger than $b$ and grow rapidly as $b$ grows.

The above argument roughly shows why the blowup equations are sufficient to determine all $F_{(n,g)}$ from $F_0$. Note that in \cite{Nakajima:2005fg}, the same method was actually already used to obtain the genus one free energy from the second order expansion of K-theoretic blowup equations. Our counting is of course far from a proof. The main loophole lies in that we did not prove the rank of the system of equations, which seems to be a daunting task. Therefore, a direct proof or test on the sufficiency for some examples are worthwhile.

%%%%%%%%%%%%%%%%%%%%%%%%%%%%%%%%%%%%%%%%%%%%%%%%%%%%%%%%%%%%%%%%%%%%%%%%%%%%%%%%%%%%%%%%%%%%%%%%%%%%%%%%%%%%%%%%%%
\subsection{Proof for resolved conifold}\label{sec:proofconi}
In this section, we will strictly prove that for a local Calabi-Yau with genus-zero mirror curve and a single K\"{a}hler moduli whose refined free energy satisfies the unity blowup equation with $r=1$, all refined BPS invariants must vanishes except $N_{0,0}^1$. If we further assume the classical input $N_{0,0}^1=1$, then the local Calabi-Yau must be the resolved conifold. Our proof seems to be not entirely trivial and delicately relies on the structure of refined BPS formulation. We regard this result at genus zero as a support for our conjecture that blowup equations can uniquely determine the whole refined free energy.

The relevant blowup equation was already displayed in (\ref{eq:blowupconi}), which can also be written as
\be\label{eq:blconi1}
F(Q/\sqrt{q_1},q_1,q_2/q_1)+F(\sqrt{q_2}Q,q_1/q_2,q_2)-F(Q,q_1,q_2)=0.
\ee
Since the refined free energy satisfies the refined BPS expansion, we have
\be
F(Q,q_1,q_2)=\sum_{\jj}\sum_{d=1}^{\infty}\frac{1}{w}\N f_{\jj}(q_1^w,q_2^w)Q^{wd},
\ee
where
\be
f_{\jj}(q_1,q_2)=\frac{\chi_{j_L}({q_L})\chi_{j_R}({q_R})}{\(q_1^{1/2}-q_1^{-1/2}\)\(q_2^{1/2}-q_2^{-1/2}\)}
\ee
Let us expand the equation (\ref{eq:blconi1}) with refined BPS formulation:
\be\label{eq:blconi2}
\sum_{\jj}\sum_{d=1}^{\infty}\frac{1}{w}N_{\jj}^d h_{\jj}(q_1^w,q_2^w)Q^{wd}=0,
\ee
where
\be
h_{\jj}(q_1,q_2)=f_{\jj}(q_1,q_2/q_1)/q_1^{d/2}+f_{\jj}(q_1/q_2,q_2)q_2^{d/2}-f_{\jj}(q_1,q_2).
\ee
It is easy to verify $h_{0,0}=0$ for $d=1$, which was what we did in section \ref{sec:zero11}. For $(\jj)\ne(0,0)$, the function $h$ is usually quite complicated. The key to derive $\Nd$ must vanish for all $(\jj)\ne(0,0)$ from equation (\ref{eq:blconi2}) lies on two points. One is the finiteness of spins $(\jj)$ at each degree $\bd$. We have shown the general formulae in the beginning of this section. However, our proof in the following does not rely on those formulae or any specific upper bound of $j_L$ and $j_R$, but only relies on the existence of the upper bound.

The other point of our proof is to define a strict total order on the finite plane $(\jj)$ such that certain terms in $h_{\jj}$ do not exist in $h_{j'_L,j'_R}$ if $(j'_L,j'_R)<(\jj)$. Then we can prove the vanishing of refined BPS invariants which are the coefficients of $h$ by descent method. Let us consider the first order of $Q$ in equation (\ref{eq:blconi2}). Obviously the coefficient only comes from the contribution of $w=d=1$. Now we want to prove if
\be\label{eq:blconi3}
\sum_{\jj}N_{\jj}^1 h_{\jj}(q_1,q_2)=0,
\ee
where
\be\label{eq:hd1}
h_{\jj}(q_1,q_2)=f_{\jj}(q_1,q_2/q_1)/\sqrt{q_1}+f_{\jj}(q_1/q_2,q_2)\sqrt{q_2}-f_{\jj}(q_1,q_2),
\ee
then for all $(\jj)\ne(0,0)$, $N_{\jj}^1=0$. Denote $x=\sqrt{q_1}$ and $y=\sqrt{q_2}$, we have
\be
h_{\jj}=\frac{\chi_{j_L}(\frac{x^2}{y})\chi_{j_R}(y)}{(x-\frac{1}{x})(\frac{y}{x}-\frac{x}{y})x}+\frac{\chi_{j_L}(\frac{x}{y^2})\chi_{j_R}(x)}{(\frac{x}{y}-\frac{y}{x})(y-\frac{1}{y})y}-\frac{\chi_{j_L}(\frac{x}{y})\chi_{j_R}(xy)}{(x-\frac{1}{x})(y-\frac{1}{y})}.
\ee
Let us get rid of the denominator by defining a new function
\be\label{eq:Hjj}
\ba
H_{\jj}=&\(x-\frac{1}{x}\)\(y-\frac{1}{y}\)\(\frac{y}{x}-\frac{x}{y}\)h_{\jj}\\
=&\frac{1}{x}\(\(\frac{x^2}{y}\)^{2j_L}+\cdots+\(\frac{x^2}{y}\)^{-2j_L}\)\(y^{2j_R+1}-\frac{1}{y^{2j_R+1}}\)\\
&-\frac{1}{y}\(\(\frac{x}{y^2}\)^{2j_L}+\cdots+\(\frac{x}{y^2}\)^{-2j_L}\)\(x^{2j_R+1}-\frac{1}{x^{2j_R+1}}\)\\
&-\(\frac{y}{x}-\frac{x}{y}\)\(\(\frac{x}{y}\)^{2j_L+1}-\(\frac{x}{y}\)^{-2j_L-1}\)\(\(xy\)^{2j_R+1}-\(xy\)^{-2j_R-1}\).
\ea
\ee
Now we need to have a close look at the structure of $H$. Consider the lowest-order terms of $x$ in $H_{\jj}$, they are
\be\label{eq:lowestx}
\left\{  
\begin{array}{ll}
x^{-2j_L-2j_R-1}\(y^{4j_L-1}-y^{2j_L-2j_R+1}\),&\quad \mathrm{for}\quad j_R>j_L,\\
x^{-4j_L-1}\(y^{4j_L+1}-y+y^{4j_L-1}-y^{-1}\),&\quad \mathrm{for}\quad j_R=j_L,\\
x^{-4j_L-1}\(y^{2j_L+2j_R+1}-y^{2j_L-2j_R-1}\),&\quad \mathrm{for}\quad j_R<j_L.
\end{array}\right.
\ee
Since $r=B=1$, according to the B field condition, every non-vanishing refined BPS invariant has $2j_L+2j_R\equiv 0\,(\mathrm{mod}\ 2)$. Assuming $j_R^{\rm max}\ge j_L^{\rm max}$,\footnote{Our proof actually does not rely on this assumption.} on the finite plane lattice with constraint  $2j_L+2j_R\equiv 0\,(\mathrm{mod}\ 2)$, we can define a strict total order by defining the subsequence (next smaller) of every spin pair $(\jj)$ as: 
\be\label{eq:defjj}
\left\{  
\begin{array}{ll}
(j_L+\half,j_R-\half),&\quad \mathrm{if}\quad j_R>j_L,\\
(j_L,j_R-1),&\quad \mathrm{if}\quad j_R\le j_L\ \mathrm{and}\ j_R\ge 1,\\
(2j_L-j_R^{\rm max}-1,j_R^{\rm max}),&\quad \mathrm{if}\quad  j_R<1\ \mathrm{and}\ 2j_L>j_R^{\rm max},\\
(0,2j_L),&\quad \mathrm{if}\quad  j_R<1\ \mathrm{and}\ 2j_L\le j_R^{\rm max}.\\
\end{array}\right.
\ee
This strict total order is demonstrated in Figure~\ref{fig:1216}.
\begin{figure}[h]
\begin{center}
\includegraphics[scale=0.5]{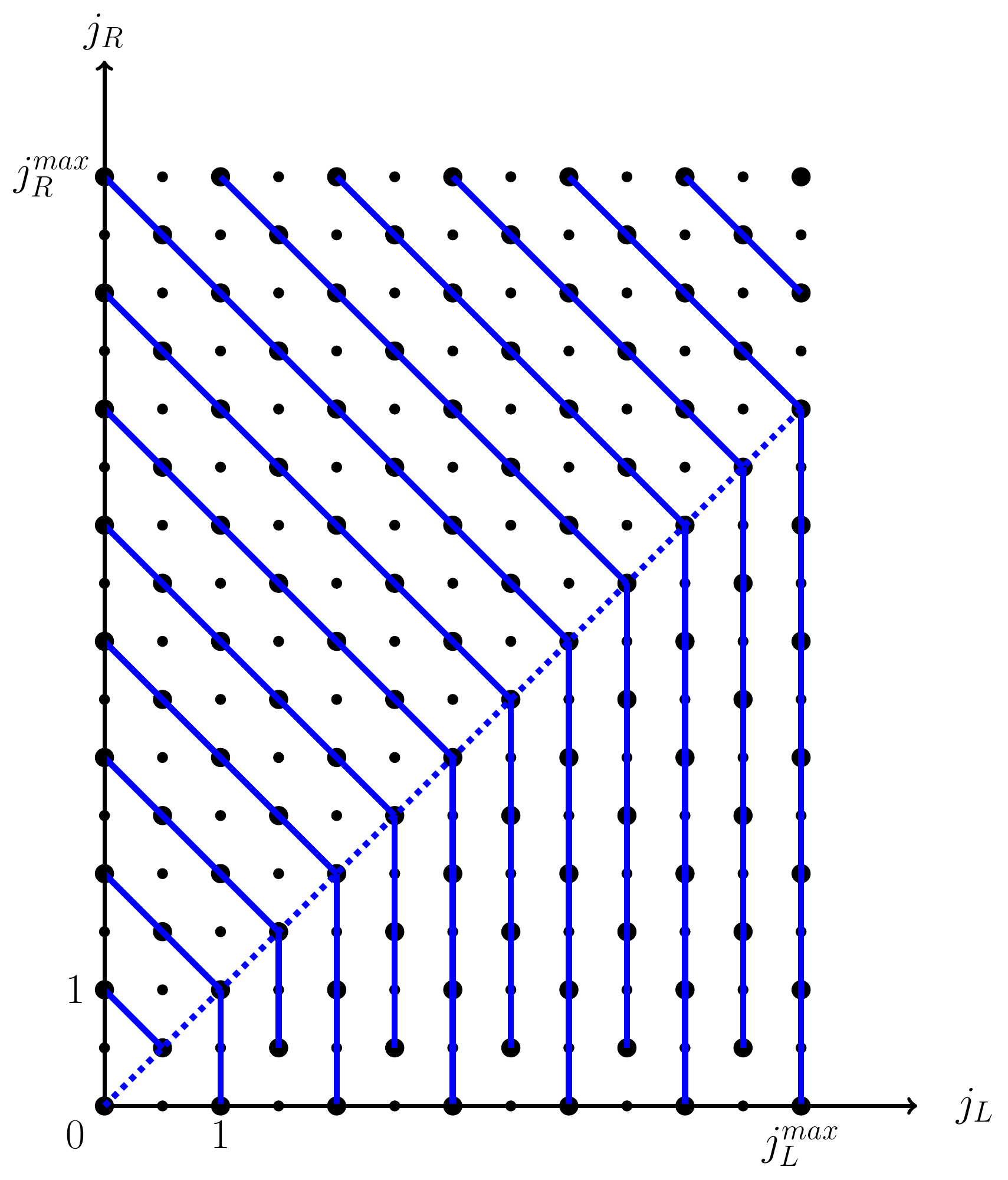}
\caption{\label{fig:1216} Strict total order for $(\jj)$ at $d=1$.}
\end{center}
\end{figure}

We call the spin-pair chain defined by the first two definitions in (\ref{eq:defjj}) as a \emph{slice}. In each slice, $(\jj)$ decreases as $j_R$ decreases and the lowest-order terms of $x$ in (\ref{eq:lowestx}) have the same order of $x$ as
 \be\label{eq:lowestorderx}
\left\{  
\begin{array}{ll}
x^{-2j_L-2j_R-1},&\quad \mathrm{if}\quad j_R+j_L>2j_L^{\rm max},\\
x^{-4J-1},&\quad \mathrm{otherwise.}
\end{array}\right.
\ee
In the otherwise case in  (\ref{eq:lowestorderx}), the slice has a `kink' and the vertical segment has $j_L=J$. The slices themselves have an strict total order compatible with the strict total order of $(\jj)$. The order of a slice can be represented by absolute value of the lowest order of $x$ in $H_{\jj}$.

Apparently $(j_L^{\rm max},j_R^{\rm max})$ is the biggest slice. Since $-2j_L^{\rm max}-2j_R^{\rm max}-1$ is lowest order of $x$ in (\ref{eq:blconi3}) and it can not be contributed from any $(\jj)$ other than $(j_L^{\rm max},j_R^{\rm max})$, then we deduce $N^1_{j_L^{\rm max},j_R^{\rm max}}$ must vanish to make the summation  (\ref{eq:blconi3}) vanish. 

Now begins the descent. We fix a slice and look at the power of $y$ in (\ref{eq:lowestx}). Certain power of $y$ in  (\ref{eq:lowestx}) of $(\jj)$ can not be contributed from the spin pairs smaller than $(\jj)$. Then from the largest to the smallest, one by one we can deduce every coefficient $\Nd$ must vanish. Then we go to the next slice and do the same. The descent finally stops at $(0,0)$ since $H_{0,0}$ always vanishes no matter what $N^1_{0,0}$ is. Therefore, only $N^1_{0,0}$ can be non-vanishing. With the classical input assuming as $N^1_{0,0}=1$, we deduce the local Calabi-Yau must be resolved conifold.

Now we amplify a bit on the above procedure. On the slice with $j_R+j_L>2j_L^{\rm max}$, apparently all ($\jj$) satisfy $j_R>j_L$. Consider the negative power of $y$ in (\ref{eq:lowestx}), which comes from $y^{2j_L-2j_R+1}$. According to the definition of the order for $j_R>j_L$ in (\ref{eq:defjj}), clearly we can see that the larger of $(\jj)$, the lower of the power of $y^{2j_L-2j_R+1}$. Then from the largest $(\jj)$ on the slice to the smallest, one by one we can rule out every $(\jj)$. Then from largest slice $(j_L^{\rm max},j_R^{\rm max})$ to the smallest slice satisfying $j_R+j_L>2j_L^{\rm max}$, one by one we can rule out every slice.

Then we come to the slices with $j_R+j_L\le 2j_L^{\rm max}$. These slices has a `kink' at $j_L=j_R=J$. Fix a slice and observe the power of $y$ in (\ref{eq:lowestx}), which is
 \be\label{eq:lowestx-ordery}
\left\{  
\begin{array}{ll}
\(y^{8J-4j_R-1}-y^{4J-4j_R+1}\),&\quad \mathrm{for}\quad J<j_R<j_R^{\rm max},\\
\(y^{4J-1}-y+y^{4J+1}-y^{-1}\),&\quad \mathrm{for}\quad J=j_R,\\
\(y^{2J+2j_R+1}-y^{2J-2j_R-1}\),&\quad \mathrm{for}\quad 0\le j_R<J.
\end{array}\right.
\ee
Assume $J>1$. For $j_R>J+\half$, again consider the negative-power term $y^{4J-4j_R+1}$ in (\ref{eq:lowestx-ordery}), we can deduce the vanishing of $N_{\jj}^1$ one by one from large $(\jj)$ to small. For $0\le j_R<J-\half$, consider the term $y^{2J-2j_R-1}$ in (\ref{eq:lowestx-ordery}), these powers of $y$ are distinct from each other and cannot be contributed from other ($\jj$) outside the range $0\le j_R<J-\half$. Therefore all of them can be ruled out. Now three only remains three contributions from $(J-\half,J+\half)$, $(J,J)$ and $(J,J-1)$, which are
\be\label{eq:threeterms}
\ba
&y^{4J-3}-y^{-1},\\
y^{4J-1}&-y+y^{4J+1}-y^{-1},\\
&y^{4J-1}-y.
\ea
\ee
Obviously they are linear independent. Thus their $N_{\jj}^1$ must vanish as well. Now we conclude all refined BPS invariants vanish on this slice. Then one by one from large slice to small slice we can prove the vanishing of all $N_{\jj}^1$.

One should be careful when $J$ becomes near zero. In the above argument, the assumption $J>1$ is necessary to make the $y^{4J-3}$ term in (\ref{eq:threeterms}) do not contact with the $y^{2J-2j_R-1}$ term in (\ref{eq:lowestx-ordery}). Luckily, for $J\le 1$ there are only a few possible $(\jj)$ so that we can easily check the linear independence of $H_{\jj}$ by hand. For $J=1$, there are four possible $(\jj)$. From large to small, they are $(0,2)$,  $(1/2,3/2)$,  $(1,1)$ and $(1,0)$. The respective terms in $H_{\jj}(x,y)$ with the lowest order of $x$ are
\be
\ba
&\frac{1}{yx^5}-\frac{1}{y^3 x^5},\\
&\ \frac{y}{x^5}-\frac{1}{y x^5},\\
\frac{y^5}{x^5}+&\frac{y^3}{x^5}-\frac{y}{x^5}-\frac{1}{y x^5},\\
&\ \frac{y^3}{x^5}-\frac{y}{x^5}.
\ea
\ee
Apparently they are linear independent. Therefore, $N^1_{0,2}=N^1_{1/2,3/2}=N^1_{1,1}=N^1_{1,0}=0$. For $J=1/2$,
\be
\ba
H_{0,1}=&-\frac{y}{x}+\frac{y^3}{x}+\frac{x}{y}-x y^3-\frac{x^3}{y}+x^3 y,\\
H_{1/2,1/2}=&\frac{y^3}{x^3}-\frac{1}{x^3 y}+\frac{1}{x y^3}-\frac{y^3}{x}-\frac{x^3}{y^3}+\frac{x^3}{y}.
\ea
\ee 
Apparently they are linear independent. Therefore, $N^1_{0,1}=N^1_{1/2,1/2}=0$. We conclude the proof for $d=1$ all refined BPS invariants $N_{\jj}^1$ vanish except $N^1_{0,0}$.

Now we move on to high degrees. For degree two of $Q$ in equation (\ref{eq:blconi2}), the possible contributions come from $w=2,d=1$ and $w=1,d=2$. Since we have proved all $N_{\jj}^1=0$ except $N_{0,0}^1$, then apparently all contributions from $w=2,d=1$ vanish. Therefore, we only need to consider the contributions from $w=1,d=2$. Now, we need to prove if
\be\label{eq:conid2}
\sum_{\jj}N_{\jj}^2 h_{\jj}(q_1,q_2)=0,
\ee
where
\be\label{eq:hd2}
h_{\jj}(q_1,q_2)=f_{\jj}(q_1,q_2/q_1)/q_1+f_{\jj}(q_1/q_2,q_2)q_2-f_{\jj}(q_1,q_2),
\ee
then for all $(\jj)$ within the upper bounds $j_L^{\rm max}$ and $j_R^{\rm max}$, $N_{\jj}^2=0$. Note here $j_L^{\rm max}$, $j_R^{\rm max}$ and $h_{\jj}$ are different with the those of degree $d=1$. However, we stick to the same symbols because they actually share parallel properties as we will see later. Note also for $d=2$, the parity of $2j_L+2j_R$ has changed. Thus all possible spin pairs satisfy $2j_L+2j_R\equiv 1\,(\mathrm{mod}\ 2)$.

Let us use induction method. Fix a degree $d$ and assume $N_{\jj}^{d'}=0$ for all $d'<d$, then all terms in blowup equation (\ref{eq:blconi2}) with degree $d$ of $Q$ contributed from $w\ne 1$ vanish. Since the blowup equation holds at $Q^d$, then the remaining contributions from $w=1$ must vanish as well:
\be\label{eq:conidd}
\sum_{\jj}N_{\jj}^d h_{\jj}(q_1,q_2)=0,
\ee
where
\be\label{eq:hdd}
h_{\jj}(q_1,q_2)=f_{\jj}(q_1,q_2/q_1)/q_1^{d/2}+f_{\jj}(q_1/q_2,q_2)q_2^{d/2}-f_{\jj}(q_1,q_2).
\ee
If we can use the above two formulae to prove all $N_{\jj}^{d}=0$, then by mathematical induction, we have proved for all $d\in\IZ$.

Now we come to the last part of our proof. To derive all $N_{\jj}^{d}=0$ from equation (\ref{eq:conidd}) for general $d>1$ is quite similar with our previous proof for $d=1$. All we need to do is to define a strict total order for the finite $(\jj)$ plane at each degree $d$. Again we define the order by defining the subsequence (next smaller) of every spin pair $(\jj)$ as
\be\label{eq:defjjd}
\left\{  
\begin{array}{ll}
(j_L+\half,j_R-\half),&\quad \mathrm{if}\quad j_R>j_L+(d-1),\\
(j_L,j_R-1),&\quad \mathrm{if}\quad j_R\le j_L+(d-1)\ \mathrm{and}\ j_R\ge 1,\\
(2j_L-j_R^{\rm max}-1+(d-1)/2,j_R^{\rm max}),&\quad \mathrm{if}\quad  j_R<1\ \mathrm{and}\ 2j_L>j_R^{\rm max}-(d-1)/2,\\
(0,2j_L+(d-1)/2),&\quad \mathrm{if}\quad  j_R<1\ \mathrm{and}\ 2j_L\le j_R^{\rm max}-(d-1)/2.\\
\end{array}\right.
\ee
It is easy to check the above definition is consistent with the $B$ field condition $2j_L+2j_R\equiv d-1\,(\mathrm{mod}\ 2)$. The order and subsequence relation for $d>1$ are demonstrated in Figure.~\ref{fig:1219}
\begin{figure}[h]
\begin{center}
\includegraphics[scale=0.5]{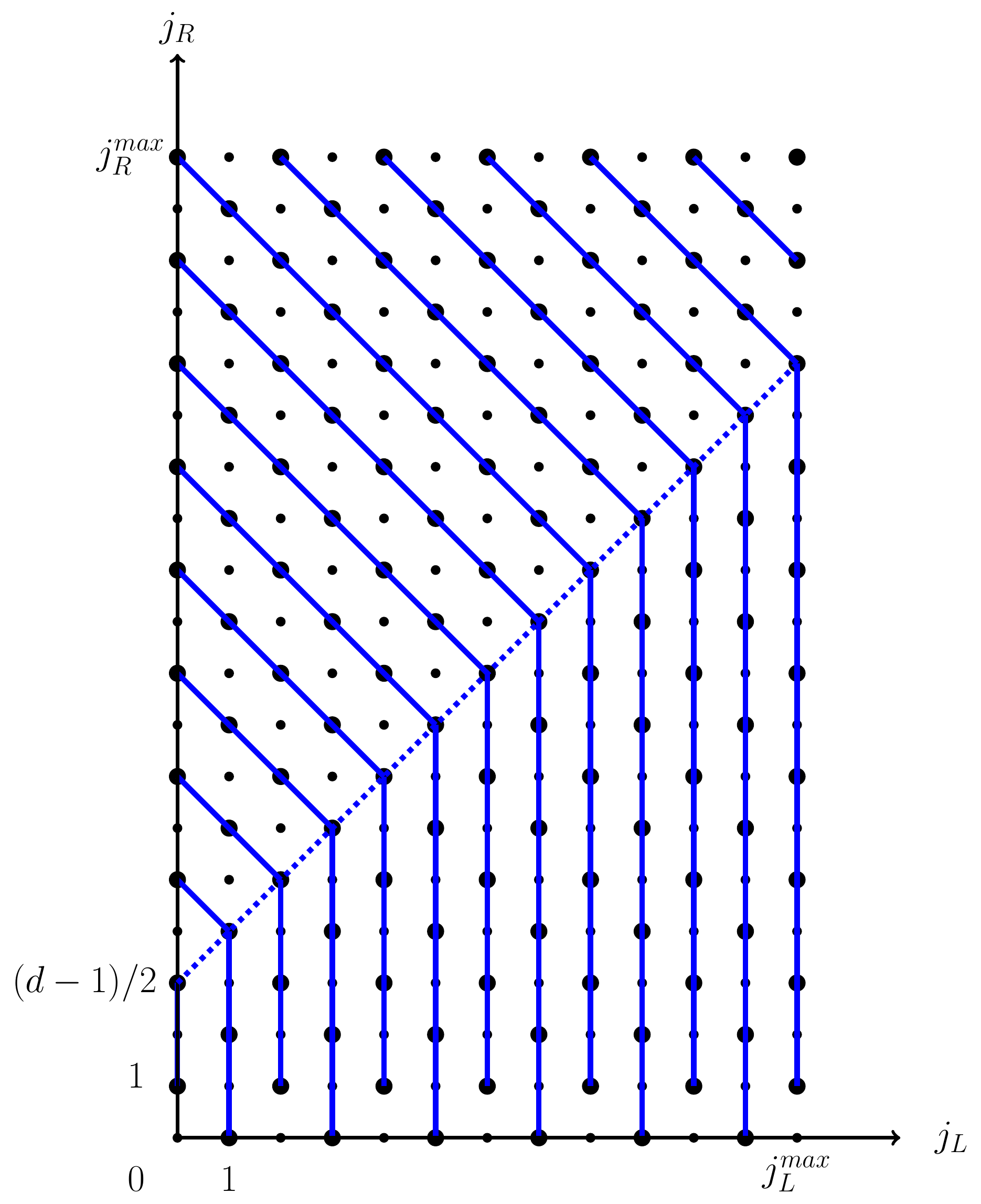}
\caption{\label{fig:1219} Strict total order for $(\jj)$ at general degree $d$.}
\end{center}
\end{figure}
They are very much similar with Figure.~\ref{fig:1216} but with a shift $(d-1)/2$ on the $j_R$ direction. Now the `kink' point of each slice lies on the line $j_{R}=j_L+(d-1)/2$. Again we consider the function $H_{\jj}$ defined as in (\ref{eq:Hjj}) with $h_{\jj}$ replacing as (\ref{eq:hdd}). On each slice, $H_{\jj}$ have the same lowest order of $x$. By analyzing the powers of $y$ in the terms with lowest order of $x$ in $H_{\jj}$, we can deduce the linear independence of all $H_{\jj}$ on the slice. Therefore, all $\Nd=0$ on the slice. Then, one slice by one slice, from large to small, we can deduce the linear independence of all $H_{\jj}$ on the whole finite plane of $(\jj)$. Therefore, all $\Nd=0$ on the finite plane. Combined with our previous result for $d=1$, we conclude that among all $\Nd$, only $N_{0,0}^1$ can be non-vanishing.

In the above proof we can only determine $N_{0,0}^1$ to be non-vanishing. It is impossible to determine the exact value of $N_{0,0}^1$ from the single unity blowup equation. This is understandable since the free energy of genus-zero models does not have polynomial part. It is interesting to consider whether the local Calabi-Yau threefolds with sole non-vanishing invariant $N_{0,0}^1=2,3,4\dots$ exist.
\subsection{A test for local $\mathbb{P}^2$}\label{sec:solvep2}
In this section, we focus on local $\IP^2$, the simplest yet  non-trivial enough local toric Calabi-Yau with genus-one mirror curve. If the blowup equations can determine the full refined BPS invariants for this geometry, it will be a strong support for arbitrary local toric geometries. In fact, we checked to high degree that for this geometry the unity blowup equation ($r=1$) itself can already determine the full refined BPS invariants. This is not very surprising from our argument in section \ref{sec:counting}. Since local $\IP^2$ has only one K\"{a}hler moduli $t$, then at each total genus $g+n$, there are $g+n+1$ independent free energy component $F_{(n,g)}$. The unity blowup equation ($r=1$) gives $g+n$ algebraic equations for $F_{(n,g)}$ at order $2(g+n)-2$ and $g+n$ first-order differential equations for $\partial_t F_{(n,g)}$ at order $2(g+n)-1$. Because there is only one moduli $t$, the first-order differential equations can be integrated to $g+n$ new algebraic equations for
  
 $F_{(n,g)}$.\footnote{The integration constants do not matter here because they do not affect the refined BPS invariants.} Since $2(g+n)\ge g+n+1$ for all $g+n\ge 1$, these normally are over-determined systems. Therefore, with the knowledge of $F_{0}$, theoretically one can recursively determine all $F_{(n,g)}$. 

Still, a concrete check in the refined BPS formulation is worthwhile, since we did not prove the rank of the $2(g+n)$ algebraic constraint equations. In the following, we explicitly show how to actually determine the refined BPS invariants of local $\IP^2$. The input $a_{ijk},b_i,\bins,C_{ij}$ are easy to computed from the traditional methods and have been shown in \ref{sec:p2}. The upper bound of $j_L^{\rm max}(d)$ and $j_R^{\rm max}(d)$ for a fixed degree $d$ can be computed using (\ref{eq:jjbound}), which are
\be
\ba
j_L^{\rm max}(d)&=\frac{(d-1)(d-2)}{4},\\
j_R^{\rm max}(d)&=\frac{d(d+3)}{4}.
\ea
\ee

Let us focus on the degree one refined BPS invariants $n^1_{\jj}$. From the above formulae, we have $j_L^{\rm max}=0$ and $j_R^{\rm max}=1$. Since for local $\IP^2$ the B field is $1$, obviously the only possible spin-pairs are $(0,0)$ and $(0,1)$. Consider the unity blowup equation with $r=1$, the perturbative contribution to the exponent is
\be
G_{\rm pert}=-\frac{n(3n+1)t}{2}-\frac{3}{2}\(n-\frac{1}{3}\)\(n+\frac{1}{6}\)\(n+\frac{2}{3}\)(\eee),
\ee
Since only degree one invariants are concerned, we only need to compute the leading and subleading expansion of the unity blowup equation, where only the terms coming from some small $n$ in the summation matter. The terms proportional to $N_{0,0}^1$ do not contribute here due to the same identity (\ref{eq:id1}) which makes unity blowup equation for resolved conifold holds. Therefore we only need to count in the terms proportional to $N_{0,1}^1$. It is also easy to see only $n=0,-1$ in the summation affects the linear order of $Q$. For $n=0$, we have
\be
G_{\rm pert}(n=0)=\frac{\eee}{18},
\ee
and
\be
\ba
G_{\rm inst}(n=0)&=N_{0,1}^1\sum_w(-1)^w\frac{Q^w}{w}\(\frac{S(q_2^{3})q_1^{-w/2}-S(q_1^{3})q_2^{-w/2}}{S(q_1)S(q_2/q_1)S(q_2)}-\frac{S(q_1^3q_2^3)}{S(q_1)S(q_2)S(q_1q_2)}\)\\
&=N_{0,1}^1\sum_w(-1)^w\frac{Q^w}{w}\(-(q_1q_2)^{w/2}\),
\ea
\ee
where we used the notation $S(q)=q^{w/2}-q^{-w/2}$. Only $w=1$ term in the $w$ summation affects the linear order of $Q$. For $n=-1$, we have
\be
G_{\rm pert}(n=-1)=-t+\frac{5}{9}(\eee).
\ee
Clearly the instanton terms from $n=-1$ result in higher order contributions. In summary, we have
\be
\ba
\sum_{n=-\infty}^{\infty}(-1)^n\re^{G_{\rm pert}+G_{\rm inst}}&=\re^{G_{\rm pert}(n=0)+G_{\rm inst}(n=0)}-\re^{G_{\rm pert}(n=-1)}+\mathcal{O}(Q^2)\\
&=(q_1q_2)^{1/18}\re^{N_{0,1}^1Q(q_1q_2)^{\frac{1}{2}}}-Q(q_1q_2)^{\frac{5}{9}}+\mathcal{O}(Q^2)\\
&=(q_1q_2)^{1/18}+N_{0,1}^1Q(q_1q_2)^{\frac{5}{9}}-Q(q_1q_2)^{\frac{5}{9}}+\mathcal{O}(Q^2)
\ea
\ee
Therefore, we obtain $N_{0,1}^1=1$. It is important to note that this quantum invariant is determined by the polynomial part of the refined topological string! This is indeed the spirit of blowup equations. This simple calculation also shows the $\Lambda$ factor here is $(q_1q_2)^{1/18}$ which is exactly we have given in section \ref{sec:p2}. As for $N_{0,0}^1$, it can be determined by the higher order $Q$ expansion of the blowup equation.

The above procedure to solve the refined BPS invariants are very straight forward and can be easily gathered into computer program. Using this method, we have checked up to degree five that the $r=1$ unity blowup equations can solve all the refined BPS invariants. 

%%%%%%%%%%%%%%%%%%%%%%%%%%%%%%%%%%%%%%%%%%%%%%%%%%%%%%%%%%
\section{Blowup equations at generic points of moduli space}\label{sec:generic}
In the above sections, we mostly study the free energy in the refined BPS expansion, which is an expansion at the large radius point of moduli space. In this section, we would like to show that at other points of the moduli space certain variant of blowup equations still holds. This shows the structure of blowup equations is general in the moduli space, just like the holomorphic anomaly equations.
%%%%%%%%%%%%%%%%%%%%%%%%%%%%%%%%%%%%%%%%%%%%%%%%%%%%%%%%%%%%%%%%%%%%%%%%%%%%%%%%%%%%%%%%%%%%%%%%%%%%%%%%%%%%%%%%%%%%%%%%%%%%%
\subsection{Modular transformation}\label{sec:modular}
Following the discussion of (\ref{subsection:modular}) we know that the blowup equations are modular invariant, this property makes it possible to write down blowup equation at conifold point and orbifold point by taking modular transformation. Especially, we write down blowup equation at conifold point with the similar form as it is in large radius point.

It is also known \cite{Sun:2016obh} that vanishing blowup equation is a compatible condition of equivalence between GHM and NS quantization. Together with the definition of $\mathbf{r}$ fields in section \ref{ch:solver}, and the proof of blowup equation in $SU(N)$ geometries\cite{Grassi:2016nnt}, we complete the proof of equivalence between GHM and NS quantization. On the other hand, from the idea of this section, we can also in principle prove the equivalence of GHM and NS quantization at half orbifold point study in \cite{Codesido:2015dia}. It is also interested to study this in detail in the future.
%%%%%%%%%%%%%%%%%%%%%%%%%%%%%%%%%%%%%%%%%%%%%%
\subsection{Conifold point}
\subsubsection{Local $\mathbb{P}^2$}
We mainly consider conifold point of this section. Our analysis should be in principle available near orbifold point, but it is unknown that how to write down a compact form of theta functions after general transformations not belong to $SL(2,\mathbb{Z})$ (but indeed keep the symplectic form, see \cite{Aganagic:2006wq} for details), we only exhibit the results for the first identity here. We now perform the modular transformations to transform the blowup equations form large radius point to the conifold point. The modular transformation is 
$$
\tau \rightarrow -\frac{1}{3\tau_c},
$$
with $2\pi i\tau_c=3\frac{\partial^2}{\partial t_c^2}F^{(0,0)}_c$. 
Using
$$
\eta( -1/3\tau_c)=\sqrt{-3i\tau_c}\eta(3\tau_c),
$$
also 
$$
\sum_n (-1)^ne^{\frac{1}{2}(n+1/6)^2 4\pi^2 /(2\pi i \tau_c)}=\sqrt{-i\tau_c}\sum_n e^{1/2 (n-1/2)^2 2\pi i \tau_c+2n\pi i(n-1/2)1/6},
$$
(\ref{first}) becomes
\be
\sum_n e^{1/2 (n-1/2)^2 2\pi i \tau_c+2n\pi i(n-1/2)1/6}=\sqrt{3}\eta(3\tau_c).
\ee
With the notation
\be
F^{(0,0)}_c=\frac{1}{12} t_c^2 \left(2 \log \left(t_c\right)-3-2 \log
   (27)\right)-\frac{t_c^3}{324}+\frac{t_c^4}{69984}+O\left(t_c^5\right)
\ee
\be\label{F10c}
F^{(1,0)}_c=\frac{1}{8}\log(3)+\frac{1}{24}\log(t_c)+\frac{7
  t_c}{432}-\frac{t_c^2}{46656}-\frac{19
   t_c^3}{314928}+O\left(t_c^4\right),
\ee
\be
F^{(0,1)}_c=\frac{1}{4}\log(3)-\frac{1}{12}\log(t_c)+\frac{5
  t_c}{216}-\frac{t_c^2}{23328}-\frac{5
   t_c^3}{157464}+O\left(t_c^4\right),
\ee
we have 
\be\label{confirst}
\sum_{n=-\infty}^{\infty}e^{\frac{1}{2}(n-1/2)^2 3\frac{\partial^2}{\partial t_c^2}F^{(0,0)}_c+2\pi i(n-1/2)*1/6}=e^{F^{(1,0)}_c-F^{(0,1)}_c}.
\ee
It can be seen that (\ref{confirst}) has exact the same structure as (\ref{first}) has, but now with 
\be\label{change}
R\rightarrow R_c=\sqrt{3} (n-1/2),\  \ (-1)^n\rightarrow e^{2\pi i(n-1/2)1/6}.
\ee
Actually, the changes of $R$ and $(-1)^n$ come from the transformation rule of the Riemann theta functions. The transformation rule of a Riemann theta function is written in section \ref{subsection:modular}, for a Riemann theta function,
\be
\Theta\left[\begin{array}{c}\bm{\alpha} \\ \bm{\beta}\end{array}\right](\bm{\tau},\bm{z})=\sum_{\bm{n}\in \mathbb{Z}^g}e^{\frac{1}{2} (\bm{n}+\bm{\alpha})\cdot\bm{\tau}\cdot(\bm{n}+\bm{\alpha})+(\bm{n}+\bm{\alpha})\cdot(\bm{z}+\bm{\beta})},
\ee
under modular transformation $\bm{\Gamma}\in Sp(2g,\mathbb{Z})$
$$
\bm{\Gamma}=\left(\bm{\begin{array}{cc} A& B \\C & D\end{array}}\right)
$$
has the transformation rule (\ref{modularrules}). The change (\ref{change}) is only a change of $\bm{\alpha,\beta}$ of theta function under a modular transformation. As we have already discussed, the whole identities consist of theta function, weight zero quasi-modular forms $F^{(n,g)}$ and their derivatives for $n+g>=2$, $\partial_t=C_{ttt}\partial_\tau$ and $\partial_t F^{(0,1)},\partial_t F^{(1,0)}$, and times to each other, add together, to get a weight 0 modular form. They all have good transformation rules, but with characteristics of the theta function change under the rule of modular transformations. So we may conclude nothing important changes, but only
\be
R\rightarrow R_c=\sqrt{3} (n-1/2),\  \ (-1)^n\rightarrow e^{2\pi i(n-1/2)1/6},
\ee
for the whole identities. For this reason, we may have a compact expression of the blowup equation at conifold point\footnote{There are constant terms appear in the exponential, which can be absorbed into $F^{(n,g)}$.}
\bdm
\sum_n\exp\left\{(F_c(\ep_1-\ep_2,\ep_2,t+i R_c \ep_2)+F_c(\ep_1,\ep_2-\ep_1,t+i R_c \ep_1)-F_c(\ep_1,\ep_2,t)+2n\pi i(n-1/2)1/6)+\frac{1}{18}(\ep_1+\ep_2)\right\}=1
\edm
where $R_c=\sqrt{3}(n-1/2)$. This has been checked up to $n+g<=4$(series expansion of $\ep_1,\ep_2$ up to order 7).

It is easy to write down the vanishing blowup equation at conifold point. Following the modular transformation rules (\ref{modularrules})
\be
\sum_R\exp\left\{(F_c(\ep_1-\ep_2,\ep_2,t+i R_c \ep_2)+F_c(\ep_1,\ep_2-\ep_1,t+i R_c \ep_1)+2n\pi i(n-1/2)1/2)\right\}=0,
\ee
where $ R_c=\sqrt{3} (n-1/2)$. We also have check it up to $n+g<=4$.

%%%%%%%%%%%%%%%%%%%%%%%%%%%%%%%%%%%%%%%%%%%%%%%%%%%%%%%%%%%%%%
\subsubsection{Local $\mathbb{P}^1\times \mathbb{P}^1$}

The basic ingredients of local $\mathbb{P}^1\times \mathbb{P}^1$ are reviewed in section \ref{sec:fn}.
The conifold "point" now is not a point, but a cycle determined by the zero loci of the discriminant, 
$$
1-8(z_1 +z_2)+16(z_1 -z_2)^2=0,
$$
we use the terminology conifold frame instead.
Choosing $z_2=z_1=z$, we may have simply one complex parameter in conifold frame
$$
z_c=1-16 z.
$$
And the K\"ahler parameter solve from the Picard-Fuchs equations is
\be
t_{c}=z_{c}+\frac{5}{8}z_{c}^2+\frac{89}{192}z_c^3+\mathcal{O}(z_{c}^4).
\ee
For later convenience, we also define $t'_c=\sqrt 2 t_c$.

Since the modular group of local $\mathbb{P}^2\times \mathbb{P}^1$ is $\Gamma(2)$, under the modular transformation 
\be\label{p1p1_modular}
\tau \rightarrow -\frac{1}{2\tau_D},
\ee
(\ref{p1p1:first}) move to conifold frame.  We can now compute the B-period from the Picard-Fuchs operators, get series expansion of $\tau_D$, but a more convenient way is to exact it from j invariant (\ref{p1p1_j}) 
\be
\tau_D=2\frac{\partial^2}{\partial {t'_c}^2} F^{(0,0)}_D=\log \left(\frac{z_c}{16}\right)+\frac{z_c}{2}+\frac{13 z_c^2}{64}+\frac{23
   z_c^3}{192}+O\left(z_c^{4}\right),
\ee
\be
F_D^{(1,0)}=-\frac{1}{6}\log(\frac{\theta_4(2\tau_D)^2}{\theta_2(2\tau_D) \theta_4(2\tau_D)})%=\frac{1}{24} \log\left(z_{c,2}\right)+\frac{z_{c,1}^2+4 z_{c,2} z_{c,1}+12 z_{c,2}^2}{192 z_{c,2}}+\mathcal{O}(z_c^3)
,
\ee
\be
F_D^{(0,1)}=-\log(\eta (2\tau_D)).
\ee

For simplicity, we only consider massless case $m=1$, in this case, $t_{c,1}=0$ and there is only one parameter. The $\mathbf{r}$ field of this case is $2(0)$ for vanishing(unity)
 blowup equation. 
Then, we write down (\ref{first}) in the conifold frame by a direct modular transformation of (\ref{p1p1:first})
\be\label{p1p1:conifoldfirst}
2^{3/8}e^{F_D^{(1,0)}-F_D^{(0,1)}}=\sum_{n=-\infty}^{\infty} e^{\frac{1}{2}(n+\frac{1}{2})^2\tau_D},
\ee
where the irrelevant coefficient $2^{3/8}$ can be absorbed into $F^{(0,1)},F^{(1,0)}$.

To check higher genus identities, we use the results in \cite{Huang:2013yta} by direct integrate holomorphic anomaly equation, e.g.
\be
\begin{split}
F^{(0,2)}_D&=-\frac{1}{240 t_c^2}-\frac{1}{1152}+\frac{53 t_c}{122880}-\frac{2221
   t_c^2}{14745600}+\frac{43 t_c^3}{1179648}-\frac{32497
   t_c^4}{9512681472}+O\left(t_c^5\right),\\
F^{(1,1)}_D&=\frac{7}{1440 t_c^2}-\frac{7}{55296}+\frac{169 t_c}{737280}-\frac{49681
   t_c^2}{176947200}+\frac{5321 t_c^3}{28311552}-\frac{819029
   t_c^4}{8153726976}+O\left(t_c^5\right),\\
F^{(2,0)}_D&=-\frac{7}{5760 t_c^2}-\frac{101}{221184}-\frac{889 t_c}{2949120}+\frac{181981
   t_c^2}{707788800}-\frac{16157 t_c^3}{113246208}+\frac{2194733
   t_c^4}{32614907904}+O\left(t_c^5\right),\\
\end{split}
\ee
to summarize, we have
\be\label{p1p1_con}
\sum_{R_c}\exp\left\{(F_c(\ep_1-\ep_2,\ep_2,t'_c+i R_c \ep_2)+F_c(\ep_1,\ep_2-\ep_1,t'_c+i R_c \ep_1)-F_c(\ep_1,\ep_2,t'_c)\right\}=1
\ee
where $R_c=\sqrt{2}(n-1/2)$. Note that here we have rescaled $t_c$ to $t'_c$ to make (\ref{p1p1_modular}) satisfied. We have checked (\ref{p1p1_con}) up to $n+g<=4$(series expansion of $\ep_1,\ep_2$ up to order 7). Since odd order of $R_c$ must come from the derivatives of $F^{(n,g)}$, half of the identities are odd order of $R_c$, we note that the blowup equation is symmetric under $R_c \rightarrow -R_c$, so half of the identities become zero trivially. Because of this reason, only one singular unity blowup equation at conifold is not enough to solve all genus free energy. It is also interesting to study modular transformation of general $\mathbf{r}$ field, and check that with all unity $\mathbf{r}$, whether we can solve all free energy or not in the future.

At the end of this section, we write down the vanishing blowup equation(without $\mathbf{r}$ field in mass parameter ) directly
\be\label{p1p1_con2}
\sum_{R_c}\exp\left\{(F_c(\ep_1-\ep_2,\ep_2,t'_c+i R_c \ep_2)+F_c(\ep_1,\ep_2-\ep_1,t'_c+i R_c \ep_1)+i\pi (n-1/2)\right\}=0,
\ee
with $R_c=\sqrt{2}(n-1/2)$. We also have checked it up to $n+g<=4$.

\subsection{Orbifold point}\label{sec:orbifold}
Topological string for local $\mathbb{P}^2$ at orbifold point is studied in \cite{Aganagic:2006wq} and also in \cite{Shen:2016}. We first have a review on its contents. Orbifold point corresponding to the point around $\frac{1}{z} \sim 0$. At this point, the mirror curve have a $Z_3$ symmetry and the geometry is simply $\mathbb{C}^3/\mathbb{Z}_3$. Define $\psi=(-\frac{1}{27z})^{1/3}$ as a natural coordinates encoding this symmetry. The period of local $\mathbb{P}^2$ can be computed from Picard-Fuchs equations. The general solutions to these Picard-Fuchs equations are hypergeometric functions, one can write down the orbifold period directly from analytic continuous of these hypergeometric functions
$$
B_{k}(\psi)=\frac{(-1)^{k\over 3}}{k}(3\psi)^k\sum_{n=0}^{\infty}\frac{\left(\left[\frac{k}{3}\right]_n\right)^3}{\prod_{i=1}^{3}\left[\frac{k+i}{3}\right]_n}\psi^{3n},
$$
for $k=1,2$ gives A-period $\sigma$ and B-period $\sigma_D$ respectively. For this choice of bases, we have the correct monodromy group at orbifold point. We can perform analytic continuation from the period at larger radius to orbifold point directly by Barnes integral method \cite{Candelas:1990rm}, and the result is \cite{Aganagic:2006wq}, 
$$
\Pi =\left(\begin{array}{ccc}
-{1\over 1-\alpha} c_2
&{\alpha\over 1-\alpha} c_1 & {1\over 3}
\\
c_2& c_1 &0 \\
0&0&1
\end{array}\right)
\left(\begin{array}{c}\sigma_D \\ \sigma\\1\end{array}\right)
$$
where 
$$
c_1= {i\over 2 \pi} {\Gamma\left({1\over 3}\right)\over
\Gamma^2\left({2\over 3}\right)}, \qquad c_2= -{i \over 2 \pi}
{\Gamma\left({2\over 3}\right)\over \Gamma^2\left({1\over 3}\right)\ }.
$$
The modular parameter at orbifold point is defined by 
\be
\tau_o=3\frac{\partial}{\partial \sigma} \sigma_D.
\ee

This "modular" transformation roughly keeps symplectic form if we add some normalization factor\cite{Aganagic:2006wq}, and the modular transformation is no longer in $SL(2,\mathbb{Z})$, but $SL(2,\mathbb{C})$\footnote{We can indeed write down a $SL(2,\mathbb{Z})$ transformation transform the large radius $\tau$ near orbifold point, we write down a blowup equation relate to the base of this transformation. However, this choice of base do not satisfy $\mathbb{Z}_3$ symmetry in an obvious way.}. So the traditional modular transformation properties are failed to be true. Because of the analytic continuation relation, we can in principle replace the modular parameter $\tau$ to a function of $\tau(\tau_o)$
\be
\tau(\tau_o)=\frac{-{1\over 1-\alpha} {c_2\over c_1} \tau_o+{\alpha\over 1-\alpha}}{{c_2\over c_1} \tau_o+1}
\ee
 to get the blowup equation at orbifold point. E.g the first identity (\ref{first:P2}) becomes
\be\label{copy_orb}
\sum_n (-1)^ne^{\frac{1}{2}(n+1/6)^2 3*2\pi i \tau(\tau_o)}=\eta(\tau(\tau_o)).
\ee
This corresponding to expand the modular form in the inner disc of moduli space. The general formula of a weight $k$ modular form $f$ expand near inner point $z=x+i y$ is given in \cite{Zagierbook} (Proposition 17)
$$
(1-w)^{-k}f\left(\frac{z-\bar{z}\omega}{1-\omega}\right)=\sum_{n=0}^{\infty} D^n f(z)\frac{(4\pi y\omega)^n}{n!},
$$
where $\partial_{k}$ act on a weight $k$ modular and preserve the modular property
$$
D^n f=\sum_{r=0}^{n} (-1)^{n-r} \left(\begin{array}{c}n \\r\end{array}\right)\frac{(k+r)_{n-r}}{(4\pi y)^{n-r}} \partial^r f.
$$
We can see that for this kind of transformations, the form of the original modular form changes, our purpose is finding a theta series like function in the orbifold point, this could be done if we could find a theta series function $g(z)=f(\tau)|_{\tau=z}$. Unfortunately, there is no systematical way of doing this. Even though, we could still try to transform it, at least partially. 

Recall that right the hand side of (\ref{copy_orb}) is $e^{F^{(1,0)}-F^{(0,1)}}$, up to constant terms, $e^{F^{(1,0)}}$ is weight $\frac{1}{2}$ and $e^{F^{(0,1)}}$ is weight 0.
With $z={\alpha\over 1-\alpha},\ \omega=-{c_2\over c_1} \tau_o$, after the transformation of (\ref{copy_orb}), we have
\be
(1+{c_2\over c_1} \tau_o)^{-\frac{1}{2}} \sum_n (-1)^n\exp\left({3\pi i(n+1/6)^2 \frac{-{1\over 1-\alpha} {c_2\over c_1} \tau_o+{\alpha\over 1-\alpha}}{{c_2\over c_1} \tau_o+1}}\right)=\eta\left(\frac{\alpha}{1-\alpha} \right)e^{F_o^{(1,0)}-F_o^{(0,1)}},
\ee
where we have set the constant mirror map of genus one free energy to be zero,
\be
\begin{split}
F_o^{(1,0)}&=\frac{\sigma ^3}{648}-\frac{\sigma ^6}{46656}+\frac{1319
   \sigma ^9}{3174474240}-\frac{10453 \sigma ^{12}}{1142810726400}+\frac{2662883
   \sigma ^{15}}{12354698200965120}+O\left(\sigma ^{17}\right),\\
F_o^{(0,1)}&=\frac{\sigma ^6}{174960}-\frac{\sigma ^9}{6298560}+\frac{13007 \sigma
   ^{12}}{3142729497600}-\frac{22951 \sigma
   ^{15}}{212134241088000}+O\left(\sigma ^{17}\right).\\
\end{split}
\ee
We can in principle repeat this process to write down higher genus identities, but since the derivatives of theta series are quasi-modular, if we write down other variables in terms of orbifold free energies $F_o^{(n,g)}$, we must add extra terms to cancel terms coming from the theta series. This become tedious and meaningless since we don't know how to write down a theta series after this $SL(2,\mathbb{C})$ transformation at current time. We hope to solve this problem in the future.
%%%%%%%%%%%%%%%%%%%%%%%%%%%%%%%%%%%%%%%%%%%%%%%%%%%%%%%%%%%%%
\section{Outlook}\label{sec:outlook}
This paper collects abundant tests on the existence, properties of blowup equations for general local Calabi-Yau and their capacity to determine the full partition function of refined topological string. Still there are many open problems which beg to be answered. The main questions are the following:
\begin{itemize}
\item How general is the blowup equations?
\item How to prove the blowup equations physically and mathematically?
\item Is the refined partition function uniquely determined by the blowup equations? How to prove?
\end{itemize}

We give some further remarks on these questions. Based on the good behaviors of toric geometries, it is plausible that the current form of blowup equations (\ref{eq:bltoge}) applies to all local toric Calabi-Yau threefolds. We have shown the toric geometry $\mathfrak{B}_3(\IP^2)$ which cannot be obtained from $X_{N,m}$ geometries satisfies the blowup equations as well. We also showed the half K3 Calabi-Yau which is $\mathfrak{B}_9(\IP^2)$ satisfied the vanishing blowup equation. This suggests the range of application of blowup equations is not necessarily toric. In fact, we conjecture the blowup equations (\ref{eq:bltoge}) exist for all local Calabi-Yau threefolds which have the refined BPS expansion (\ref{eq:F-inst}). This includes a lot of geometries that have 5d or 6d gauge theory correspondence. However, another large class of geometries, the Dijkgraaf-Vafa geometries which have direct link to matrix model ($\beta$-ensembles), may not have refined BPS formulation. Some of DV geom
 etries correspond to the 4d $N=2$ gauge theories. Since the Nekrasov partition function for 4d $N=2$ gauge theories satisfy the original blowup equations, certain form of blowup equations should also exist for DV geometries. We would like to deal with such cases in a separate publication.

The blowup equations for the refined topological string on local Calabi-Yau threefolds are rigorously defined in the mathematical sense. However, a complete proof for them seems far beyond the current reach. Due to the GNY equations for general $m\le N$, by the rigorously defined geometric engineering, the blowup equations for all $X_{N,m}$ geometries and their reductions are proved. It may be further possible to construct the blowup equations and $\md r$ fields for local Calabi-Yau on blowup surface $\widehat{S}$ from those for local $S$. This `blowup' has totally different meaning from the one in our current setting. By blowing up the toric surface of $X_{N,m}$ geometries, one can obtain a huge class of local Calabi-Yau threefolds. This is of course not a direct path. To prove the blowup equations once and for all requires a deeper understanding on the structure of the refined partition function for general local Calabi-Yau. Nevertheless, a physical proof should be possible using t
 he brane interpretation in \cite{Aganagic:2011mi}. We would like to address these issues in the future.

Since the blowup equations correspond to the target physics in topological string while refined holomorphic anomaly equations correspond to the worldsheet physics, one may expect some connections can be established between the two systems of equations. At first, one can ask whether the full anholomorphic refined free energy $\mathcal{F}(t,\bar{t})$ satisfies certain form of blowup equations. A naive thought that is simply replacing the holomorphic $F(t)$ to $\mathcal{F}(t,\bar{t})$ does not work, as can be seen from leading equation in the $\epsilon$ expansion where $F_{(0,1)}$ has anholomorphic part while $F_{(0,0)}$ and $F_{(1,0)}$ do not. The anholomorphic form of blowup equations, if exists, should involve the theory of Maass wave forms, which is very much worthwhile to explore. Besides, one can also consider in holomorphic polarization where the holomorphic anomaly equations becomes modular anomaly equations. It may be possible to draw some relations between modular anomaly equa
 tions and blowup equations. As we emphasized in the paper, unlike the anomaly equations which are differential equations with respect to the anholomorphic or non-modular generators, the blowup equations are functional equations that do not suffer from the integral ambiguities. Therefore, it seems unlikely to derive the blowup equations from the refined anomaly equations since the blowup equations contain more information at least on the appearance. However, it may be still possible to derive
\be
\frac{\d}{\d E^{IJ}}\(\sum_{\md n \in \mb Z^{g}}  (-1)^{|\md n|} \widehat{Z}_{\rm ref}\( \md t+ \epsilon_1\bR, \epsilon_1, \epsilon_2 - \epsilon_1 \) \widehat{Z}_{\rm ref}\( \md t+ \epsilon_2\bR, \epsilon_1 - \epsilon_2, \epsilon_2 \)/ \widehat{Z}_{\rm ref}\( \md t, \epsilon_1, \epsilon_2\)\)=0
\ee 
from the modular anomaly equations, where $E^{IJ}(\tau)$ are the non-modular generators for genus $g$ mirror curve defined in \cite{Klemm:2015iya} which is the high dimension analogy of Eisenstein series $E_2(\tau)$.

In section \ref{sec:vanish} and \ref{sec:unity}, we proposed some procedures to determine the $\md r$ fields and $\Lambda$ factor from the polynomial part of refined topological string. Although such procedures passed the tests in all models studied in this paper, it is worthwhile to quest for a strict proof. As for the capacity of blowup equations to determine the refined partition function, we presented supports from three different viewports in section \ref{sec:solve}.  To confirm our conjecture that the blowup equations combined together can determine the full refined partition function still requires much endeavors. From section \ref{sec:counting} we know the blowup equations are normally an over determined system, it is interesting to ask whether part of blowup equations can already determine the full refined partition function, in particular, whether only the unity blowup equations can determined the full refined partition function.\footnote{It is easy to check that only the v
 anishing blowup equations are not enough to determine the full refined partition function.} At least for resolved conifold and local $\IP^2$, we have seen in section \ref{sec:proofconi} and \ref{sec:solvep2} that the unity blowup equations are enough to determine the full refined partition function. 

The vanishing blowup equations of a local Calabi-Yau are related to the exact quantization of its mirror curve. It is intriguing to ask whether there is some physical background hidden behind the unity blowup equations. The Grassi-Hatsuda-Mari\~no conjecture connects the topological string and spectral theory by identifying the Fredholm determinant of the inverse operator of quantum mirror curve with certain infinite summation of partition function of topological string, which contains much more information than just the quantization of mirror curve. As was shown in \cite{Sun:2016obh}, the GHM conjecture actually contains many different phases labelled by the vanishing $\md r$ fields. Thus it is interesting to consider if the unity $\md r$ fields can result in certain analogies of GHM conjecture. This probably has relations with the Nekrasov-Okounkov partition \cite{Nekrasov:2003rj}. It is also intriguing to ask whether there exist some refined version of NS quantization and GHM conj
 ecture such that their equivalence is guaranteed by the complete vanishing blowup equations rather than their NS limit. In the recent paper \cite{Bershtein:2016aef}, some conjecture arising from Painleve $\tau$ function and $q$-deformed conformal blocks was shown to be quite similar to the K-theoretic blowup equations of Nekrasov partition function, which may led some light on the full refined problem, see also \cite{Bonelli:2017gdk}\cite{Bershtein:2017swf}.

In this paper, we only considered the vanishing blowup equation of E-strings (local half K3). It is natural to consider other local geometries with elliptic singularities, in particular the minimal 6d SCFTs, which have a one dimensional tensor branch and are non-Higgsable \cite{Haghighat:2014vxa}. Recently there were much progress on the computation of their refined partition function, or elliptic genus in other word \cite{Kim:2016foj}\cite{DelZotto:2016pvm}\cite{Haghighat:2017vch}\cite{Hayashi:2017jze}\cite{AGHKZ}. If the blowup equations exist for all such geometries, it would be a strong support for the universality. Besides, the elliptic genera of general 6d (1,0) SCFT are normally difficult to compute due to the lack of Lagrangian. Sometimes, one need to tackle the theories one by one. With the validity of blowup equations, it should be effortless to determine their refined BPS invariants from the classical settings. We would like to report the results in the near future \cite{H
 SW}. 

We have also studied the modular transformation of blowup equations. In particular we argue that the $\Lambda$ factor is a modular invariant. However, for $\mathbf{r}_m$ field appear in mass parameter, we do not have a clearly understanding of the $\tau$ parameter. Further study is needed in this direction. It is also interest to consider what is the meaning of vanishing blowup equations at conifold point for GHM and NS quantization condition. 

The G\"{o}ttsche-Nakajima-Yoshioka K-theoretic blowup equations only concern the 5d $N=1$ theories with gauge group $SU(r)$. It was shown in \cite{Keller:2012da} that blowup equations actually exist for gauge theories with general Lie group $G$, including exceptional group. It is interesting to consider the topological string theories corresponding to the gauge theory with gauge group other than $SU(r)$, see some recent works \cite{Hayashi:2017jze}. This again enlarges the range of application of blowup equations.

Although it is commonly believed that only local Calabi-Yau threefolds exhibit refined formulation, in the recent paper \cite{Huang:2015sta}, it was observed that certain compact elliptic Calabi-Yau threefolds like elliptic $\IP^2$ may also have a well-defined refinement. It is interesting to see if the refined partition function of those compact Calabi-Yau satisfies some blowup equations as well and if the blowup equations can determine all the GV invariants. If so, it would be a huge progress since no known method can determine the all-genus all-degree invariants of such compact Calabi-Yau.

Another possible direction is to incorporate knots and links. It is known the refined Chern-Simons invariants of a knot or link are related to the refined BPS states in the resolved conifold, which is the generalization of the LMOV invariants in refined topological string theory. In fact, it was recently realized that various types of knot invariants can be expressed in terms of characteristics of a moduli space of representation of certain quivers. In particular, the LMOV invariants of a knot can be expressed in terms of motivic Donaldson-Thomas invariants of corresponding quiver \cite{Kucharski:2017poe}\cite{Kucharski:2017ogk}. See some other recent developments on the refined BPS invariants of knot/link in \cite{Diaconescu:2017tga}\cite{Kameyama:2017ryw}. It is interesting to consider whether blowup equations exist for such circumstances and whether they are able to determine the knot invariants.

In this paper, we obtain the blowup equations for refined topological string by putting M-theory on $\mathrm{CY3}\times S^1\ltimes\widehat{\IC^2}_{\eq,\et}$. It is interesting to further consider M-theory on $\mathrm{CY3}\times S^1\ltimes S_{\eq,\et}$, where $S$ is a general toric surface. It should be possible to obtain more intricate functional equations, where the structure in \cite{Nekrasov:2003} is expected to emerge.

\section*{Acknowledgements}
We thank Giulio Bonelli, Peng Gao, Alba Grassi, Babak Haghighat, Tomoyoshi Ibukiyama, Qing-Yuan Jiang, Amir-Kian Kashani-Poor, Albrecht Klemm, Si Li, Guglielmo Lockhart, Andrei Losev, Hiraku Nakajima, Yi-Wen Pan, Yong-Bin Ruan, Alessandro Tanzini, Don Zagier, Hong Zhang, Lu-Tian Zhao and Jie Zhou for discussion. MH is supported by the ``Young Thousand People" plan by the Central Organization Department in China,  Natural Science Foundation of China grant number 11675167, and CAS Center for Excellence in Particle Physics (CCEPP). KS is grateful for the hospitality of ICTP and YMSC where part of this work was done.  XW is grateful for the hospitality of Nankai University, UIUC and YMSC where part of this work was done.
\appendix

%\section{Jacobi form of higher degrees}

%\section{Refined BPS invariants}

\end{document}